\documentclass[12pt]{article}
   \usepackage{epsfig}
      \usepackage{amsmath}
      \usepackage{amssymb}
      \usepackage{caption}
\usepackage{subcaption}
\usepackage{cite}
\usepackage{mathrsfs}

\usepackage{mylatex}

\usepackage{epstopdf}

\usepackage{amsmath}

\usepackage{setspace}
  \usepackage{graphicx}
  \usepackage{color}

 \newcommand{\be}{\begin{equation}}
\newcommand{\bea}{\begin{eqnarray}}
\newcommand{\eea}{\end{eqnarray}}
\newcommand{\beq}{\begin{equation}}
 \newcommand{\ee}{\end{equation}}

\begin{document}
  \renewcommand{\theequation}{\thesection.\arabic{equation}}

\begin{titlepage}

  \bigskip\bigskip\bigskip\bigskip

  \bigskip

\centerline{\Large \bf {Semi-Classical Analysis of the String
    Theory Cigar}} 

    \bigskip

  \begin{center}

 \bf { Daniel Louis Jafferis and Elliot Schneider}
  \bigskip \rm
\bigskip

{\it  Center for the Fundamental Laws of Nature, Harvard University, Cambridge, MA, USA}
\smallskip

\vspace{1cm}
  \end{center}

  \bigskip\bigskip

 \bigskip\bigskip
  \begin{abstract}

We study the semi-classical limit of the reflection
coefficient for the $\mrm{SL}(2,\reals)_k/\mrm{U}(1)$ CFT. For large
$k$, the CFT describes a string in a Euclidean black hole of
2-dimensional dilaton-gravity, whose target space is
a cigar with an asymptotically linear dilaton.
This sigma-model description is weakly coupled in the large $k$ limit,
and we investigate the saddle-point expansion of the functional
integral that computes the reflection coefficient.
As in the semi-classical limit of Liouville CFT studied in
\cite{Harlow:2011ny}, we find that one must complexify the
functional integral and sum over complex saddles to reproduce
the limit of the exact reflection coefficient.
Unlike Liouville, the $\mrm{SL}(2,\reals)_k/\mrm{U}(1)$ CFT
admits bound states 
that manifest as poles of the reflection coefficient. To
reproduce them in the semi-classical limit, we find that one must sum
over configurations that hit
the black hole singularity, but nevertheless contribute to the
saddle-point expansion with finite action.

 \medskip
  \noindent
\end{abstract}

  \end{titlepage}

  \tableofcontents

\section{Introduction and Overview}
\label{sec:intro}

The $\mrm{SL}(2,\reals)_k/\mrm{U}(1)$ CFT has been a subject of
great interest for almost 30 years, since it was shown in
\cite{Witten:1991yr} that for large $k$ it describes a string in
a Euclidean black hole of 2-dimensional dilaton-gravity.
Much is known about the CFT, thanks in
particular to its simple relation to the $\mrm{SL}(2,\reals)_k$
and $\mrm{SL}(2,\bbC)_k/\mrm{SU}(2)$ WZW models, which are themselves
well-studied CFTs
\cite{Dijkgraaf:1991ba,Maldacena:2000hw,Maldacena:2000kv,Maldacena:2001km,Gawedzki:1991yu,Teschner:1997ft,Teschner:1999ug,Hanany:2002ev,Giveon:1999tq}.

In recent years there has also been renewed interest in the
saddle-point expansions of functional integrals, inspired in
large part by
\cite{Witten:2010zr,Witten:2010cx,Harlow:2011ny,Balian:1978et,Voros1983}. In these and
related contexts
\cite{Cherman:2014ofa,Behtash:2015zha,Behtash:2015loa,Fujimori:2016ljw,Behtash:2018voa},
it has been shown that in order to compute the saddle-point
expansion of a functional integral one must, in general,
complexify the integral and sum over complex saddles. In
particular, in \cite{Harlow:2011ny} the saddle-point expansions
for the
two and three-point functions of Virasoro primaries $V_\alpha(z, \bar
z)$ in Liouville CFT were studied for general complex values of
$\alpha$. By comparing to the exact correlation functions, known from
\cite{Dorn:1994xn,Zamolodchikov:1995aa}, the authors of
\cite{Harlow:2011ny} identified the saddles that
contribute to the corresponding functional integrals. Even for
real values of $\alpha$, the functional integral receives
contributions from complex saddle-points.\footnote{Except for a special
range of $\alpha$'s in the case of the three-point function.}

In this paper, we similarly investigate the 
saddle-point expansion of the two-point function for the
$\mrm{SL}(2,\reals)_k/\mrm{U}(1)$ CFT. The exact answer is again
known
\cite{Teschner:1997ft,Teschner:1999ug,Giveon:1999tq,Maldacena:2001km},
and by comparing to its semi-classical limit we identify the
saddles that contribute to the functional integral. We again
find that one must sum over complex saddles to reproduce the
known result. In fact, as has long been known in Liouville CFT
\cite{Seiberg:1990eb} and is also the case in the
$\mrm{SL}(2,\reals)_k/\mrm{U}(1)$ CFT, the functional integral
over real fields for the two-point function is
divergent.\footnote{As is the partition function.} Instead, the
functional integral for these and related (asymptotic)
linear-dilaton backgrounds should in general be defined by an
integral over a contour in complexified field space
\cite{Harlow:2011ny}. By identifying the complex saddles that
contribute to the functional integral, one may in fact define the
appropriate integration cycle by the sum of steepest-descent
contours attached to the contributing saddles
\cite{Harlow:2011ny}.

The necessity of complexification is of course encountered
already in finite-dimensional integrals. For example, when evaluating
the asymptotic expansion of a real integral
$\int_{\cC=\reals} \diff X 
\,e^{-k \til S[X]}$, one typically continues $\til S[X]$ to a
holomorphic function 
on the complex $X$-plane, identifies its saddle-points $\til
S'[X_n] = 0$, constructs the steepest-descent contours $\cC_n$
attached 
to each saddle, and deforms the original integration contour
into the sum of steepest-descent contours $\cC=\sum_{n \in D}
\cC_n$ that is 
Cauchy-equivalent to the original contour. In the $k\to \infty$
limit, the integral along a steepest-descent contour $\cC_n$ is
dominated by the contribution from its saddle, $e^{-k
  \til S[X_n]}$. Thus, in such favorable circumstances the asymptotic
expansion of the original integral is given by the sum
$\sum_{n \in D} e^{-k \til S[X_n]}$ of contributions from the subset of
saddles that lie on the deformed integration contour.

One may apply analogous methods to extract the
asymptotic expansions of functional integrals
\cite{Witten:2010zr,Witten:2010cx,Harlow:2011ny, Balian:1978et,Voros1983}. The main
complication in the infinite-dimensional case is that it is
challenging to derive from first principles which sum of
steepest-descent contours is equivalent to the original
contour, and therefore which subset of saddles one should sum
over in computing the asymptotic expansion
\cite{Witten:2010cx}. As cited above, however, 
since the exact result for the $\mrm{SL}(2,\reals)_k/\mrm{U}(1)$
CFT is known, we may take its semi-classical limit and identify
the set of saddles that reproduce it.

\begin{figure}[t]
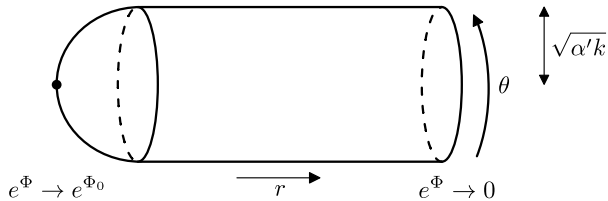

  \centering
  \ig{0cm}{width=\linewidth/2}{cigar_pic}
  \caption{\footnotesize\bd{The Cigar Background.} The cigar sigma-model is a
  weakly-coupled Lagrangian description of the
  $\mrm{SL}(2,\reals)_k/\mrm{U}(1)$ CFT when $k$ is large.
  For large $r$, the geometry is a cylinder of radius $\sqrt{\alpha'
    k}$, and as $r \to 0$ the cylinder smoothly caps off. The
  dilaton is a monotonically decreasing function of $r$. Its
  maximal value  $\Phi_0$ is attained at the tip, and at large
  $r$ it falls off linearly as $-r$. Although a string in the
  weak-coupling region appears to be able to wind around the
  cylinder, there is no conserved topological charge because the
  string can unwind at the tip.}  
  \label{fig:cigar}
\end{figure}

The $\mrm{SL}(2,\reals)_k$ WZW model describes a string in
$\mrm{AdS}_3 \simeq \mrm{SL}(2,\reals)$, where $k =
l_\mrm{AdS}^2/l_s^2$ sets the $\mrm{AdS}$ length. In global
coordinates, $\mrm{AdS}_3$ is a solid cylinder with Lorentzian
time running along its length. The
$\mrm{SL}(2,\reals)_k/\mrm{U}(1)$ coset gauges the time
translation isometry, yielding a unitary CFT. At large $k$, it
admits a weakly-coupled Lagrangian 
description given by a sigma-model into a cigar-shaped geometry
with an asymptotically linear dilaton, pictured in
Fig. \ref{fig:cigar} \cite{Witten:1991yr}. With the compact
coordinate  $\theta \sim \theta+ 2\pi$ interpreted as Euclidean
time, one obtains a two-sided black hole when the geometry is
continued to Lorentzian signature \cite{Witten:1991yr}, with its
horizon at $r=0$ where the $\theta$ circle
shrinks.

The Virasoro primaries $\cO_{jnw}$ of the CFT are labeled by
three quantum numbers (c.f. Eqn. \ref{eqn:coset-spectrum}). $n$ and $w$ are
integers, and  $j$ is a complex number that is constrained in
the normalizable spectrum of the CFT, though we consider its
analytic continuation to general complex values.
In the asymptotic $r \to \infty$ region, the cigar approaches a
free $\text{linear-dilaton}\times S^1$ background, and the
abstract primaries may be expanded in $\text{linear-dilaton}\times
S^1$ primaries \cite{Dijkgraaf:1991ba}:
\begin{align}
  \label{eqn:intro-limit}
  \cO_{jnw}\overright{r \to \infty}
  \lp e^{-2(1-j) r} + R(j,n,w) e^{-2jr} \rp
  e^{i \lp n - k w \rp  \theta_\mrm{L} + i \lp n + k w \rp\theta_\mrm{R}}.
\end{align}
$\theta_\mrm{L}(z)$ and $\theta_\mrm{R}(\bar z)$ are the
chiral components of the asymptotic circle, $n$ is the momentum
number around the circle, and $w$ is (minus) the winding
number. $j$ and its reflection $1-j$ are meanwhile momenta
along the asymptotic linear-dilaton direction.
The reflection coefficient $R(j,n,w)$, which is known exactly
(c.f. Eqn. \ref{eqn:r}) 
\cite{Teschner:1999ug,Giveon:1999tq,Maldacena:2001km},
is the amplitude for a string sent in from the asymptotic region to
reflect in the interior and return to infinity. 

The normalization chosen in Eqn. \ref{eqn:intro-limit} is not
canonical, and the two-point function of $\cO_{jnw}$ and
$\cO_{j,-n,-w}$ is proportional to $R(j,n,w)$. It is $R$
that we wish to compute by a saddle-point expansion. 
As an abstract CFT quantity, it
characterizes a redundancy in the space of CFT operators
$\cO_{jnw}$ when analytically continued to the complex
$j$-plane; operators labeled by $j$ and $1-j$ are identical, up
to rescaling by the reflection coefficient.

The redundancy under $j \to 1-j$ is a reflection about $j =
\frac{1}{2}$. To avoid double-counting operators, one restricts
the domain to $\mrm{Re}(j) > \frac{1}{2}$ or $j \in \frac{1}{2}
+ i \reals_+$. This is the $\mrm{SL}(2,\reals)_k/\mrm{U}(1)$
version of the Seiberg bound for Liouville CFT
\cite{Seiberg:1990eb}. The set $j \in\frac{1}{2} + i \reals_+$
corresponds  to delta-function normalizable scattering states on the
cigar \cite{Dijkgraaf:1991ba}, analogous to the spectrum of
Liouville. When $\mrm{Re}(j)>0$, on the other hand,
$e^{-2(1-j)r}$ is exponentially  
dominant over $e^{-2jr}$ in Eqn. \ref{eqn:intro-limit}, and the
wavefunction of the associated state generically diverges in the
asymptotic region.\footnote{Due to the background-charge
  contribution from the dilaton, the radial wavefunction at large $r$
  differs from Eqn. \ref{eqn:intro-limit} by an additional
  factor of $e^r$,
  \begin{align}
    \Psi_{jnw}(r)\overright{r \to \infty}
    e^{-2\lp\frac{1}{2}-j\rp r} + R(j,n,w) e^{-2\lp j-
    \frac{1}{2} \rp r},
  \end{align}
  as reviewed in Secs. \ref{sec:asymptotic-conditions} and
  \ref{sec:cigar}.}

At special real values of $j_N>\frac{1}{2}$, however, $R(j_N,n,w)$
has simple poles. Then it is the otherwise sub-leading term
$e^{-2j_Nr}$ that dominates in the asymptotic region, and one
obtains normalizable bound states defined as the residue of
$\cO_{jnw}$ as $j \to j_N$
\cite{Dijkgraaf:1991ba,Aharony:2004xn}. This discrete 
spectrum of bound states is 
a remarkable feature of the $\mrm{SL}(2,\reals)_k/\mrm{U}(1)$
CFT that has no analog in Liouville.
To reproduce these poles of the reflection coefficient in the
saddle-point expansion, we find that we must sum 
over configurations that hit the singularity of the black
hole in the complexified target space.\footnote{Similar complex
  contours hitting the black hole singularity were considered in
  Hartle and Hawking's path integral derivation of Hawking radiation
  \cite{Hartle:1976tp}.}

We are interested in the reflection coefficient for ``heavy''
operators, for which $j \equiv \frac{k \eta}{2}$ is order $k$,
and therefore contributes at the same order as the leading terms
in the action in the $k \to \infty$ limit. 
For simplicity, we restrict our attention to the
pure-winding sector where $n=0$. In this sector,
we argue that the only
classical solution for $\theta(\rho,\phi)$ that contributes to
the saddle-point expansion is the simplest winding solution
$\theta = - w \phi$, where $\rho$ and $\phi \sim \phi + 2\pi$
are Euclidean worldsheet cylinder coordinates. Then we show that
the integral over $r(\rho,\phi)$ reduces to a quantum
mechanics problem in the potential $V(r) = -\frac{1}{2} w^2
\sech^2(r)$, with action (c.f. Eqn. \ref{eqn:qm-action})
\begin{align}
  \til S[r]=\int\limits_{-L}^L \diff \rho\,
  \bigg(
  \frac{1}{2} \lp \td{r}{\rho} \rp^2
  + V(r)
  \bigg)
  -\eta(r(L)+r(-L))+ \eta^2L.
\end{align}
Here $\rho \in [-L,L]$ is an interval of length $2L$, which
is taken to infinity. The action is defined by this
limiting procedure so that the boundary terms $-\eta (r(L) +
r(-L))$ insert in the far past and future 
the operators with $ j = \frac{k\eta}{2}$ whose
two-point function we wish to compute
\cite{Zamolodchikov:1995aa,Harlow:2011ny}. The last term $\eta^2
L$ is a counterterm that renders the on-shell action finite in
the $L \to \infty$ limit. Using the energy conservation
equation in the inverted-potential $-V(r)$, the on-shell action
may  be written as a contour integral in the $r$-plane:
\begin{align}
  \label{eqn:intro-contour}
  \til S[r] = \int_\mathscr{C} \diff r\, \sqrt{\eta^2 + 2 V(r)}
  - \eta \lp r(L) + r(-L) \rp.
\end{align}

This quantum mechanics may in fact be solved exactly, as we
review in Appendix \ref{sec:cigar-qm}. There we show that the
semi-classical limit of the exact quantum mechanics reflection
coefficient reproduces the semi-classical limit of the exact CFT
reflection coefficient at order $e^k$
(c.f. Eqn. \ref{eqn:half-semi}). The two differ beginning at
order one.
Thus, the saddle-point
expansion of the reflection coefficient of the CFT in a
pure-winding sector reduces to a saddle-point
expansion of a quantum mechanics path integral.\footnote{The
  same reduction to quantum mechanics occurs in the saddle-point
  expansion of the Liouville reflection coefficient, though the
  calculation is presented from a slightly different perspective in
  \cite{Harlow:2011ny}.} Note that the quantum mechanics is
defined on a half-line, the radial cigar coordinate being
non-negative. Its reflection coefficient may be obtained from the
difference of the reflection and transmission coefficients for
the quantum mechanics on the full-line.

Our task therefore reduces to computing the 
saddle-point expansions of the reflection and transmission
coefficients for the complexified quantum mechanics on
the full $r$-plane. As expected, one must sum over complex
solutions, even when $\eta$ is real.\footnote{Except for the
  transmission coefficient when $\eta > w$, as discussed in Sec.
  \ref{sec:complex-t}. In that case a single real
  saddle is sufficient, corresponding to a particle that rolls over the
  inverted-potential hill $-V(r)$. For $\eta < w$
  there is likewise a real saddle of the reflection coefficient
  corresponding to a particle that rolls partway up the hill and
  then rolls back to infinity. But in that case one must also sum over infinitely
  many complex saddles to reproduce the limit of the exact
  reflection   coefficient.} 
Rather more surprising, however, is that one must also sum over
singular configurations that hit the poles of $V(r)$. The
potential has double poles at $r = \frac{\pi i}{2} + \pi i
\bbZ$, which coincide with the physical singularities of the
Lorentzian black hole in the continued geometry. Though
singular, we argue that these trajectories contribute to the
functional integral with finite action. In the neighborhood of a
pole $z = r - \frac{\pi i}{2}$, the equation of motion is
$\td{z}{\rho} \propto \frac{1}{z}$
(c.f. Eqn. \ref{eqn:singular-eom}). The speed-squared of the
solution has a $\frac{1}{\rho}$ singularity at the pole, which
is integrable up to an ambiguity in its imaginary part by $2\pi
i \bbZ$. The ambiguity amounts to the choice of deformation of
the contour $\mathscr{C}$ in Eqn. \ref{eqn:intro-contour} along
which one evaluates the on-shell action to avoid the pole.
Singular
saddles were similarly necessary to understand the three-point
function in Liouville in \cite{Harlow:2011ny}, and in related
contexts in \cite{Behtash:2015loa}.

Thus, one obtains families of saddles with common real part, but
whose imaginary parts differ by integer multiples. 
With these
ingredients, we are able to construct contours $\mathscr{C}$
for which the saddle-point expansions reproduce the
semi-classical limits of the exact results. The full answer should be a sum over the Borel resummation of the perturbative expansions around each of these saddles. The agreement we find further demonstrates that the 1-loop determinants around all of the contributing saddles are equal\footnote{Potentially up to sign.} so that their sum with equal weights is not corrected at leading order in $k$. This is not unexpected given that all of the saddles we discuss are related by varying an impact parameter at infinity, and we note that it also applies to the singular saddles.

The outline of the remainder of the paper is as follows. In
Sec. \ref{sec:asymptotic-conditions} we review the formulation
of asymptotic conditions in linear-dilaton theories, by which
operator insertions in the functional integral may be described
by cutting out the neighborhood of the insertion and adding an
appropriate boundary term to the action. In
Sec. \ref{sec:cigar} we review the cigar sigma-model background
that describes the $\mrm{SL}(2,\reals)_k/\mrm{U}(1)$ CFT at
large $k$, the operator spectrum of the CFT,
and the associated asymptotic conditions that describe operator
insertions in the cigar background. In
Sec. \ref{sec:semi-classical} we come to the main calculation of
the paper, where we show that the large $k$ limit of the exact
reflection coefficient of winding operators in 
the $\mrm{SL}(2,\reals)_k/\mrm{U}(1)$
CFT may be reproduced by a saddle-point expansion 
in complexified field space. Finally, in Sec. \ref{sec:sL} we
discuss the $k \to 2$ limit of the
$\mrm{SL}(2,\reals)_k/\mrm{U}(1)$  CFT, which according to the
FZZ duality admits a dual Lagrangian description given by the
sine-Liouville background. We argue that the saddle-point
expansion in this limit is again given by a sum over complex
cycles, and we reproduce the poles of the reflection coefficient
in this limit. Appendix \ref{sec:cigar-qm} reviews the exact solution
of the quantum mechanics that describes the pure-winding sector
of the cigar, which is relevant to the calculation of the
saddle-point expansion in Sec. \ref{sec:semi-classical}.

\section{Asymptotic Conditions in Linear-Dilaton Backgrounds}
\label{sec:asymptotic-conditions}

Before coming to the $\mrm{SL}(2,\reals)_k/\mrm{U}(1)$ CFT, in
this section we review some 
aspects of linear-dilaton theories that will be important in
what follows, especially the
formulation of ``asymptotic conditions.''
These provide a convenient description of 
operator insertions in the functional integral via boundary
modifications of the action.\footnote{This section is included
  to establish conventions and keep the paper
  self-contained. Experienced readers may wish to go directly to
Sec. \ref{sec:cigar}.}

\subsection{Asymptotic Conditions in Free Theory}
\label{sec:asympt-cond-free}

A local operator insertion in a functional integral produces a
delta-function source in the 
equations of motion, which, semi-classically, requires that the
saddles  behave as the associated
Green function in the neighborhood of the insertion. The
resulting saddles are singular at the insertion point. 
With an asymptotic condition, this singular behavior is
regulated by cutting out the neighborhood of the insertion from
the worldsheet and introducing an appropriate boundary
action there. The boundary term is chosen such that 
the boundary equations of motion impose the required Green
function behavior. In the limit
that the cut-out neighborhood shrinks away, the functional
integral defined by the action with boundary reproduces the
functional integral with the operator insertion.

To understand how to apply this procedure in practice, consider first the
free theory of a non-compact boson $X(z,\bar z)$ with a
linear-dilaton\footnote{We take $Q$ 
real and positive, such that the effective 
string coupling $e^{\Phi(X)}$ decays at $X \to \infty$ and diverges
at $X \to - \infty$, which we refer to as the weak and strong
coupling regions.}
$\Phi(X) = - QX$.  The action in locally flat complex coordinates is
\begin{align}
  \label{eqn:X-action}
  S  = \frac{1}{2\pi \alpha'} \int \diff^2z\, \partial X
  \bpartial X + \cdots,
\end{align}
up to boundary terms due to the dilaton, which
we will account for shortly. The Virasoro
primaries may be written as
$V_\alpha(z,\bar z) = e^{-2\alpha X(z,\bar z)}$,
where we will consider the analytic continuation of $\alpha$ to
general complex values. They are scalars, of conformal weights
\begin{align}
  h_\alpha = \bar h_\alpha = \alpha' \alpha (Q - \alpha)
\end{align}
with respect to the holomorphic stress tensor
\begin{align}
  T(z) = -\frac{1}{\alpha'} (\partial X)^2 - Q \partial^2 X
\end{align}
and its anti-holomorphic counterpart. The stress tensor
satisfies the Virasoro algebra with central charge $c_X = 1 + 6
\alpha' Q^2$.

Note that the conformal weights are symmetric under reflection about
$\alpha = \frac{Q}{2}$: $h_{\alpha} = h_{Q - \alpha}$. In the
free theory, $\alpha$ and $Q-\alpha$ label
independent operators, though in the interacting theories of
interest they will in fact correspond to the same operator.
Note also that the weights are real when $\alpha \in \reals$ or
$\alpha \in \frac{Q}{2} + i \reals$, which are referred to as
the real and complex branches of operators. On the complex
branch, $h_\alpha = \alpha' |\alpha|^2 \geq \alpha'
\frac{Q^2}{4}$ is always positive. On the real branch, the weight is only
positive in the window $0 < \alpha < Q$, its maximal value
coinciding with the minimal weight on the complex branch.

Inserting $V_\alpha(z',\bar z')$ in the functional integral,
\begin{align}
  \int D X\, &e^{-S } V_\alpha(z', \bar z')\\
  &= \int DX\, \exp \lc
  -\frac{1}{2\pi \alpha'} \int \diff^2z\, \lp \partial X \bpartial X
  +4\pi\alpha' \alpha \delta(z-z',\bar z-\bar z')X(z,\bar z) \rp\rc\nt,
\end{align}
introduces a source term in the bulk equation of motion, 
\begin{align}
    \label{eqn:source}
  \partial \bpartial X = 2\pi \alpha' \alpha \delta(z-z',\bar z
  - \bar z'). 
\end{align}
Recalling the Green function for the 2-dimensional
wave equation, $\partial \bpartial \log (z \bar z) = 2\pi 
\delta(z,\bar z)$,
we find that in the neighborhood of the insertion point on the worldsheet
the solution of the equation of motion is
\begin{align}
  \label{eqn:free-green}
  X(z,\bar z) \overright{|z-z'|\to 0} 2\alpha' \alpha \log|z -  z' | + \cO(1),
\end{align}
or $X(\rho,\phi) \overright{\rho \to -\infty} 2 \alpha' \alpha \rho
+ \cO(1)$ 
in local cylinder coordinates $z - z'  \equiv e^{\rho + i
  \phi}$. Thus, the operator insertion requires that the
solution is asymptotically linear in $\rho$, with $\partial_\rho
X \to 2\alpha' \alpha$ as $\rho \to -\infty$. 

Let us therefore
cut out a small disk $d_\vep$ of radius $|z-z'|=\vep$ surrounding the
insertion point and deform the action by a boundary term
\cite{Zamolodchikov:1995aa,Harlow:2011ny}: 
\begin{align}
  \label{eqn:boundary-term}
  S (\vep) = S  + 2\alpha \int_{\partial d_\vep} \frac{\diff
  \phi}{2\pi} \, X -2\alpha' \alpha^2 \log(\vep), 
\end{align}
where $\diff \phi = \frac{1}{2i} \lp \frac{\diff z}{z-z'} - \frac{\diff
\bar z}{\bar z - \bar z'} \rp$.
Then the boundary variation $-\frac{1}{2\pi \alpha'} \int_{\partial
  d_\vep} \diff \phi\, \delta X\, \partial_\rho X$ of $S $
and the variation of the boundary term $2\alpha \int_{\partial
  d_\vep}\frac{\diff \phi}{2\pi} \delta X$
yield the desired boundary equation of motion,
\begin{align}
  \partial_\rho X\big|_{\rho=\log(\vep)} = 2 \alpha' \alpha. 
\end{align}
In the limit $\vep \to 0$, one expects the functional integral weighted by
the deformed action $e^{-S (\vep)}$ to reproduce the functional
integral weighted by $e^{-S }e^{-2\alpha X(z',\bar z')}$. The
counterterm $-2\alpha' \alpha^2 \log(\vep)$ is included to
render the on-shell action finite. 

\subsection{Background-Charge Operators}
\label{sec:background-charge-}

Even in the absense of any operator insertions, however, the
action Eqn. \ref{eqn:X-action} is supplemented by boundary terms
due to the non-trivial dilaton. To understand these terms,
consider the more general sigma-model action on a worldsheet
$\Sigma$ with metric $h$, 
\begin{align}
  \label{eqn:sigma-model}
  S[X;h]
  =& \frac{1}{4\pi \alpha'} \int_\Sigma \diff^2 \sigma\,
  \sqrt{h}    h^{ab} \partial_a X \partial_b X \\
  &-Q \lp \frac{1}{4\pi}\int_\Sigma \diff^2 \sigma \,\sqrt{h}
  \cR  X
  + \frac{1}{2\pi} \int_{\partial \Sigma} \diff \phi
  \,\sqrt{\gamma}  \cK X \rp.\nt
\end{align}
Here, $\cR $ is the scalar curvature of $h$, $\cK$ is the trace
of the extrinsic curvature of the boundary (if present), $\gamma$ is the induced
metric of the boundary, and $\diff \phi\, \sqrt{\gamma}$ is the
induced volume form. 

The sigma-model is Weyl-invariant up to a field-independent
anomaly, provided that $X$
simultaneously transforms as a Goldstone boson,
\begin{align}
  S\lb X+\alpha' Q \omega; e^{2\omega} h\rb
  = S \lb X; h\rb - S\lb-\alpha' Q \omega; h\rb.
\end{align}
The dilaton violates 
the target translation symmetry of the kinetic
term, $S[X+\vep;h] =S[X;h]-\vep Q \chi$, with
$\chi$ the Euler
characteristic of $\Sigma$, which implies the
anomalous conservation law
\begin{align}
  \label{eqn:anomaly}
  \sum_{j}\alpha_j = \frac{1}{2}Q \chi
\end{align}
for a correlation function of operators $\prod_j
V_{\alpha_j}$. The bulk equation of motion,
\begin{align}
  \label{eqn:linear-dilaton-eom}
 \del^2 X = -\frac{1}{2} \alpha' Q \cR[h],
\end{align}
likewise reflects the anomalous
conservation of the currents $\partial X$ and $\bpartial X$. 

We are interested in the theory at string tree-level, for which
$\Sigma$ has the topology of a sphere and $\chi = 2.$
Eqn. \ref{eqn:X-action} was written in locally flat coordinates
$\diff s^2 = \diff z\, \diff \bar z$,
and would follow from Eqn. \ref{eqn:sigma-model} by
discarding the curvature terms. Globally, however, there does
not exist a flat metric on the sphere.
In particular,
the coordinate $z$ does not cover the neighborhood of the
point-at-infinity, and the ``flat'' metric is singular there:
$\diff s^2 = \frac{\diff u\, \diff \bar u}{(u \bar u)^2}$, with a
local coordinate $u = \frac{1}{z}$. This singularity contributes
a delta-function source of curvature, $\cR  = 16\pi (u \bar u)^2
\delta (u, \bar u) = 16 \pi 
\delta(z-z_\infty, \bar z - \bar z_\infty)$,
as required to reproduce the Euler characteristic of the
sphere.

To avoid this singular behavior, one could choose instead the
round metric, 
$\diff s^2 = \frac{4}{\lp 1+ z \bar z \rp^2} \diff z \,\diff
\bar z$, for which  $\cR =2$ is a constant. However, the
dilaton term then produces a linear potential, which is slightly
awkward. It is instead common practice to work with the plane metric
$\diff z\, \diff \bar z$ or the cylinder metric 
$\frac{\diff z\,\diff \bar z}{z \bar z}$, which are related to
the round metric by singular Weyl  
transformations that push all the
curvature of the sphere to the point-at-infinity or the two ends of
the cylinder. 

With the plane metric, the effect of the curvature singularity
is to shift Eqn. \ref{eqn:X-action} by $-2 Q
X(z_\infty, \bar z_\infty)$, which may be thought of as an
insertion of 
$V_{-Q}(z_\infty,\bar z_\infty)$ at the point-at-infinity. Thus,
one can study the linear-dilaton background in flat coordinates,
provided that one keeps track of this so-called background-charge
operator. It inserts a fixed source in the equation of motion, 
\begin{align}
  \partial \bpartial X = -2\pi \alpha' Q \delta (z - z_\infty,
  \bar z - \bar z_\infty),
\end{align}
again demonstrating the anomalous
conservation of the $\Lu_1$ currents in the presence of the
linear-dilaton, and imposing the asymptotic condition 
\begin{align}
  \label{eqn:plane-asymptotic}
  X(z, \bar z) \overright{|z|\to \infty} 2 \alpha' Q \log|z| +
  \cO(1). 
\end{align}
The Green function $2\alpha'
\sum_{j}\alpha_j\log|z-z_j|$ in the presence of  operator
insertions $\prod_j V_{\alpha_j}(z_j, \bar z_j)$ satisfies the
asymptotic condition only provided
\begin{align}
  \sum_{j}\alpha_j =  Q,
\end{align}
reproducing Eqn. \ref{eqn:anomaly}.
Note also that the 1-point function of the operator $V_Q(z,\bar
z)$ is not required to vanish, which is compatible with 
conformal symmetry because $h_Q = \bar h_{Q} = 0$.

By the earlier discussion, the background-charge insertion at the
point-at-infinity may be replaced by
excising its neighborhood 
and introducing a boundary term as in
Eqn. \ref{eqn:boundary-term}. The action for the
linear-dilaton on the plane is therefore given by the $R\to \infty$ limit
of \cite{Zamolodchikov:1995aa,Harlow:2011ny,Teschner:2001rv}
\begin{align}
  \label{eqn:plane-action}
  S  = \frac{1}{2\pi \alpha'} \int_{D_R}
  \diff^2z\, \partial X \bpartial X
  -2Q \int_{\partial D_R} \frac{\diff\phi}{2\pi} X
  +2 \alpha' Q^2 \log(R),
\end{align}
with $D_R$ a disk of radius $R$. Note that the prescription
amounts to cutting out the source and doubling the extrinsic
curvature term in Eqn. \ref{eqn:sigma-model} on the resulting
boundary, which ensures that 
$2 \times\frac{1}{2\pi} \int_{\partial \Sigma} \diff \phi\, \sqrt{\gamma}
\cK = 2$ produces
the Euler characteristic of the sphere rather than the disk. 
As before, one may introduce
additional insertions $V_{\alpha_j}(z_j,\bar z_j)$ by cutting
out disks at $(z_j, \bar z_j)$ and including additional boundary
terms as in Eqn. \ref{eqn:boundary-term}. 

For the most part, we will actually work on the cylinder rather
than the plane, corresponding to the metric $\diff s^2 =
\frac{\diff z\, \diff \bar z}{z \bar z}$. The cylinder metric is
singular at both its ends, 
$\cR  = 8\pi z \bar z\lp \delta(z,\bar z) + \delta(z-z_\infty,
  \bar z - \bar z_\infty) \rp,$ and
the background-charge is now split symmetrically
between them with insertions $V_{-Q/2}(0)$ and
$V_{-Q/2}(z_\infty, \bar z_\infty)$. The equation of motion becomes
\begin{align}
  \label{eqn:cyl-eom}
  \partial\bpartial X = -\pi \alpha' Q \lp \delta(z,\bar z)
+ \delta(z-z_\infty, \bar z-\bar z_\infty)\rp, 
\end{align}
and the asymptotic conditions are
\begin{align}
  \label{eqn:cyl-asymptotic}
  X(\rho,\phi) \overright{\rho \to \pm \infty}\pm \alpha' Q \rho
  + \cO(1),
\end{align}
where $z = e^{\rho + i \phi}$.
The action is given by the $L \to \infty$ limit of
\begin{align}
  \label{eqn:cylinder-action}
 S 
  = &\frac{1}{4\pi \alpha'}
      \int\limits_{-L}^L \diff \rho \int\limits_0^{2\pi}\diff\phi\, 
      \lp (\partial_\rho X)^2 +
      (\partial_\phi X)^2 \rp
  - Q \int\limits_0^{2\pi} \frac{\diff
  \phi}{2\pi} \lp X|_{\rho =L} + X|_{\rho=-L} \rp + 
    \alpha' Q^2 L.
\end{align}

Suppose a primary $e^{-2\alpha X}$ is inserted in the far past
on the cylinder. The asymptotic condition is
\begin{align}
  \label{eqn:shifted-asymp}
  X(\rho,\phi) \overright{\rho \to -\infty}
  -2\alpha'\lp \frac{Q}{2} -  \alpha \rp\rho+ \cO(1). 
\end{align}
Note that for $\mrm{Re}(\alpha) < \frac{Q}{2}$, the asymptotic
condition sends $X$ to the weak-coupling region, whereas for
$\mrm{Re}(\alpha) > \frac{Q}{2}$ it  is mapped to the
strong-coupling region.
Comparing to the usual mode-expansion 
$X =  X_0-i\alpha'P_0 \rho + \cdots$, one finds that the
operator insertion $e^{-2\alpha X}$ prepares a state on the
cylinder of momentum $P_0 = -2i \lp \frac{Q}{2} - \alpha \rp.$
Note that the reflection $\alpha \to Q - \alpha$ flips the sign
of $P_0$.

The zero-mode wavefunction $e^{i P_0 X_0}$ of the state prepared by $e^{-2
  \alpha X}$ is then
\begin{align}
\Psi_\alpha(X_0) =  e^{2   \lp \frac{Q}{2} - \alpha \rp X_0}.
\end{align}
It differs from $e^{-2\alpha X}$ by the background-charge
operator $e^{Q X}$, which is fixed on the end of the cylinder. 
More generally, the target space string-frame effective
action is multiplied by an overall factor of $e^{-2\Phi}.$
Extracting the target space wavefunction from $e^{-2\Phi}(\del_X
\til \Psi(X))^2$ requires
rescaling $\til \Psi(X) \to \Psi(X) =e^{-\Phi} \til \Psi(X)$.
For the linear-dilaton, the necessary factor is again $e^{-\Phi} = e^{QX}.$

For $\alpha \in \frac{Q}{2} + i\reals$, the zero-mode wavefunction
is oscillatory and delta-function normalizable. Otherwise it is
non-normalizable, exponentially 
diverging either at 
$X \to \infty$ for $\mrm{Re}(\alpha)
< \frac{Q}{2}$, or $X \to -\infty$ for  $\mrm{Re}(\alpha) >
\frac{Q}{2}$, in correspondence with the sign of the asymptotic
condition. 

\subsection{Asymptotic Conditions in Liouville}
\label{sec:asympt-cond-liouv}

The above discussion was confined to the
free linear-dilaton theory. However, the free linear-dilaton is
not a unitary CFT.\footnote{The operators which map to
  delta-function normalizable 
  states have $\alpha \in \frac{Q}{2} + i \reals$, but the OPE
  generates operators with $\alpha$ outside this range.}
As a string background it is
clearly ill-behaved because 
the string coupling $e^{\Phi}$ diverges exponentially as $X
\to -\infty$. This pathological behavior may be regulated by
turning on a potential barrier $\mu e^{-2b_\mrm{L} X}$, with
$\mu > 0$ and $\mrm{Re}(b_\mrm{L}) > 0$, that suppresses
string configurations that extend too deeply into the
strong-coupling region. $b_\mrm{L}$ is fixed by demanding that
the potential is marginal, $\alpha' b_\mrm{L}(Q - b_\mrm{L}) = 1$.
The result is the Liouville CFT, with
bulk action
\begin{align}
  \label{eqn:liouville-action}
  S = \frac{1}{2\pi \alpha'} \int \diff^2 z
  \lp \partial X \bpartial X + \pi \mu e^{-2b_\mrm{L} X} \rp + \cdots. 
\end{align}
Much of the preceding machinery of the free theory continues to apply,
with a few caveats that we describe now. The cigar CFT, discussed
in the next section, will be closely analogous.

As $X \to \infty$, the Liouville potential vanishes and the free
linear-dilaton is recovered. One therefore again has scattering
solutions of the zero-mode quantum mechanics that behave as
plane waves $e^{\pm i P_0 X_0}$ in the free-field
region. They are no longer independent, however.
The solutions which decay under the potential in the
strong-coupling region behave as
linear combinations of incoming and reflected waves in the
free-field region,
$\Psi(X_0) \overright{X_0 \to \infty}e^{i P_0 X_0}+R(P_0) e^{-i
  P_0 X}$, with $R(P_0)$ the reflection coefficient. 

In the CFT, one correspondingly has operators $V_\alpha(z, \bar
z)$ behaving asymptotically as\footnote{With this choice of
  normalization, the 2-point function of Liouville primaries is
  proportional to $R(\alpha)$. A canonically normalized 2-point
  function is obtained by rescaling the primaries by
  $R(\alpha)^{-1/2}$.}
\begin{align}
  V_\alpha \overright{X \to \infty}e^{-2 \alpha X} + R(\alpha)
  e^{-2(Q-\alpha)X}.
\end{align}
The exact reflection coefficient is \cite{Dorn:1994xn,Zamolodchikov:1995aa}
\begin{align}
  \label{eqn:R-liouville}
  R(\alpha) = -\lp \pi \mu \frac{\Gamma(b_\mrm{L}^2)}{\Gamma(1-b_\mrm{L}^2)}
  \rp^{\frac{2}{b_\mrm{L}}\lp 
  \frac{Q}{2}-\alpha \rp}
  \frac{\Gamma \lp 1 - \frac{2}{b_\mrm{L}}\lp \frac{Q}{2}-\alpha \rp
  \rp}
  {\Gamma \lp 1 + \frac{2}{b_\mrm{L}}\lp \frac{Q}{2}-\alpha \rp \rp}
  \frac{\Gamma \lp 1 - 2b_\mrm{L}\lp \frac{Q}{2}-\alpha \rp
  \rp}
  {\Gamma \lp 1 + 2b_\mrm{L}\lp \frac{Q}{2}-\alpha \rp \rp}.
\end{align}
It satisfies $R(\alpha)R(Q-\alpha) = 1$. 
Whereas $\alpha$ and $Q - \alpha$
labeled independent operators of identical conformal weights in
the free theory, they now label two components 
of the same operator due to reflection off the potential.
One therefore labels Liouville operators by $\alpha$ 
satisfying $\mrm{Re}(\alpha) \leq \frac{Q}{2}$, and moreover
$\mrm{Im}(\alpha) > 0$ if $\mrm{Re}(\alpha) = \frac{Q}{2}$, in
order to avoid double-counting.
With this convention, $e^{-2\alpha X}$ is the
exponentially dominant term at infinity, except when
$\mrm{Re}(\alpha) = \frac{Q}{2}$,
in which case neither term dominates the other. It is impossible
to have an operator that asymptotes to
$e^{-2\alpha X}$ with $\mrm{Re}(\alpha) > \frac{Q}{2}$
because it is sub-dominant to its reflection $e^{-2(Q-\alpha)X}$, 
and both terms are required to obtain a non-singular solution in
the interior \cite{Seiberg:1990eb}. Equivalently, one
may allow all values of $\alpha$, in which case the CFT
operators $V_\alpha$ and $V_{Q-\alpha}$ are identical up to
rescaling by the reflection coefficient, $V_\alpha = R(\alpha)
V_{Q-\alpha}$. The complex and real  
branches of operators with non-negative conformal weights are 
labeled by $\alpha \in \frac{Q}{2} + i \reals_+$ and $\alpha \in
\lb 0, \frac{Q}{2} \rb$.

Consider the worldsheet neighborhood of an operator insertion
$V_\alpha(z',\bar z')$. Suppose $\mrm{Re(\alpha)}
< \frac{Q}{2}$, such that operator at large $X$ is dominated
by $e^{-2\alpha X}$, as in the free theory.
If one further requires $\mrm{Re}(\alpha)<0$,
then the free-field Green function Eqn. \ref{eqn:free-green}
remains a self-consistent solution of the Liouville equation of
motion, since it maps the neighborhood
of the insertion to the free-field region where the potential is
sub-leading. As before, one can cut out the insertion and
replace it with the boundary action
Eqn. \ref{eqn:boundary-term}. 

The same considerations as in the free theory require the
asymptotic conditions Eqn. \ref{eqn:plane-asymptotic} on the
plane or Eqn. \ref{eqn:cyl-asymptotic} on the cylinder. These
likewise map $X\to \infty$ where the potential is sub-leading,
and the free-field results remain consistent. 

Inserting $V_\alpha$ with $\mrm{Re}(\alpha)<
\frac{Q}{2}$ in the far past on the cylinder 
imposes the same asymptotic condition as before,
Eqn. \ref{eqn:shifted-asymp}. 
Indeed, $X$ is sent to the free-field region when
$\mrm{Re}(\alpha) < \frac{Q}{2}.$ One may similarly describe a
complex branch insertion by shifting $\alpha \to \alpha -
\vep$ with a small regulator $\vep > 0$, such that the
asymptotic condition again sends $\mrm{Re}(X)\to \infty$.

The zero-mode wavefunction for the state prepared by inserting
$V_\alpha$ in the far past behaves at large $X$ as
\begin{align}
  \Psi_\alpha(X_0) \overright{X_0 \to \infty}
  e^{2 \lp \frac{Q}{2} - \alpha \rp X_0}
  + R(\alpha) e^{-2\lp \frac{Q}{2} - \alpha\rp X_0}.
\end{align}
On the complex branch it is oscillatory and delta-function
normalizable, corresponding to a scattering state with
asymptotic momentum $2 \mrm{Im}(\alpha)$, while on the real branch
it is exponentially divergent
at weak-coupling and therefore non-normalizable.
The Hilbert space of normalizable states is then spanned by the
complex branch $\alpha \in \frac{Q}{2} + i \reals_+$
\cite{Seiberg:1990eb}.

With insertions of $V_\alpha$ at both ends of the cylinder, one
obtains the following action
\begin{align}
  S_\alpha
  =& \frac{1}{4\pi\alpha'} \int\limits_{-L}^L \diff \rho\,
     \int\limits_0^{2\pi}\diff \phi\,
  \lp (\partial_\rho X)^2 + (\partial_\phi X)^2+ 4\pi \mu e^{-2b_\mrm{L}X}
  \rp\\
   &-2\lp\frac{Q}{2} - \alpha\rp \int\limits_0^{2\pi}\frac{\diff
     \phi}{2\pi} \lp X|_{\rho=L} +   X|_{\rho = -L} \rp
     +4\alpha'\lp \frac{Q}{2} -\alpha \rp^2 L, \nt
\end{align}
whose functional integral computes the Liouville
reflection coefficient $R(\alpha)$ in the limit $L \to \infty$.
The saddle-point expansion of this integral in the
semi-classical ($b_\mrm{L} \to 0$) limit was computed in 
\cite{Harlow:2011ny}, and matched to the limit of the exact
reflection coefficient.

Our objective in Secs. \ref{sec:cigar} and
\ref{sec:semi-classical} is to treat the large $k$ limit of the
$\mrm{SL}(2,\reals)_k/\mrm{U}(1)$ CFT similarly.

\section{Review of the $\mrm{SL}(2,\reals)_k/\mrm{U}(1)$ CFT}
\label{sec:cigar}

The $\mrm{SL}(2,\reals)_k/\mrm{U}(1)$ CFT is a coset of the
$\mrm{SL}(2,\reals)_k$ WZW model. The latter describes a string
propagating\footnote{By $\mrm{AdS_3} = \mrm{SL}(2,\reals)$ we
  mean the simply-connected covering-space.} in Lorentzian $\mrm{AdS}_3 =
\mrm{SL}(2,\reals)$, where the WZW level $k$ sets the
$\mrm{AdS}$-length, $l_\mrm{AdS}^2 = k l_\mrm{s}^2$.
$\mrm{AdS}_3$ may be described as a solid cylinder, and
the coset is defined by 
gauging the timelike isometry along its length, producing a
target space of Euclidean signature.
Its central charge is
\begin{align}
  \label{eqn:central-charge}
  c = \frac{3k}{k-2} - 1, 
\end{align}
where $k$ is a real number greater than 2, which
need not be an integer.

We will primarily be
interested\footnote{The $k \to 2$ limit of the CFT is also
  interesting, and we discuss it briefly in Sec. \ref{sec:sL}.}
in the semi-classical limit of the CFT at large $k$,
which describes a string propagating in
a 2-dimensional Euclidean black hole \cite{Witten:1991yr}. 
In this section we review the Lagrangian description of the CFT
in this limit, given by the cigar sigma-model background. Then
we review the operator spectrum of the CFT and the associated
asymptotic conditions in the cigar description. Finally, we 
write the action for the 2-point function of primaries on the
cylinder, which we will use in Sec. \ref{sec:semi-classical} to
compute the saddle-point expansion of the reflection coefficient.

\subsection{The Cigar Sigma-Model}
\label{sec:cigar-sigma-model}

For large $k$, the coset admits a Lagrangian description given
by the following sigma-model background:
\begin{subequations}
  \label{eqn:cigar-bkgd}
  \begin{align}
    \label{eqn:metric}
    &\diff s^2 = \alpha' k \lp  \diff r^2 + \tanh^2(r)
      \diff\theta^2\rp \\
    \label{eqn:dilaton}
    &\Phi = -\log \cosh(r)+ \Phi_0.
  \end{align}
\end{subequations}
The action on a closed worldsheet $\Sigma$ is
\begin{align}
  S
  =& \frac{k}{4\pi} \int_\Sigma \diff^2 \sigma \,\sqrt{h}
     \lc
     (\del r)^2 + \tanh^2(r) (\del \theta)^2
     + \frac{1}{k} \cR[h] \lp \Phi_0 - \log \cosh r \rp\rc,
\end{align}
with  equations of motion
\begin{subequations}
  \label{eqn:cigar-eom}
  \begin{align}
    \label{eqn:cigar-radial}
  & \del^2 r - \tanh(r) \sech^2(r) (\del \theta)^2 +
    \frac{1}{2k}\cR[h]\tanh(r) = 0\\
  & \del^2 \theta + 2 \sech(r) \csch(r) h^{ab}\del_a r \del_b
    \theta =0.
\end{align}
\end{subequations}

The target space has the topology of a disk, with coordinates $r
\in [0,\infty)$ and $\theta \sim \theta + 2\pi$.
The dilaton is a monotonically decreasing function of $r$,
with the constant $\Phi_0$ setting its maximal value 
at the origin: 
$\Phi|_{r = 0} = \Phi_0$. In that neighborhood the geometry is
simply $\reals^2$ in polar coordinates:
\begin{subequations}
\begin{align}
  &\diff s^2 = \alpha' k\lp \diff r^2 + r^2 \diff \theta^2
    \rp + \cO(r^3)\\
  & \Phi = \Phi_0 - \frac{1}{2} r^2 + \cO(r^3). 
\end{align}
\end{subequations}
At large $r$, on the other hand, the geometry 
approaches a cylinder of radius $\sqrt{\alpha' k}$, with a
linearly decreasing dilaton along its length:
\begin{subequations}
  \label{eqn:large-r}
\begin{align}
  &\diff s^2 = \alpha' k (\diff r^2 + \diff
    \theta^2) + \cO\lp e^{-2r} \rp\\
  &\Phi = -r 
    +\cO(1).
\end{align}
\end{subequations}
The target space therefore resembles a cigar,
with its asymptotic cylinder at large $r$ and its tip at $r =
0$, where the $\theta$ circle shrinks to a point, as pictured in
Fig. \ref{fig:cigar}.

For large $k$ the cigar is large and weakly-curved,
\begin{align}
  \label{eqn:curvature}
  \cR = \frac{4}{\cosh^2(r)} \frac{1}{\alpha' k},
\end{align}
and the sigma-model is weakly-coupled in the $\alpha'$ sense. 
Meanwhile, the string coupling  $e^\Phi$ attains its
maximal value $e^{\Phi_0}$ at the tip of the cigar and decays
to zero at large $r$. The parameter $\Phi_0$ is a modulus of the
theory. It reflects the usual freedom to shift the dilaton by a
constant, the only effect being to shift the action by 
$\Phi_0 \chi$, with $\chi$ the Euler characteristic of $\Sigma.$

The rather exotic coupling to curvature  represented by the
dilaton Eqn. \ref{eqn:dilaton} is required to satisfy the 1-loop
beta function equation \cite{Witten:1991yr}
\begin{align}
  \label{eqn:metric-beta}
  \beta_{IJ}(G) = \alpha' \lp R_{IJ} + 2 \del_I \del_J\Phi \rp
  + \cO(\alpha'^2)= \cO(\alpha'^2),
\end{align}
which implies conformal invariance of the sigma-model to leading
order in the large $k$ limit.

\subsection{Operator Spectrum}
\label{sec:operator-spectrum}

Although the above sigma-model is a valid description of the
coset only at large $k$, the exact spectrum of the CFT is known
in light of its relation to the $\mrm{SL}(2,\reals)_k$ WZW model
via the coset construction
\cite{Dijkgraaf:1991ba,Maldacena:2000hw,Hanany:2002ev}.
The Virasoro primaries $\cO_{jnw}(z,\bar z)$ of the coset are labeled by
integers $n$ and $w$ and a complex number $j$, taking the
following values:
\begin{subequations}
\label{eqn:coset-spectrum}
  \begin{align}
  \label{eqn:complex-branch}
  &(\mrm{\bd{i}}) \quad j= \frac{1}{2}+i s,\quad s \in
    \reals_+\\
  \label{eqn:real-branch}
  &(\mrm{\bd{ii}})\quad  j_N = \frac{k|w|-|n|}{2}-N \in
    \lp \frac{1}{2}, \frac{k-1}{2}\rp,\quad N \in \bbN.
\end{align}
\end{subequations}
We refer to these two sets as the
complex and real branches of primaries based on the value of
$j$. As in the free linear-dilaton and Liouville theories
reviewed in the previous section,
the complex branch primaries correspond to scattering states  on
the cigar with
momentum proportional to $s$. On the other hand, the real branch
primaries with $j=j_N$ correspond to bound states living at the
tip of the cigar.
One may also consider real branch operators where $j$ is not
valued in this discrete set, which 
map to non-normalizable states. The integers $n$ and $w$,
meanwhile, correspond to the momentum and winding numbers around
the asymptotic cylinder at large $r$.

Note that on the complex branch the value of $j$ is independent
of the integers $n$ and $w$, whereas on the real branch
$n$ and $w$ determine the allowed values of $j$ up
to shifts by the natural number $N$, constrained to lie within the
interval $\frac{1}{2} < j < \frac{k-1}{2}$. That
lower-bound implies there may only exist real branch primaries
with $k |w|  - |n| > 1$. In particular, there are none with $w = 0$. 

These primaries carry conformal weights
\begin{align}
  \label{eqn:coset-weights}
  h_{jnw}= -\frac{j(j-1)}{k-2} +
  \frac{(n-kw)^2}{4k},\quad\quad
  \bar h_{jnw} =-\frac{j(j-1)}{k-2} +
  \frac{(n+kw)^2}{4k}.
\end{align}
Note that the quantity $-j(j-1)$ is a real number not only on
the real branch, but also on the complex branch where $-j(j-1) =
\frac{1}{4} + s^2$. On the real branch, it is non-negative for
$\frac{1}{2} \leq j \leq 1$ and negative thereafter, its maximal
value coinciding with the minimal value on the complex
branch. The total conformal weight is non-negative, however.

Since the sigma-model reduces to the free 
$\text{linear-dilaton}\times S^1$ background at large $r$, 
the abstract primaries $\cO_{jnw}$ may be expanded in free-field
primaries in that limit. To compare with the formulas from
the previous section, one may define a canonically normalized
field $\hat r \equiv \sqrt{\alpha' k} r$, and likewise $\hat
\theta = \sqrt{\alpha' k} \theta$, in terms of which the
asymptotic background Eqn. \ref{eqn:large-r} is
\begin{subequations}
  \label{eqn:linear-dilaton-circle}
  \begin{align}
  &\diff s^2 \overright{\hat r \to \infty} \diff \hat r^2 +
    \diff \hat \theta^2\\
  &\Phi \overright{\hat r \to \infty} - Q \hat r,
\end{align}
\end{subequations}
with
\begin{align}
  \label{eqn:large-k-Q}
  Q = \frac{1}{\sqrt{\alpha'k}}.
\end{align}
Note that $Q$ goes to zero in the large $k$ limit, in contrast
to Liouville where $Q \overright{b \to 0} \frac{1}{\alpha' b}$
diverged in the semi-classical limit. In the semi-classical
limit of the cigar, the dilaton contribution  is sub-leading
to the metric.

The Virasoro primaries of the free theory are 
\begin{align}
  \label{eqn:primaries}
  \cV_{\alpha p_\mrm{L} p_\mrm{R}}(z, \bar z) =
  e^{-2\alpha \hat  r(z, \bar z)}
  e^{i p_\mrm{L} \hat\theta_\mrm{L}(z) + i p_\mrm{R} \hat \theta_\mrm{R}(\bar z)},
\end{align}
where $p_\mrm{L}$, $p_\mrm{R}$ are valued in the lattice
\begin{align}
  \label{eqn:momentum-lattice}
  &p_\mrm{L} = \frac{n}{\sqrt{\alpha' k}} -
    \sqrt{\frac{k}{\alpha'}}w,
    \quad\quad
  p_\mrm{R} = \frac{n}{\sqrt{\alpha' k}} +
    \sqrt{\frac{k}{\alpha'}}w,\quad \quad
    n,w \in \bbZ. 
\end{align}
$n$ is the momentum number around the cylinder and $w$ is (minus) the
winding number. Their conformal weights with respect to the free
theory stress tensor
\begin{align}
  T(z) = - \frac{1}{\alpha'} (\partial \hat r)^2 - Q \partial^2
  \hat r - \frac{1}{\alpha'} (\partial \hat \theta)^2
\end{align}
are
\begin{align}
  \label{eqn:free-weights}
  &h_{\alpha p_\mrm{L} p_\mrm{R}} = \alpha' \alpha(Q - \alpha) 
    +\alpha' \frac{p_\mrm{L}^2}{4},
    \quad\quad
  \bar h_{\alpha p_\mrm{L} p_\mrm{R}} = \alpha' \alpha(Q - \alpha) 
  +\alpha' \frac{p_\mrm{R}^2}{4}.
\end{align}
The central charge of the Virasoro algebra is
\begin{align}
  c_{\mrm{LD}\times S^1}=2 + 6 \alpha' Q^2.
\end{align}

The large $\hat r$ expansion of   the coset primary $\cO_{jnw}(z,\bar z)$ 
is\footnote{\label{foot:Q} The finite $k$
  corrections to the background Eqn. \ref{eqn:cigar-bkgd}
  imply $Q = \frac{1}{\sqrt{\alpha'(k-2)}}$. In the remainder of
  this subsection, we use this value of $Q$ so that the formulas for
  the asymptotic operators and wavefunctions are
  valid at finite $k$.  
  Note that with the corrected  value of $Q$,
  $c_{\mrm{LD}\times S^1}$ reproduces the exact central charge
  Eqn. \ref{eqn:central-charge}. Likewise the conformal weights
  Eqn. \ref{eqn:free-weights} reproduce the exact weights
  Eqn. \ref{eqn:coset-weights} with the dictionary discussed
  below.}
\footnote{This choice of operator normalization does not produce
a canonically normalized two-point function. Rather, the
two-point function of $\cO_{jnw}$ and $\cO_{j,-n,-w}$ is
proportional to $R(j,n,w)$. Note that $R$ given in
Eqn. \ref{eqn:r} is appropriately even in both $n$ and $w$.}
 \cite{Dijkgraaf:1991ba} 
\begin{align}
  \label{eqn:asymptotic}
  \cO_{jnw}\overright{\hat r \to \infty}
  \lp e^{-2Q(1-j)\hat r} + R(j,n,w) e^{-2Qj\hat r} \rp
  e^{ip_\mrm{L}  \hat\theta_\mrm{L} + i p_\mrm{R} \hat \theta_\mrm{R}},
\end{align}
where $R(j,n,w)$ is the reflection coefficient
\cite{Teschner:1999ug,Giveon:1999tq,Maldacena:2001km}:
\begin{align}
  \label{eqn:r}
  R(j,n,w)
  =& 
     \lp  \nu(k)\rp^{2j-1}
     \frac{\Gamma \lp 1 - \frac{2j-1}{k-2} \rp}
     {\Gamma \lp 1 + \frac{2j-1}{k-2} \rp}\\
   &\times
     4^{2j-1}
     \frac{\Gamma\lp 1 - 2j \rp}{\Gamma\lp 2j - 1\rp}
     \frac{\Gamma\lp j+\frac{|n|-kw}{2}\rp
     \Gamma\lp j + \frac{|n|+kw}{2}\rp}
     {\Gamma\lp 1-j+\frac{|n|-kw}{2}\rp
     \Gamma\lp 1-j+\frac{|n|+kw}{2}\rp}.\nt
\end{align}
It satisfies $R(1-j,n,w) R(j,n,w) = 1$. $\nu(k)$ is a
$j$-independent function, analogous to the prefactor $\pi \mu
\Gamma(b_\mrm{L}^2)/\Gamma(1-b_\mrm{L}^2)$ appearing in the Liouville reflection
coefficient Eqn. \ref{eqn:R-liouville}. We will set it to one in
what follows. 

As recalled in the previous section, the zero-mode wavefunction
for the state prepared by inserting $\cO_{jnw}$ in the far past
on the cylinder is obtained after rescaling by $e^{-\Phi} =
e^{-\Phi_0} \cosh(r)$. For large $r$, the radial wavefunction is then
\begin{align}
  \Psi_{jnw}(\hat r_0) \overright{\hat r_0 \to \infty}
  \frac{1}{2} e^{-\Phi_0}
  \lp e^{2Q\lp j-\frac{1}{2}\rp \hat r_0} + R(j,n,w) e^{-2Q \lp j-\frac{1}{2}
  \rp \hat r_0}\rp.
\end{align}
With $j \in \frac{1}{2} + i \reals$, neither exponential
dominates the other, the asymptotic radial wavefunction is oscillatory,
and one obtains a delta-function normalizable state. The
asymptotic operator is identified with the linear-dilaton
primary $e^{-2\alpha \hat r}$ with $\alpha = Q(1-j)$ plus its
reflection $e^{-2(Q-\alpha)\hat r}$, together with a compact
boson primary of momentum $n$ and winding $-w$.

For real $j$, on the other hand, the reflected term is
exponentially sub-dominant for $j > \frac{1}{2}$. Then,
generically, the wavefunction diverges
exponentially at weak-coupling,
and the associated state is non-normalizable.

There is an important exception, however,
when $R(j,n,w)$ is singular, and this is the manifestation of
the bound states. Indeed, on the
real branch with $j=j_N$ given by 
Eqn. \ref{eqn:real-branch}, one of the two Gamma functions
$\Gamma \lp j + \frac{|n| \pm 
  kw}{2} \rp$ in Eqn. \ref{eqn:r} has a simple pole, depending on the sign of
$w$. For $w>0$ one has
\begin{align}
  j_N + \frac{|n| - k w}{2} = -N
\end{align}
and therefore
\begin{align}
  \label{eqn:poles}
  \Gamma \lp j + \frac{|n|-kw}{2} \rp
  \overright{j\to j_N} \frac{1}{j-j_N} \frac{(-)^{N}}{N!}
  + \cO(1). 
\end{align}
$\Gamma \lp j_N + \frac{|n|+kw}{2} \rp$ is similarly singular for $w
< 0$. The remaining Gamma functions have additional
singularities, but they are not associated to bound states
\cite{Aharony:2004xn}. 

Thus, for $j=j_N$, it is the reflected component $R(j_N,n,w)
e^{-2Qj_N \hat r}$ that dominates in the asymptotic region.
One obtains a discrete set of operators $\til \cO_{j_N n w}$
defined as the residue of $\cO_{jnw}$ as $j \to j_N$
\cite{Aharony:2004xn}.  
The zero-mode radial wavefunction
decays in the weak-coupling region, corresponding to a 
normalizable bound state with wavefunction proportional to
\begin{align}
  \label{eqn:bound-wavefunction}
  \Psi_{j_N n w}(\hat r_0) \underset{\propto}{\overright{\hat r_0 \to \infty}}
  e^{-2Q \lp j_N-\frac{1}{2}\rp \hat  r_0}.
\end{align}

The simplest pair of bound states have $n=0$, $w = \pm 1,$
and $N = 1$, such that $j_N = \frac{k}{2} -1$. 
The asymptotic form of these operators is
\begin{align}
  \label{eqn:sl-operator}
  \frac{1}{R}\cO_{j = \frac{k}{2}-1,n=0,w=\pm 1}
  \overright{\hat r \to \infty}
  e^{-\sqrt{\frac{k-2}{\alpha'}}\hat r} e^{\mp i
  \sqrt{\frac{k}{\alpha'}}(\hat \theta_\mrm{L}-\hat
  \theta_\mrm{R})}.
\end{align}
The sum of these two winding operators is called
the sine-Liouville operator \cite{Giveon:2016dxe}. It defines a
normalizable, marginal bound state for $k > 
3$. For $k < 3$, $j = \frac{k}{2} - 1$ falls
below the lower-bound $j>\frac{1}{2}$. Then the zero-mode wavefunction
Eqn. \ref{eqn:bound-wavefunction} 
diverges at large $\hat r$ and the state becomes
non-normalizable.

\subsection{Asymptotic Conditions in the Cigar}
\label{sec:cigar-asymptotics}

As in the linear-dilaton theories discussed in the
Sec. \ref{sec:asymptotic-conditions}, the
dilaton's coupling to curvature produces source terms in the
equations of motion on $S^2$ with a singular metric.
Choosing the cylinder metric, the radial equation of motion
Eqn. \ref{eqn:cigar-radial} in the absence of any insertions may
be written
\begin{align}
  \partial \bpartial r
  - \tanh(r) \sech^2(r) \partial \theta\bpartial \theta
  =-\frac{\pi}{k}
  \lp \delta(z,\bar z) + \delta(z-z_\infty,\bar z-\bar z_\infty)
  \rp
  \tanh(r).  
\end{align}
At large $r$, we recover the free linear-dilaton equation of
motion Eqn. \ref{eqn:cyl-eom},
\begin{align}
  \partial \bpartial \hat r 
  =-\pi \alpha' Q
  \lp \delta(z,\bar z) + \delta(z-z_\infty,\bar z-\bar z_\infty)\rp,
\end{align}
with Green functions
\begin{align}
  \hat r(\rho,\phi) \overright{\rho \to \pm \infty}\pm \alpha' Q \rho
  + \cO(1),
\end{align}
where $z = e^{\rho+i \phi}$. These free Green functions
are self-consistent solutions of the full cigar equations of
motion in the neighborhood of the source terms
because the ends of the cylinder are mapped to the
asymptotic region, where the corrections to the
$\text{linear-dilaton}\times S^1$ equations of motion are
exponentially sub-leading. Then just as before we may write a
regulated action for the cigar on a cylinder worldsheet with
linear boundary terms at its ends:
\begin{align}
 S
  = &\frac{k}{4\pi}
      \int\limits_{-L}^L \diff \rho \int\limits_0^{2\pi}\diff\phi\, 
      \lp (\partial_\rho r)^2 +
      (\partial_\phi  r)^2
      + \tanh^2(r)
      \lp  (\partial_\rho  \theta)^2 + (\partial_\phi  \theta)^2 \rp
      \rp\\
  &-  \int\limits_0^{2\pi} \frac{\diff
      \phi}{2\pi} \lp  r|_{\rho =L} +  r|_{\rho=-L} \rp + 
       \frac{L}{k}.\nt
\end{align}
Next consider the equations of motion in the neighborhood of an
insertion $\cO_{jnw}(z',\bar z')$ away from the curvature
singularities. Suppose 
$\mrm{Re}(j) > \frac{1}{2}$, with $R(j,n,w)$ regular,
such that the operator is dominated by
$\cV_{Q(1-j)p_\mrm{L}p_\mrm{R}}(z',\bar z')$ at large $\hat
r$. The Green functions of the asymptotic
$\text{linear-dilaton}\times S^1$ background in the presence of
this source are
\begin{subequations}
  \label{eqn:ld-circle-green}
\begin{align}
  &\hat r(z,\bar z) \overright{|z-z'|\to 0}
    2\alpha' Q(1-j) \log |z-z'| + \cO(1)\\
  &\hat \theta(z,\bar z)\overright{|z-z'|\to 0}
    -\frac{i}{2} \sqrt{\frac{\alpha'}{k}}
    \lp
    2n \log |z-z'| - kw \log  \frac{z-z'}{\bar z-\bar z'}
    \rp+ \cO(1). 
\end{align}
\end{subequations}
Suppose furthermore that $\mrm{Re}(j)>1$. Then the neighborhood
of the insertion is mapped to $\hat r
\to \infty$, and once again one obtains a
self-consistent solution of the cigar equations of motion.

By contrast, if $j=j_N$, such that $R(j,n,w)$ is singular and
the operator approaches $\cV_{Qj,p_\mrm{L}p_\mrm{R}}$ at large
$\hat r$, then the free radial Green function in the neighborhood of the
(appropriately normalized) operator insertion is
\begin{align}
  \hat r(z,\bar z) \overright{|z-z'|\to 0}
    2\alpha' Qj \log |z-z'| + \cO(1).  
\end{align}
In this case, even if $\hat r$ begins in the asymptotic region,
as one approaches the insertion point on the worldsheet $\hat r$
is mapped out of the free-field region, and one no longer has a
self-consistent solution. 

Away from these discrete values, however, the appropriate
asymptotic conditions are obtained from the free-field Green
functions, and correlation functions with bound state insertions
may be obtained by computing the functional integral for generic
$j$ and then taking the residue of the result as $j \to
j_N$. It is nevertheless interesting to identify an asymptotic
condition that describes a bound state insertion directly,
rather than as the residue of an ordinary
insertion. We return to this problem in Sec. \ref{sec:bound-states}.

If a generic operator is inserted in the far past on the cylinder,
then the asymptotic conditions including the effect of the
background-charge are
\begin{subequations}
  \label{eqn:ld-circle-green}
  \begin{align}
    \label{eqn:r-green}
  &\hat r(\rho,\phi) \overright{\rho\to-\infty}
    -2\alpha' Q\lp j-\frac{1}{2} \rp \rho + \cO(1)\\
    \label{eqn:theta-green}
  &\hat \theta(\rho,\phi)\overright{\rho\to-\infty}
    -i\sqrt{\frac{\alpha'}{k}}
    \lp n \rho - ikw \phi \rp+ \cO(1). 
\end{align}
\end{subequations}
For $\mrm{Re}(j) > \frac{1}{2}$, the solution is consistent. A
complex branch operator may be similarly described by perturbing $j \to j
+ \vep$ by a small positive regulator. 

The $\rho$ dependence of the asymptotic conditions may be
enforced as before by linear boundary terms. The $\phi$
dependence of $\hat \theta$, on the other hand, may be
implemented using Lagrange multipliers $\sigma_\pm$. For example, the regulated
action for the 2-point function of $\cO_{jnw}$ and
$\cO_{j,-n,-w}$ is given by 
\begin{align}
  \label{eqn:action-cigar}
 S_{jnw}
  = &\frac{k}{4\pi}
      \int\limits_{-L}^L \diff \rho \int\limits_0^{2\pi}\diff\phi\, 
      \bigg( (\partial_\rho r)^2 +
      (\partial_\phi  r)^2
      + \tanh^2(r)
      \lp  (\partial_\rho  \theta)^2 + (\partial_\phi  \theta)^2 \rp
      \bigg)\\
    &+ 2 \lp \frac{1}{2} -j \rp \int\limits_0^{2\pi} \frac{\diff
      \phi}{2\pi} \lp  r|_{\rho =L} +  r|_{\rho=-L} \rp
      +in \int\limits_0^{2\pi}\frac{\diff \phi}{2\pi}
      \lp \theta|_{\rho=L} - \theta|_{\rho = -L} \rp      
      \nt\\
    &+k\int\limits_0^{2\pi} \frac{\diff \phi}{2\pi}
      \bigg(
      \sigma_+ \lp \partial_\phi \theta|_{\rho = L} + w \rp
      +\sigma_- \lp \partial_\phi \theta|_{\rho = -L} + w \rp
      \bigg)\nt\\
    &+ 4\frac{L}{k}\lp j-\frac{1}{2}\rp^2- k w^2 L  - \frac{L}{k} n^2
      \nt.
\end{align}
Note that the imaginary boundary term for the momentum mode of
$\theta$ ensures invariance of $e^{-S_{jnw}}$ under
$\theta\sim \theta + 2\pi$, where $n \in \bbZ$.

The boundary equations of motion obtained by varying $r$ and
$\sigma_\pm$ are
\begin{subequations}
\begin{align}
  \label{eqn:r-bc}
  &\partial_\rho r|_{\rho=\pm L} =\pm \frac{2}{k}\lp j
    -\frac{1}{2}\rp\\
  \label{eqn:theta-bc}
  &\partial_\phi \theta|_{\rho = \pm L} = -w,
\end{align}
\end{subequations}
while the variation of $\theta$ gives
\begin{align}
  \label{eqn:sigma-bc}
  & \pm \partial_\phi \sigma_\pm|_{\rho = \pm L}
    =\frac{in}{k}+\tanh^2(r) \partial_\rho \theta |_{\rho =\pm L}.    
\end{align}
In the large $L$ limit, Eqn. \ref{eqn:r-bc} implies the asymptotic condition
\begin{align}
  r \overright{\rho \to \pm \infty} \pm \frac{2}{k} \lp j -
    \frac{1}{2} \rp \rho,
\end{align}
as in Eqn. \ref{eqn:r-green}. Eqn. \ref{eqn:theta-bc}
requires that $\theta \to - w\phi + \theta_0(\rho)$, where
$\theta_0(\rho)$ is the zero-mode in the Fourier expansion of
$\theta(\rho,\phi)$ around the $\phi$ circle. This zero-mode is
meanwhile fixed by the integral of Eqn. \ref{eqn:sigma-bc},
\begin{align}
  \int\limits_0^{2\pi} \diff \phi\, \tanh^2(r) \partial_\rho
  \theta|_{\rho = \pm \infty} =-\frac{2\pi i n}{k}.
\end{align}
Note that $\tanh^2(r)|_{\rho = \pm  \infty}$ goes to one in
large $L$ limit. The asymptotic condition on $\theta$ is then
\begin{align}
  \theta \overright{\rho \to \pm \infty} -i\frac{n}{k} \rho - w \phi,  
\end{align}
reproducing Eqn. \ref{eqn:theta-green}.

In other words, Eqn. \ref{eqn:theta-bc} is
a Dirichlet condition that requires the non-zero-modes of
$\theta$ to vanish at the boundaries,
while the integral of Eqn. \ref{eqn:sigma-bc} is a Neumann condition
on the zero-mode. One solves the bulk equations of motion with
these boundary conditions, together with the Neumann condition
Eqn. \ref{eqn:r-bc} on $r$. The Lagrange multipliers are then
determined by Eqn. \ref{eqn:sigma-bc} up to a zero-mode,
which we discard.

The functional integral weighted by $e^{-S_{jnw}}$ computes the reflection
coefficient $R(j,n,w)$ in the $L \to \infty$ limit. In the next
section, we compute the saddle-point expansion of this
integral, restricted to the pure-winding sector for
simplicity, and show that there exists a set of saddles that 
reproduces the semi-classical limit of the exact reflection coefficient.

\section{Semi-Classical Limit}
\label{sec:semi-classical}

In this section we compute the semi-classical limit of the cigar
reflection coefficient by a saddle-point expansion, and compare to
the large $k$ limit of the exact reflection coefficient
Eqn. \ref{eqn:r}. As in the analogous calculation for Liouville
\cite{Harlow:2011ny}, doing so requires summing over complex
saddles, even for real branch operators. We will also find that
saddles which hit the black hole singularity contribute
to the saddle-point expansion with finite action, and are
important for recovering the real branch bound states. 

\subsection{The Reflection Coefficient in the Large $k$ Limit }
\label{sec:refl-coeff-large}

As reviewed in Sec. \ref{sec:operator-spectrum}, the exact
reflection coefficient $R(j,n,w)$ of the
$\mrm{SL}(2,\reals)_k/\mrm{U}(1)$ CFT is known thanks to work on
the $\mrm{SL}(2,\reals)_k$ WZW model (and its Euclidean
continuation $\mrm{SL}(2,\bbC)_k/\mrm{SU}(2)$), and its relation to
$\mrm{SL}(2,\reals)_k/\mrm{U}(1)$ via the coset construction
\cite{Teschner:1999ug,Giveon:1999tq,Maldacena:2001km}. $R(j,n,w)$ 
defines the normalization of the 
2-point function of the coset primaries $\cO_{jnw}$ and
$\cO_{j,-n,-w}$, with the operator normalization chosen in
Eqn. \ref{eqn:asymptotic}.
Physically, it
is the amplitude for a string
sent from the weak-coupling 
region to reflect off the tip of the cigar.

We will focus for
simplicity on the pure-winding sector, where $n=0$. Then
Eqn. \ref{eqn:r} simplifies to
\begin{align}
  \label{eqn:winding-r}
  R(j,w)
  =& 
    4^{2j-1}
     \frac{k-2}{
     \gamma \lp \frac{2j-1}{k-2} \rp}
     \frac{1}{\gamma(2j)}
     \gamma\lp j + \frac{1}{2}kw\rp
     \gamma\lp j - \frac{1}{2}kw\rp,
\end{align}
where $\gamma(z) \equiv \Gamma(z)/\Gamma(1-z).$

The cigar-sigma model description of the CFT is weakly-coupled
for large $k$, and our goal in this section is to
compute $R(j,w)$ by a saddle-point expansion in the $k \to
\infty$ limit. In order to 
compare with the same limit of the exact result, let us first
determine the large $k$ asymptotics of Eqn. \ref{eqn:winding-r}. 
To do so, we must first decide how $j$ scales with $k$. 
We will restrict our attention to ``heavy'' operators, whose
insertions 
contribute at the same order in $k$ as the leading terms
in the action. We therefore define
\begin{align}
  j = \frac{k \eta}{2},
\end{align}
with $\eta = \cO(k^0)$. Imposing $\mrm{Re}(j) > \frac{1}{2}$
requires $\mrm{Re}(\eta) > \frac{1}{k}$, which relaxes to
$\mrm{Re}(\eta) > 0$ in the large $k$ limit.

The asymptotic behavior of the Gamma function for large complex
values of its argument depends on the direction in the complex
plane in which the limit is taken. To $e^{\cO(z^{-1})}$, it
is given by
\cite{BerryM.V.1991IMSS,Pasquetti:2009jg,Harlow:2011ny}  
\begin{align}
  \label{eqn:Gamma-asymp}
  \Gamma(z) \overright{|z|\to\infty}
  \begin{dcases}
    e^{\lp z - \frac{1}{2} \rp \log(z) - z+ \frac{1}{2} \log(2\pi)
      +\cO(z^{-1})}& \mrm{Re}(z) >0\\ 
    \csc(\pi z)
    e^{\lp z -\frac{1}{2}\rp\log(-z) - z + \frac{1}{2} \log \lp
      \frac{\pi}{2} \rp+ \cO(z^{-1})} 
    & \mrm{Re}(z)<0.
  \end{dcases}
\end{align}
The first line is the usual Stirling approximation, and the
second follows from the first in combination with the identity
$\Gamma(z) \Gamma(-z) = -\frac{\pi}{z} \csc(\pi z).$ The
asymptotics of $\gamma(z)$ are then
\begin{align}
  \label{eqn:gamma-asymp}
  \gamma(z) \overright{|z|\to\infty}
  \begin{dcases}
    \frac{2\sin(\pi z)}{z}
    e^{2z \lp \log(z) - 1 \rp 
      +\cO(z^{-1})}& \mrm{Re}(z) >0\\ 
    -\frac{\csc(\pi z)}{2z}
    e^{ 2z\lp\log(-z) - 1 \rp + \cO(z^{-1})} 
    & \mrm{Re}(z)<0.
  \end{dcases}
\end{align}
Assuming without loss of
generality that $w > 0$, we obtain
\begin{align}
  \label{eqn:cigar-semi-r}
  R(\eta,w) \overright{k\to\infty}
  & \eta^{-2k \eta}(w+\eta)^{k(w+\eta)}
   \csc\lp \pi k \eta\rp
    \sin \lp \frac{\pi}{2} k (w+\eta) \rp\\
  &\times
    \begin{dcases}
    \frac{1}{2}(w-\eta)^{k(\eta-w)}
    \csc \lp \frac{\pi}{2} k (w-\eta) \rp
    &0< \mrm{Re}(\eta) < w\\
    2(\eta-w)^{k(\eta-w)}
   \sin \lp \frac{\pi}{2} k (\eta-w) \rp
   &\mrm{Re}(\eta) > w
 \end{dcases}\nt\\
  &\times  \frac{\eta}{(\eta^2-w^2)\gamma(\eta)}\nt.
\end{align}
We have
kept terms to order $k^0$ in the exponent.
Note
that the bound states now correspond to the poles of $\csc \lp
\frac{\pi}{2} k(w-\eta) \rp$ at $\eta_N=w - \frac{2N}{k}$.
When we compute the
saddle-point expansion, we expect the  contribution
$\sum e^{-S + \cO(k^0)}$ from the order $k$ action 
evaluated on its saddles to reproduce 
the first two lines.
The last line is order $e^{k^0}$, which we expect to arise from the
fluctuation determinant, as well as the
order $k^0$ corrections to the on-shell action.\footnote{It has
  been suggested that the cigar sigma-model is supplemented even at
  large $k$ by a potential that modifies the background in the
  neighborhood of the tip \cite{Kutasov:2005rr,Giveon:2015cma,Giveon:2013ica}.
  It would be interesting to see if the third line of
  Eqn. \ref{eqn:cigar-semi-r}, in particular the factor of
  $\gamma(\eta)$ which originated in the factor of $\gamma \lp
  \frac{2j-1}{k-2} \rp$ in Eqn. \ref{eqn:winding-r}, is
  correctly reproduced by the 1-loop calculation in the pure
  cigar background. We have not attempted to compute this
  determinant, however.}

\subsection{Saddle-Point Expansion}
\label{sec:saddle-point}

We now turn to the calculation of the
semi-classical reflection coefficient by a 
saddle-point expansion of the functional integral 
\begin{align}
  \label{eqn:r-integral}
  R(\eta,w) = \int\limits_{\cC(\eta)} Dr D\theta D\sigma_\pm\, e^{-k
  \til S_{jw}}, 
\end{align}
with action $k\til S_{jw} \equiv  S_{j,0,w}$  (c.f. Eqn. \ref{eqn:action-cigar}):
\begin{align}
  \label{eqn:scaled-cigar-action}
  \til S_{jw}
  = &\frac{1}{4\pi}
      \int\limits_{-L}^L \diff \rho \int\limits_0^{2\pi}\diff\phi\, 
      \bigg( (\partial_\rho r)^2 +
      (\partial_\phi  r)^2
      + \tanh^2(r)
      \lp  (\partial_\rho  \theta)^2 + (\partial_\phi  \theta)^2 \rp
      \bigg)\\
    &-  \lp \eta-\frac{1}{k} \rp \int\limits_0^{2\pi} \frac{\diff
      \phi}{2\pi} \lp  r|_{\rho =L} +  r|_{\rho=-L} \rp
    + L\lp  \eta-1/k\rp^2 -  w^2 L
      \nt\\
    &+\int\limits_0^{2\pi} \frac{\diff \phi}{2\pi}
      \bigg(
      \sigma_+ \lp \partial_\phi \theta|_{\rho = L} + w \rp
      +\sigma_- \lp \partial_\phi \theta|_{\rho = -L} + w \rp
      \bigg)\nt.
\end{align}
As discussed in Sec. \ref{sec:cigar-asymptotics}, the maps $r$
and $\theta \sim \theta + 2\pi$ are defined on a worldsheet cylinder
$[-L,L]\times S^1$ whose length $2L$ is taken to infinity.
In this limit, the boundary terms insert the operators
$\cO_{j=\frac{k\eta}{2},n=0,\pm w}$, as well as the
background-charge contributions, on opposite  ends of the
cylinder, and the functional integral computes the reflection
coefficient. In the $k \to \infty$ limit, we would like to
evaluate this integral by a saddle-point expansion 
\begin{align}
  R(\eta,w) \overright{k \to \infty}
  \sum_{r_i,\theta_i} e^{-k \til
  S_{jw} + \cO(k^0)}, 
\end{align}
where $\{r_i,\theta_i\}$ are a subset of solutions
of the equations of motion
\begin{subequations}
\begin{align}
  &  (\partial_\rho^2+\partial_\phi^2)r
    - \tanh(r) \sech^2(r)
    \lp  (\partial_\rho  \theta)^2 + (\partial_\phi  \theta)^2 \rp
    =0\\
  \label{eqn:bulk-theta}
  &  (\partial_\rho^2+\partial_\phi^2)\theta
    +2 \sech(r)\csch(r)\lp  \partial_\rho r \partial_\rho \theta
    +\partial_\phi r \partial_\phi \theta\rp
    =0
\end{align}
\end{subequations}
in the bulk and
\begin{subequations}
  \begin{align}
      &\partial_\rho r|_{\rho=\pm L} = \pm \eta \\
      &\partial_\phi \theta |_{\rho = \pm L} = -w\\
      &\int\limits_0^{2\pi} \diff\phi\, \tanh^2(r)
        \partial_\rho\theta|_{\rho = \pm L} = 0
  \end{align}
\end{subequations}
on the boundaries.\footnote{As explained in
  Sec. \ref{sec:cigar-asymptotics},  the Lagrange multipliers
  are then determined on-shell
  from $\partial_\phi \sigma_\pm = \pm \tanh^2(r) \partial_\rho
  \theta|_{\rho = \pm L}$ up to a $\phi$ zero-mode, which we
  discard. Evaluated on $\theta = -w\phi$, one obtains
  $\partial_\phi \sigma_\pm = 0$, and we then gauge-fix
  $\sigma_\pm = 0$.} 
We have discarded here the contributions from the $\cO(k^{-1})$ 
terms in $\til S_{jw}$, which contribute to the order one
corrections to the saddle-point expansion.

In the limit $L \to \infty$, the boundary equations of motion for
$r$ impose the asymptotic conditions
\begin{align}
  \label{eqn:r-asymptotic}
  r \overright{\rho \to \pm \infty} \pm \eta \rho + a_\pm,
\end{align}
which sends $r$ to the weak-coupling region since
$\mrm{Re}(\eta)>0$. The integration 
constants $a_\pm$ are sub-leading, but control the asymptotic
separation $r(\infty) - r(- \infty) = a_+ - a_-$.

\begin{figure}[h]
  \centering
  \ig{0cm}{width=\linewidth/3}{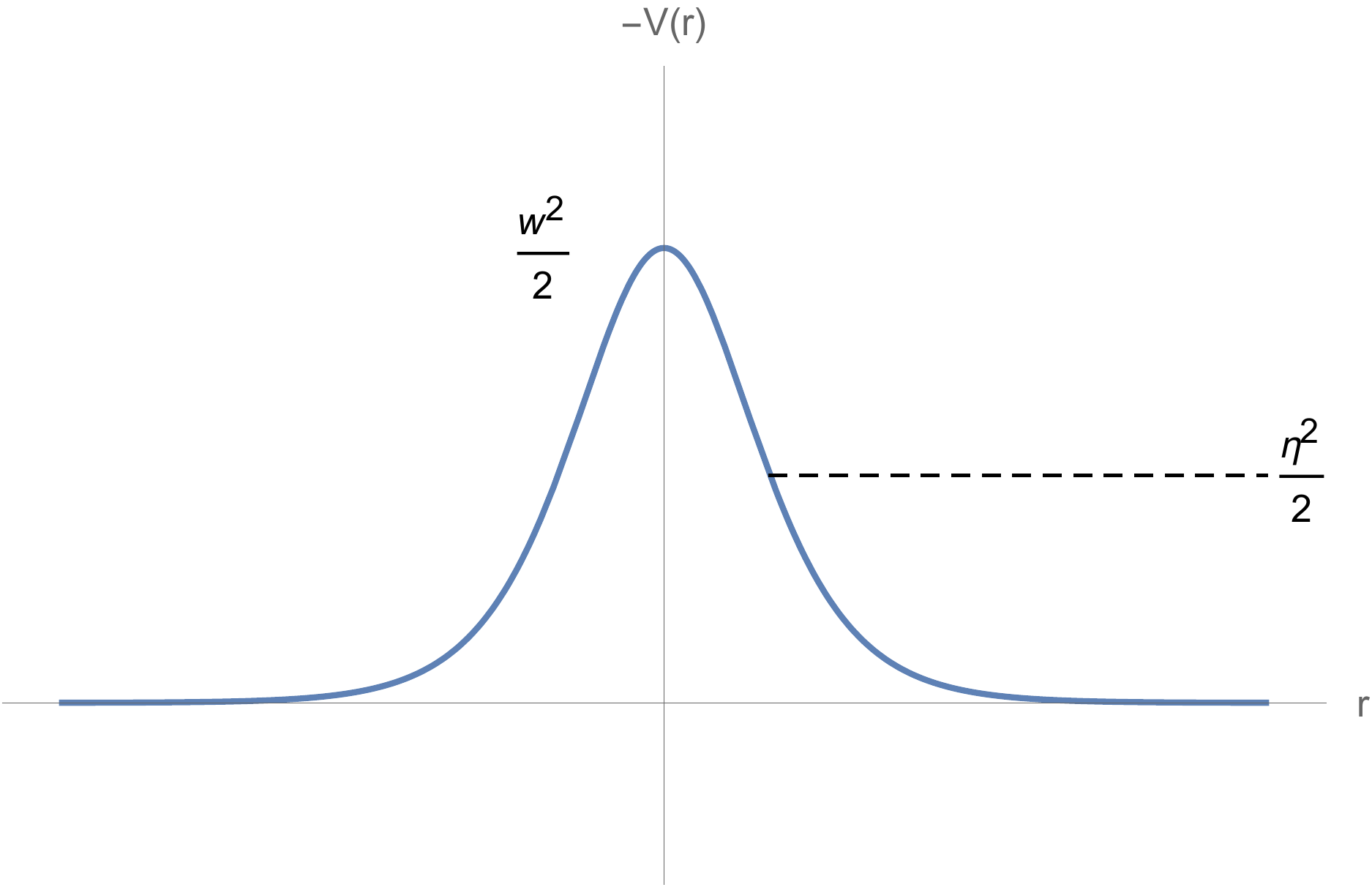}
  \caption{\footnotesize \bd{Inverted Potential}. Restricted to a pure-winding
  solution $\theta = - w \phi$, the cigar equations of motion
  describe the mechanics of a particle moving in the
  inverted potential shown. For $\eta$ real and less than $w$,
  there is a real solution that describes a particle that comes in
  from $r \to \infty$, rolls partway up the potential hill
  until it stops at the turning point, and then rolls back out
  to infinity.
  The cigar geometry is defined for
  $r\geq 0$, but to compute the saddle-point expansion of the
  functional integral we will continue $r$ to the complex
  plane. Here we draw the potential for the real $r$ slice.}
  \label{fig:inverted-potential}
\end{figure}

Applying Eqn. \ref{eqn:r-asymptotic} in the boundary $\theta$ equations
of motion allows us to discard the factor of $\tanh^2(r)$ in the
large $L$ limit. Then these equations imply the asymptotic
condition
\begin{align}
  \theta \overright{\rho \to \pm \infty}
  -w \phi,
\end{align}
which demands that $\theta$ have winding $-w$ around the ends of
the cylinder. The simplest solution sets
$\theta = - w \phi$ everywhere, on which
the equations of motion reduce to
\begin{subequations}
  \begin{align}
    \label{eqn:r-bulk}
  &\partial_\rho^2r - w^2\tanh(r) \sech^2(r) =0\\
  &\partial_\phi r =0\\
  &\partial_\rho r|_{\rho=\pm L} = \pm \eta.
\end{align}
\end{subequations}
Thus, in this pure-winding
sector the theory reduces to a quantum mechanics problem for
$r(\rho)$ with action
\begin{align}
  \label{eqn:qm-action}
  \til S[r]=\int\limits_{-L}^L \diff \rho\,
  \bigg(
  \frac{1}{2} \lp \td{r}{\rho} \rp^2
  + V(r)
  \bigg)
  -\eta(r(L)+r(-L))+ \eta^2L ,
\end{align}
where
\begin{align}
  \label{eqn:qm-potential}
  V(r) =-\frac{1}{2} w^2\sech^2(r).
\end{align}
The bulk equation of motion Eqn. \ref{eqn:r-bulk} may be written
\begin{align}
  \label{eqn:bulk-eom}
  \tdt{r}{\rho} = V'(r),
\end{align}
describing the mechanics of a particle in the inverted
potential $-V(r)$, pictured in
Fig. \ref{fig:inverted-potential}. The quantum mechanics of a
particle in this potential can in fact be solved exactly, as we
review in Appendix \ref{sec:cigar-qm}. There we show that the
semi-classical limit of the exact reflection coefficient for the
quantum mechanics, Eqn. \ref{eqn:half-semi}, reproduces that
of the CFT, Eqn. \ref{eqn:cigar-semi-r}, at order $e^k$. We
conclude that the saddles with $\theta(\phi) = - w\phi$ and
$r=r(\rho)$ are sufficient for reproducing the saddle-point
expansion of the coset reflection coefficient, and we do not
need to consider more complicated solutions $\theta(\rho,\phi)$
and $r(\rho,\phi).$

Since we allow for complex values of $\eta$, clearly
the saddles of the functional integral will in general be
complex.
In fact, even for real values of $\eta$ we will see that one
must sum over complex saddles.
The necessity of complexification is familiar from 
applications of the saddle-point method to asymptotic expansions
of ordinary integrals over real variables, where the original real
integration contour is typically deformed into a homotopically
equivalent sum of steepest-descent contours passing through
complex critical points.

The asymptotic expansion of
the Gamma 
function in Eqn. \ref{eqn:Gamma-asymp} may itself be understood
as a finite-dimensional  
example of the saddle-point expansion, since $\Gamma(z)$ may
be defined  by the integral
\begin{align}
  \Gamma(z) = \int\limits_{\cC(z)} \diff  X \, e^{-(-z X +e^ X )}.
\end{align}
We refer the reader to Appendix C of \cite{Harlow:2011ny} for a
self-contained review, because the problem for the functional
integral is in many ways analogous. Briefly, the saddle-points
of the ``action'' $S[ X ] = -z  X  + e^ X $ are given by
$ X _N = \log(z) + 2\pi i N$, with $N \in \bbZ$.
The contour $\cC(z)$ is given by the real axis for 
$\mrm{Re}(z)>0$, though it must be deformed for $\mrm{Re}(z)<
0$ to preserve convergence of the integral. For
$\mrm{Re}(z)>0$, $\cC(z)$ is homotopic to the steepest-descent
contour $\cC_0$ through $ X _0$, and one recovers the Stirling
formula $e^{-S[ X _0]}=e^{z\log(z)-z}$.

Along the imaginary
$z$-axis, however, one encounters what is known as a Stokes
wall. There the steepest-descent contour of any
saddle-point, which otherwise varies smoothly with $z$, collides
with a neighboring saddle-point. As a result, for values of
$z$ just to either side of the imaginary axis, the
steepest-descent contour jumps discontinuously. 
The integration contour $\cC(z)$ itself varies
smoothly with $z$, but its expansion in steepest-descent contours
changes abruptly upon crossing the Stokes wall, and
therefore its asymptotic expansion changes as well. For
$\mrm{Re}(z)<0$ and $\mrm{Im}(z)>0$, $\cC(z)$ is instead
homotopic to the sum of steepest-descent contours
$\sum_{N=0}^\infty \cC_N$ passing
through the saddles $ X _N$. Then the saddle-point
expansion yields
\begin{align}
  \sum_{N=0}^\infty e^{-S[ X _N]}
  =& e^{-S[ X _0]} \sum_{N=0}^\infty e^{2\pi i z N }\\
  =&\csc(\pi z)e^{z\log(-z)-z+\cO(z^0)},
\end{align}
as in Eqn. \ref{eqn:Gamma-asymp}. For $\mrm{Im}(z)< 0$, the
relevant contours are instead $\sum_{N=0}^\infty \cC_{-N}$, as
required for the geometric series to converge.

Thus, the poles of the Gamma
function on the negative real axis may be understood as the
divergence of the geometric series $\sum_{N=0}^\infty e^{2\pi i
  z N }$ that results from summing over a family of saddles
related by complex shifts $ X  \to  X  + 2\pi i$. We will
see that the poles of $\csc(\pi k \eta)$ in
Eqn. \ref{eqn:cigar-semi-r} are of similar origin. The
poles of $\csc\lp \frac{\pi}{2}k(w-\eta) \rp$, meanwhile, will
be attributed to summing over a family of singular saddles
with finite action. In Sec. \ref{sec:sL}, we will also see that
the poles of the prefactor $\gamma \lp \frac{2j-1}{k-2}\rp$ in
Eqn. \ref{eqn:winding-r} are due to a complex shift symmetry of
the dual sine-Liouville description of the coset CFT in the
$k\to 2$ limit.

As in the finite-dimensional problem,
the saddle-point expansion of a
functional integral is performed by deforming the
integration contour into a sum of complex cycles
\cite{Witten:2010cx,Harlow:2011ny}. Unlike for a
finite-dimensional integral, however, for a functional integral
it is in general very challenging to determine the
set of steepest-descent cycles that are homotopic to the
original contour. In other words, it is a hard problem to
derive from first principles which complex saddles one should sum
over in computing the saddle-point expansion, especially since
the necessary set of saddles can 
jump upon crossing Stokes walls in the parameter
space.
Since for the problem at hand we know the 
exact answer and its semi-classical limit
Eqn. \ref{eqn:cigar-semi-r}, the approach we take 
here is to show that there exists 
a consistent set of solutions for
which the saddle-point expansion reproduces the known answer.

In light of the abrupt change in the semi-classical
limit of $R(\eta,w)$ across the line $\mrm{Re}(\eta) = w$, we
anticipate that a Stokes wall in the $\eta$-plane is found there. This is
not unreasonable, since for $\eta$ real and less than $w$ there exists
a real solution of the equations of motion describing a particle
that comes in from $r \to \infty$, rolls partway up the
potential until it stops at the turning point, and then rolls
back out to infinity. For $\eta$ greater than $w$, on the other
hand, the particle rolls over the potential and continues to $r
\to -\infty$ in the continued field space.
At the crossover point, the particle has just
enough energy to (asymptotically) reach the top of the
hill.

Thus, in computing the saddle-point expansion
we expect that we will need to address the domains
for $0< \mrm{Re}(\eta) < w$ and $\mrm{Re}(\eta)> w$ separately, and
that we will find a different set of contributing saddles in
each. 

Finally, we have not yet specified the contour $\cC(\eta)$ in field space
along which the functional integral Eqn. \ref{eqn:r-integral} is
to be performed. In the example of the Gamma function, one
starts with a contour 
along the real axis when $z$ is a positive real
number. Then one deforms it as necessary for complex values of $z$
to preserve convergence of the integral and produce an analytic
function of $z$.

By contrast, even for real $\eta$ the
functional integral Eqn. \ref{eqn:r-integral} over real fields
diverges. Borrowing an argument from\footnote{In
  \cite{Harlow:2011ny} it is 
  similarly shown that the functional integral over real fields
  for the Liouville 2-point function diverges. In
  \cite{Seiberg:1990eb}, this divergence was dealt with by a
  fixed-area prescription, while in \cite{Harlow:2011ny} it was
  interpreted as an indication that  the functional integral
  must instead be defined over a complex cycle.}
\cite{Harlow:2011ny}, consider a 
finite-action configuration of $(r,\theta,\sigma_\pm)$. Now consider
another configuration with $r \to r + a$ shifted by a large,
positive real number. Then the action of the latter
configuration is given by
\begin{align}
  \til S_{jw}[r+a]
  \overright{a \to \infty}
  -2\eta a
  +
  \til S_{jw}[r]
  +
  \frac{1}{4\pi} \int \diff\rho\, \diff \phi\,
  \sech^2(r) (\del \theta)^2
  +\cO(e^{-2a},k^{-1}).
\end{align}
Since $\mrm{Re}(\eta)>0$, by making $a$ arbitrarily large
the action may be made arbitrarily negative. Thus, we have
identified a region of real field space where $e^{-S} \to
\infty$, and therefore the functional integral over real fields
cannot converge.

Instead, $\cC(\eta)$
must be chosen to be an appropriate complex cycle. By
identifying the set of saddles with which the semi-classical limit
of the exact result is reproduced, the appropriate contour may
in principle be defined by the sum of steepest-descent contours
associated to these complex saddles. Away from the Stokes walls,
the steepest-descent contours themselves vary smoothly with
$\eta$, as does $\cC(\eta)$ in turn. As in the
finite-dimensional case,
even though the steepest-descent contours jump when $\eta$
crosses a Stokes wall, $\cC(\eta)$ is expected to vary
smoothly. Its expansion as a sum of steepest-descent contours changes
across the Stokes wall, but the summed contours on either side
of the wall should be equivalent up to Cauchy deformation.

\subsubsection{Complex Quantum Mechanics}
\label{sec:complex-qm}

On the pure-winding solution $\theta = -w\phi$, the cigar action
has reduced to the complex quantum mechanics in Eqn. \ref{eqn:qm-action}.
Since the original real coordinate $r \geq 0$ was valued in a
half-line, the relevant complexification is the complex
$r$-plane quotiented by $r \sim - r$. Note that this is a
symmetry of the $\sech^2(r)$ potential.

Alternatively, one may compute the saddle-point expansion for
the reflection and transmission coefficients of the 
quantum mechanics before the quotient,
and then take their difference to obtain the reflection
coefficient in the half-space. This is the approach that we will take here.
Thus, we regard the action $\til S[r]$ in
Eqn. \ref{eqn:qm-action} as a holomorphic 
functional of maps $r \cn \reals \to \bbC$ into the complex
$r$-plane and identify its critical points.\footnote{The
  transmission coefficient is computed similarly, 
  but with $r(L)$ replaced by $-r(L)$ to
  fix the momentum at late times to $-\eta$ rather than $\eta$.}

The semi-classical limits of the reflection and transmission
coefficients obtained from the exact solution of the full-space
quantum mechanics are (c.f. Eqns. \ref{eqn:qm-r-semi} and
\ref{eqn:qm-t-semi}) 
\begin{align}
  \label{eqn:r-qm-limit}
  &R_\mrm{QM}(\eta)\\
  &\underset{\propto}{\overright{k\to\infty}}
    \begin{dcases}
      \eta^{-2k \eta}( w +\eta)^{k( w +\eta)}
      ( w -\eta)^{k(\eta- w )}
      \sin(\pi k  w )
      \csc\lp \pi k \eta\rp
      \csc \lp \pi k ( w  -\eta) \rp
      &0< \mrm{Re}(\eta) <w\\
      \eta^{-2k \eta}( w +\eta)^{k( w +\eta)}(\eta- w )^{k(\eta- w )}
    \sin(\pi k  w )
    \csc\lp \pi k \eta\rp
    &\mrm{Re}(\eta) > w
    \end{dcases}\nt
\end{align}
and
\begin{align}
  \label{eqn:t-qm-limit}
  T_\mrm{QM}(\eta) \underset{\propto}{\overright{k\to\infty}}
  \begin{dcases}
    \eta^{-2k \eta}( w +\eta)^{k( w +\eta)}  ( w -\eta)^{k(\eta- w )}
    \csc \lp \pi k ( w  -\eta) \rp
    &0< \mrm{Re}(\eta) <  w \\
    \eta^{-2k \eta}( w +\eta)^{k( w +\eta)}  (\eta- w )^{k(\eta- w )}
   &\mrm{Re}(\eta) >  w. 
 \end{dcases}
\end{align}
It is shown in Appendix \ref{sec:cigar-qm} that
the reflection coefficient of the half-space quantum mechanics,
$R_\mrm{QM}(\eta) - T_\mrm{QM}(\eta)$, reproduces the cigar
reflection coefficient at order $e^k$.
Thus, our task is reduced to reproducing
Eqns. \ref{eqn:r-qm-limit} and \ref{eqn:t-qm-limit} by 
saddle-point expansions for the infinite-space quantum mechanics.
The rest of this section is devoted to that calculation. In the
remainder of the present sub-section, we discuss some
generalities about the complex quantum mechanics and its saddles.
In the following two sub-sections, we compute the
saddle-point expansions of the reflection and transmission
coefficients.

The bulk equation of motion Eqn. \ref{eqn:bulk-eom} describes
a particle moving in an inverted potential $-V(r)$.
One therefore obtains the energy conservation equation,
\begin{align}
  \frac{1}{2} \dot{r}^2 -  V ( r ) =
  \frac{\eta^2}{2},
\end{align}
where $\dot{r} = \td{ r }{\rho}.$
The conserved energy is indeed $\frac{\eta^2}{2}$, as is clear
by evaluating the equation at $\rho \to \pm \infty$ and imposing
the asymptotic conditions.

Observe that, as a holomorphic function on the complex
$ r $-plane, the potential is periodic in $\pi i$:
\begin{align}
   V ( r  + \pi i) =  V ( r ).
\end{align}
The turning points, where
$- V ( r _\pm)=\frac{\eta^2}{2}$ and therefore $\dot{r} = 0$, are given by
\begin{align}
  \label{eqn:turning-points}
   r _\pm = \pm \cosh^{-1} \lp \frac{ w }{\eta}\rp,
\end{align}
as well as all shifts thereof by $\pi i \bbZ$.

There are also singular points where the potential diverges. 
$ V ( r )$ has a double-pole at $ r  =
\frac{\pi i}{2}$,
\begin{align}
  \label{eqn:potential-singularity}
   V ( r ) \overright{\, r \to\frac{\pi i}{2}\vphantom{\big(}\,} 
  \frac{ w ^2}{2} \frac{1}{\lp  r  - \frac{\pi i}{2}
  \rp^2} + \cO(1),
\end{align}
and likewise at all points $\frac{\pi i }{2} + \pi i \bbZ$. We
point out that $r=\pm \frac{\pi i}{2}$ coincide with the
physical singularities of the Lorentzian black hole after
continuing $\theta$ to Lorentzian time.

To compute the saddle-point expansion, we must evaluate the
action on the solutions of the equations of motion.
We will obtain these explicit solutions momentarily, but
it is not actually necessary to solve the equations of motion in
order to compute the on-shell action.
Using the energy conservation equation, we may write 
Eqn. \ref{eqn:qm-action} as 
\begin{align}
  \til{S}[ r ] = \int_{-L}^L \diff \rho\,
  \lp \td{ r }{\rho} \rp^2 
  -\eta( r (L)+ r (-L)). 
\end{align}
Letting $\mathscr{C}$ denote the contour traced by the solution
in the complex $r$-plane, we may write the action
as a contour integral:
\begin{align}
  \label{eqn:r-contour}
  \til{S}[ r ]  = \int_\mathscr{C} \diff  r \,
  \sqrt{\eta^2+2 V ( r )}
  -\eta( r (L)+ r (-L)). 
\end{align}
Note that the integrand $\sqrt{\eta^2+2V(r)} = \dot{r}$ is the
velocity function, and one should pick an appropriate branch of the
square-root such that the velocity has the correct sign.
The turning points $ r _\pm + \pi i \bbZ$ are branch points of the
square-root. The double-poles of the potential, meanwhile,
lead to simple-poles of the integrand of residue $\pm w $:
\begin{align}
  \pm\sqrt{\eta^2 + 2  V ( r )}
  \overright{\, r \to\frac{\pi i}{2}\vphantom{\big(}\,} 
  \pm  \frac{   w  }{ r  - \frac{\pi i}{2}} + \cO(1).
\end{align}
The explicit solutions of the bulk equation of motion may be
obtained by separating and integrating the energy conservation
equation. One finds
\begin{align}
  \label{eqn:qm-solution}
  r (\rho) = \sinh^{-1} \lp
  \sqrt{ \frac{w^2}{\eta^2}-1} \cosh(\eta(\rho+i\rho_0)) \rp
  + \pi i N_1,
\end{align}
where $\rho_0$ is a complex number and $N_1$ is an
integer. $\rho_0$ is the integration constant that arises in
integrating the energy conservation equation. In the limit $L
\to \infty$, the real part of $i \rho_0$ is merely a
reparameterization of $\rho$; we therefore take $i \rho_0$
to be pure imaginary. The freedom to shift any solution by $\pi
i N_1$ arises from the periodicity of the
potential. For each $N_1$, the continuous modulus $\rho_0$
parameterizes a family of 
solutions. The on-shell action is necessarily the same for all trajectories in
such a family, unless in varying $\rho_0$ one encounters a
singular solution.\footnote{The divergent sum over $\rho_0$ is
  attributed to the infinite $\delta(j-j)$ factor in the
  2-point function of $\cO_{jnw}$ and $\cO_{j,-n,-w}.$ \cite{Harlow:2011ny}}
The on-shell action does depend on the
discrete parameter $N_1$, however, through the boundary terms.

Since $\sinh^{-1}(z) = \log \lp z + \sqrt{z^2+1} \rp$ is a
multi-valued function, one has to pick a branch to define the
trajectory. The $\sqrt{z^2+1}$ term leads to square-root branch
points at $z = \pm i$, and there is
a logarithmic branch point at infinity. On the
principal branch, the cuts
extend along the imaginary axis from $i$ to $i \infty$ and from
$-i$ to $-i \infty$, though other choices are convenient
depending on the values of the parameters.\footnote{
  In particular, for complex $\eta$
  the argument of the $\sinh^{-1}$ in Eqn. \ref{eqn:qm-solution}
  behaves as a spiral at large $|\rho|$ since $\cosh(\eta \rho)
  \sim e^{\eta |\rho|}$. In that case one has to pick
  more complicated spiral branch cuts.}

Eqn. \ref{eqn:qm-solution} solves the bulk equation of motion,
but it remains to check if it satisfies the boundary equations.
The velocity function is
\begin{align}
  \label{eqn:rdot}
  \dot{r}= \eta \frac{ \sqrt{\frac{w^2}{\eta^2}-1} \sinh(\eta(\rho+i\rho_0))}
  {\sqrt{ \lp
  \frac{w^2}{\eta^2}-1\rp \cosh^2(\eta(\rho+i\rho_0))+1}},
\end{align}
which indeed asymptotes to $\pm \eta$ as $|\rho| \to
\infty$. The sign, however, depends on the branch of the
square-root in the denominator, which coincides with the branch
of the square-root in $\sinh^{-1}(z).$ Depending on the values
of the parameters, one obtains either a reflected or transmitted
solution.

\begin{figure}[H]
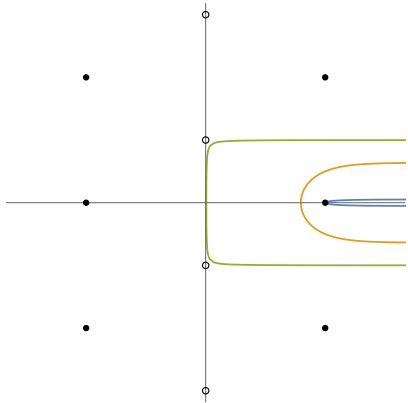

  \centering
  \ig{0cm}{width=\linewidth/3}{underbarrier}
  \caption{\footnotesize \bd{A Family of Reflected Solutions}. Pictured here are
    several solutions with $\eta$ real and less than $w$, $0< \rho_0 <
    \frac{\pi}{2\eta}$, and $N_1 = 0$, obtained from
    Eqn. \ref{eqn:qm-solution} with the principal 
    branch of $\sinh^{-1}$. 
    The solid disks indicate the turning points and the open
    circles indicate the singularities. The 
    solution hugging the real axis has $\rho_0$ just above zero,
    corresponding to the real solution that rolls up and down
    the same side of the inverted-potential. All trajectories
    related by continuously dialing $\rho_0$ have the same action, unless one
    hits a singular trajectory in the process.  The green
    solution pictured is nearly singular, with $\rho_0$ just below
    $\frac{\pi}{2\eta}$. At that value the trajectory will hit
    the poles of the potential.}
  \label{fig:underbarrier}
\end{figure}

For example, several 
trajectories with $\eta$ real and less than $w$, $0< \rho_0 <
\frac{\pi}{2\eta}$, and $N_1 = 0$
are plotted in Fig. \ref{fig:underbarrier}.
The solid disks indicate the turning points $r_\pm + \pi i
\bbZ$, and the open circles indicate the singularities
$\frac{\pi i}{2} + \pi i \bbZ$.
The blue trajectory that hugs the real axis
corresponds\footnote{In the figure, $\rho_0$ is deformed slightly away from
  0 so that the incoming and outgoing segments of the trajectory
  do not overlap.}
to the real solution with $\rho_0 = 0$ for a particle that rolls
up and down the potential hill, turning around at $r_+$.
For the green trajectory nearly hitting the poles, on the other
hand, $\rho_0$ is just below $\frac{\pi}{2\eta}$.

\begin{figure}[h]
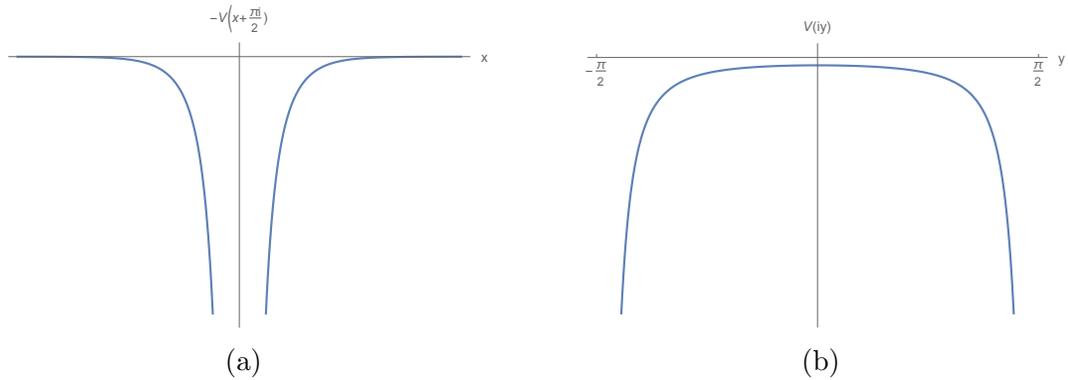

  \centering
  \begin{subfigure}[t]{.4\textwidth}
    \centering
    \ig{0cm}{width=\linewidth}{x-potential}
    \caption{\footnotesize }
    \label{fig:x-potential}
  \end{subfigure}
\hspace{1cm}
\begin{subfigure}[t]{.4\textwidth}
  \centering
  \ig{0cm}{width=\linewidth}{y-potential}
  \caption{\footnotesize }
  \label{fig:y-potential}
\end{subfigure}
\caption{\footnotesize \bd{Singular Potentials}. On the slices $\mrm{Im}(r) =
  \frac{\pi}{2}$ (left) and $\mrm{Re}(r) = 0$ (right), the
  potential experienced by the particle falls to $-\infty$ at
  the singular points.}
\label{fig:singular-potentials}
\end{figure}

Upon reaching $\rho_0 = \frac{\pi}{2\eta}$, the asymptotic
imaginary part of the trajectory reaches $\pm \frac{\pi}{2}$, and
the saddle becomes singular. The argument of the $\sinh^{-1}$ in
Eqn. \ref{eqn:qm-solution}
hits the branch points at $\pm i$ at finite $\rho$.  Indeed, the
inverted potential on 
the real slice $\mrm{Im}( r ) = \pm \frac{\pi}{2}$ is an infinite
well, $- V \lp  x  \pm \frac{\pi i}{2}\rp =
-\frac{1}{2}  w ^2 \csch^2( x )$, pictured in
Fig. \ref{fig:x-potential}, and a particle kicked to the left
from $x > 0$ falls down the well and hits the
singularity. Similarly, on the imaginary axis
$ r  = i y,$ the potential experienced by $y$
is\footnote{Note 
  that the potential for the imaginary part of the complex
  coordinate has a relative minus sign, due to the combined factors of
  $i$. The equation of motion is $\ddot{y} = \td{}{y}\lp
  \frac{1}{2} w^2 \sec^2(y) \rp$.} 
$V(iy)=-\frac{1}{2} w ^2 \sec^2(y)$, which is again singular, as pictured
in Fig. \ref{fig:y-potential}.
Remarkably, we will see that the singular
trajectories carry finite action and must be included
in the saddle-point expansion to correctly reproduce the
semi-classical reflection coefficient. The importance of singular
saddles was discussed in closely related contexts in
\cite{Harlow:2011ny,Behtash:2015loa}.

\subsubsection{Reflection Coefficient on the Complex $r$-Plane}
\label{sec:complex-r}

\begin{figure}[h]
  \centering
  \ig{0cm}{width=\linewidth/3}{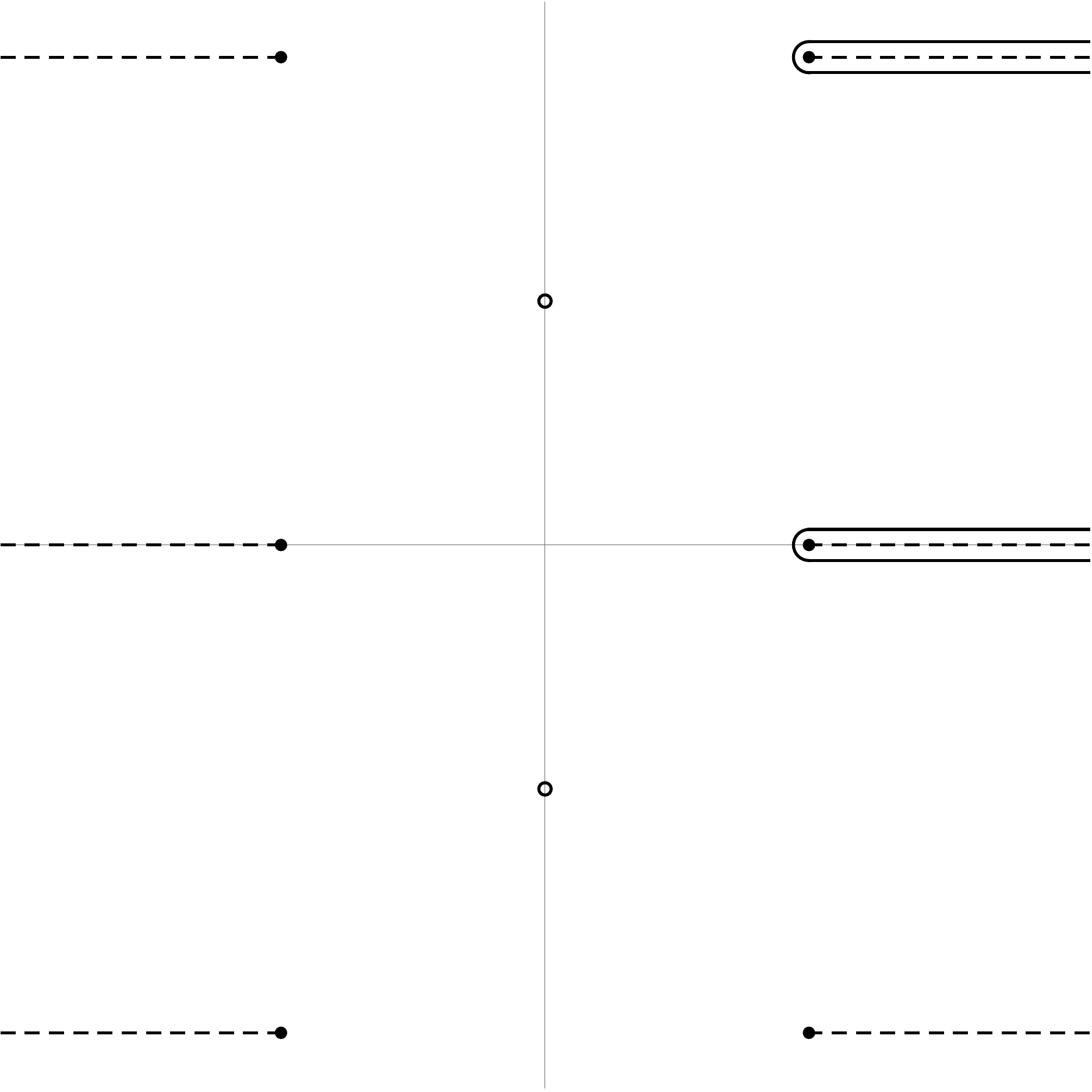}
  \caption{\footnotesize \bd{Reflected Saddles with Constant Imaginary
      Part}. For $\eta$ real and less than 
  $w$, there is a real saddle corresponding to a particle that
  comes in from $r \to \infty$, rolls partway up the hill until
  it stops at the turning point, and then rolls back out to
  infinity. The corresponding contour in the complex $r$-plane
  is pictured here, slightly deformed away from being pure real
  so that the incoming and outgoing segments of the contour do
  not overlap. Due to the $r \to r + \pi i$ shift symmetry of
  the complexified problem, one likewise has shifted contours
  with constant imaginary part $\pi N_1$, such as the second
  contour in the figure with $N_1 = 1$. The dashed lines
  represent branch cuts of the square-root in
  Eqn. \ref{eqn:r-contour}. }
  \label{fig:real-reflected}
\end{figure}

We now consider the saddle-point expansion of the reflection
coefficient for the complex quantum mechanics on the full
$r$-plane. Let us begin by computing the action of the real
saddle that exists for $0 < \eta < w$. We need to evaluate 
Eqn. \ref{eqn:r-contour} for the  
contour around the positive real axis in
Fig. \ref{fig:real-reflected}, which has been 
slightly deformed away from the real axis so that its incoming
and outgoing legs do not overlap. The dashed lines represent
a convenient choice of branch cuts of the square-root in
the integrand. As before, the solid disks indicate
the turning points and the open circles indicate the singularities.

The contribution to the contour integral of the small arc around
the turning point vanishes because the integrand is zero
there. The contributions of the remaining half-lines above and
below the real axis are identical because they sit on opposite
sides of the branch cut and have opposite orientations,
corresponding to the particle coming in from infinity and then
going back out to infinity. Each of these two integrals
contributes
\begin{align}
  \int\limits_{r_+}^{r(L)}\diff  r \, \sqrt{\eta^2 + 2
  V( r )}
  = \eta \log(\eta) + 
  \frac{ w -\eta}{2}\log( w -\eta)
  -\frac{w +\eta}{2}\log( w +\eta)
  + \eta  r (L),
\end{align}
in the limit $L \to \infty$. Combined with the boundary term
that cancels the linear divergence $\eta  r (L)$, we obtain
the on-shell action
\begin{align}
  \label{eqn:real-action}
  \til{S}_0  =
  2\eta \log(\eta) + 
  ( w -\eta)\log( w -\eta)
  -( w +\eta)\log( w +\eta). 
\end{align}
The contribution of this solution to the saddle-point expansion
is then
\begin{align}
  \label{eqn:real-contr}
  e^{-k \til{S}_0}
  =
  \eta^{-2k\eta} ( w +\eta)^{k( w +\eta)}
  ( w -\eta)^{k(\eta- w)}. 
\end{align}
This accounts for the first half of Eqn. \ref{eqn:r-qm-limit}
with $0< \mrm{Re}(\eta) < w$, leaving 
the three trigonometric factors still
to be explained.

The simplest of these three factors to understand is $\csc(\pi k
\eta)$. It is due to the shift symmetry of the potential
\cite{Harlow:2011ny}. 
Even when $\eta$ is real, one has
complex solutions shifted by $\pi i N_1$, for any integer $N_1$.
The contour with $N_1 = 1$ is also shown in Fig. \ref{fig:real-reflected}.
Each of these shifted saddles 
has action $\til{S}_0 -2\pi i \eta N_1$, due to the shift
of the boundary terms. One should not sum over all of them,
however; the appropriate set depends on the sign of
$\mrm{Im}(\eta)$, which determines whether one obtains a
convergent geometric series when $N_1 \in \bbZ_{\geq 0}$ or when $N_1 \in
\bbZ_{\leq 0}$. For $\mrm{Im}(\eta) > 0$ one finds
\begin{align}
  \label{eqn:shift-contr}
  \sum_{N_1\in \bbZ_{\geq 0}} e^{2\pi i k \eta N_1}
  =\frac{i}{2}e^{-\pi i k \eta} \csc(\pi  k \eta),
\end{align}
while for $\mrm{Im}(\eta) < 0$
\begin{align}
  \label{eqn:shift-contr2}
  \sum_{N_1\in \bbZ_{\leq 0}} e^{2\pi i k \eta N_1}
  =-\frac{i}{2}e^{\pi i k \eta} \csc(\pi  k \eta).
\end{align}
In fact, since the reflection coefficient has poles on
the real $\eta$-axis, one should always give $\eta$ a non-zero phase
in computing the saddle-point expansion. We see here that
depending on whether $\eta$ lies above or below the real axis,
we must pick one or the other half-infinite set of shifted
saddles.\footnote{The same was true for the Gamma function example
discussed earlier. $\Gamma(z)$ has poles on the negative real
axis, and depending on whether $\mrm{Im}(z)>0$ or $\mrm{Im}(z) <
0$, the integration contour is deformed into the
steepest-descent contours for saddles in either the upper or
lower-half $X$-plane.}

Next consider the factor  of $\csc(\pi k ( w  -\eta))$, which
accounts for the bound states of the infinite-space quantum
mechanics. We now argue that it is the singular saddles 
that are responsible for this factor. 
This requires some explanation, since one ordinarily expects
singular configurations to have infinite action and therefore to make
no contribution to the functional integral. In contrast, the
singular saddles in this complexified problem have finite
action, and are essential to reproducing the correct
semi-classical limit \cite{Harlow:2011ny,Behtash:2015loa}. 

\begin{figure}[h]
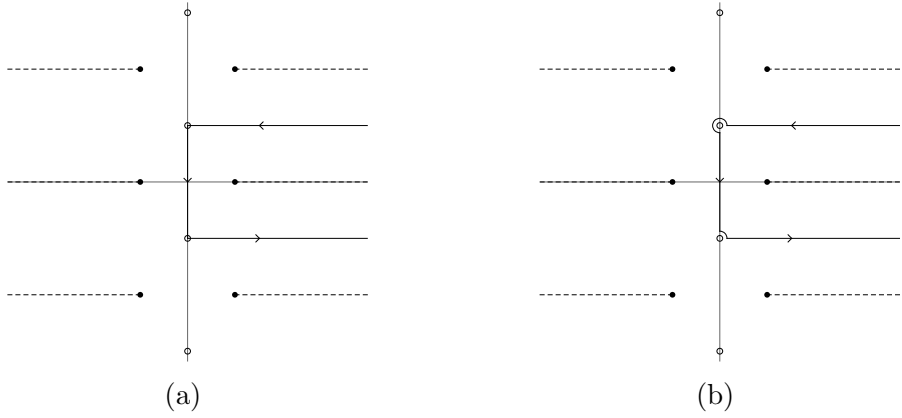

  \centering
  \begin{subfigure}[t]{.3\textwidth}
    \centering
    \ig{0cm}{width=\linewidth}{singular}
    \caption{\footnotesize }
    \label{fig:real-singular}
  \end{subfigure}
\hspace{2cm}
\begin{subfigure}[t]{.3\textwidth}
  \centering
  \ig{0cm}{width=\linewidth}{singular1}
  \caption{\footnotesize }
  \label{fig:real-singular-deformed}
\end{subfigure}
\caption{\footnotesize \bd{Singular Contours}. When the asymptotic
  imaginary part of the reflected contour shown in
  Fig. \ref{fig:underbarrier} reaches $\pm \frac{\pi}{2}$, the
  contour hits the poles of the potential and becomes
  singular (left).
  The singular contour 
  must be deformed around the poles of the
  potential in order to define the action integral
  Eqn. \ref{eqn:r-contour}. The action of the singular saddle
  then differs from Eqn. \ref{eqn:real-action} by residues. The
  deformation is not unique, and one must sum over an
  appropriate set of singular saddles to reproduce the correct
  semi-classical limit. An example deformation is shown on the right,
  with action $\til S_0 -2\pi i w$.}
\label{fig:deformed-contours}
\end{figure}

Consider the singular saddle discussed 
in the previous sub-section, obtained as the 
$\rho_0 \to \frac{\pi}{2\eta}$ limit of
Eqn. \ref{eqn:qm-solution}, with $N_1 = 0$. The contour is
shown in Fig. \ref{fig:real-singular}. As 
pictured in Fig \ref{fig:singular-potentials}, on the real and
imaginary slices 
$\mrm{Im}( r ) = \pm\frac{\pi}{2}$ and $\mrm{Re}( r ) =
0$, the effective 1-dimensional potential is an infinite well 
in the neighborhood of the singular points. The speed of the
particle subsequently diverges there. However, the remarkable
feature of the complexified problem is that the divergent
contributions to the action are equal-but-opposite on the real and
imaginary segments of the trajectory, the speed of the particle
being pure real and pure imaginary in the two cases. 
One may therefore define the total action by a 
principal-value type limit.  

\begin{figure}[H]
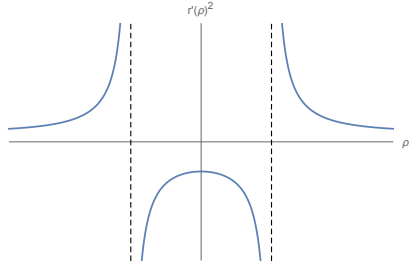

  \centering
  \ig{0cm}{width=\linewidth/3}{singular_square}
  \caption{\footnotesize \bd{Singular Speed-Squared}. The speed-squared
    $\dot{r}^2$ is plotted here for the singular saddle shown in
    Fig. \ref{fig:real-singular}. It has $\frac{1}{\rho}$ type
    singularities where the contour hits the poles of the
    potential. The action, which is the integral of $\dot{r}^2$,
    is finite because $\int_{-\vep}^\vep\frac{\diff \rho}{\rho}$
    may be defined by continuation around the pole. The
    imaginary part of the integral is 
    ambiguous up to shifts by $2\pi i 
    \bbZ$, however,  corresponding to the choice of how to deform
    the contour around the poles.} 
  \label{fig:singular-speed-squared}
\end{figure}

More precisely, let $z =  r  - \frac{\pi i}{2}$ be a local
coordinate in the neighborhood of the singularity at $ r  =
\frac{\pi i}{2}$. From Eqn. \ref{eqn:potential-singularity}, the
potential has a double-pole there, $V(z) = \frac{
  w^2}{2} \frac{1}{z^2} + \cO(1)$, and therefore the energy
conservation equation in this neighborhood becomes
\begin{align}
  \label{eqn:singular-eom}
  \td{z}{\rho} = - \frac{ w }{z} + \cdots,
\end{align}
the minus sign corresponding to the orientation chosen in
Fig. \ref{fig:real-singular}. 
The solution near the upper pole is then
$z(\rho) = -i\sqrt{2  w  (\rho+\rho_1)}$, hitting the pole at $\rho
= - \rho_1$. The speed-squared is 
\begin{align}
  \lp\td{z}{\rho}\rp^2 = - \frac{ w }{2}
  \frac{1}{\rho + \rho_1} + \cdots. 
\end{align}
The integrand of the on-shell action thus has a $\frac{1}{\rho}$
type singularity near $ r  = \frac{\pi i}{2}$, and the
integral is\footnote{The integral $\int_{-a}^a 
  \frac{\diff \rho}{\rho}$ may be defined by continuing $\rho$
  to the complex plane and deforming the contour off the real
  axis:
  \begin{figure}[H]
    \centering
    \ig{0cm}{width=\linewidth/3}{contour-deformation}
  \end{figure}
  The integral over the counter-clockwise semi-circle about the
  pole is $\pi i$. Discarding it defines the principal-value of
  the integral, which is zero in this symmetric case. Of course,
  the deformation of the contour is not unique. One could just as
  well have deformed it into a clockwise arc above the pole which
  would instead yield $-\pi i$, or an arc that encircles the
  pole any number of times. The integral is therefore only defined up to
  shifts by $2\pi i \bbZ$, which is equivalently the ambiguity
  in the choice of branch of $\int \frac{\diff \rho}{\rho} =
  \log (\rho)$.
}
\begin{align}
  \label{eqn:principal-value}
  \int_{-\rho_1 - \vep}^{-\rho_1 + \vep} \diff \rho\,
  \lp \td{z}{\rho} \rp^2
  =
  -\frac{ w }{2} \lp \pi i + 2\pi i N_2 \rp,
\end{align}
with $N_2$ an integer. 
The ambiguity in $2\pi i \bbZ$ amounts to
the choice of branch of $\log \rho = \int \frac{\diff
  \rho}{\rho}.$
One likewise has a $\frac{1}{\rho}$
singularity in the neighborhood of the lower pole.
$\dot{r}^2$ for the complete singular solution with real $\eta$
is plotted in Fig. \ref{fig:singular-speed-squared}. 

In the formulation of the action as 
a contour integral in the $ r $-plane, the
ambiguity in the action of the singular saddle arises because
one must deform the contour in Fig. \ref{fig:real-singular} away
from the poles at  
$r = \pm \frac{\pi i}{2}$, and the deformation is not
unique. The same integral Eqn. \ref{eqn:principal-value}
in the neighborhood of the pole may be written
\begin{align}
  \int_{-\rho_1 - \vep}^{-\rho_1 + \vep} \diff \rho\,
  \lp \td{z}{\rho} \rp^2
  = -\int_{\mathscr{C}_\vep} \diff z\, \frac{ w }{z}
\end{align}
where $\mathscr{C}_\vep$ is a contour that avoids the pole. For example,
the integral around the pole at $\frac{\pi i}{2}$ along the contour
deformation shown in Fig. \ref{fig:real-singular-deformed} 
is $-\frac{3\pi i}{2}w$, corresponding to $N_2 = 1$
in Eqn. \ref{eqn:principal-value}. 

\begin{figure}[h]
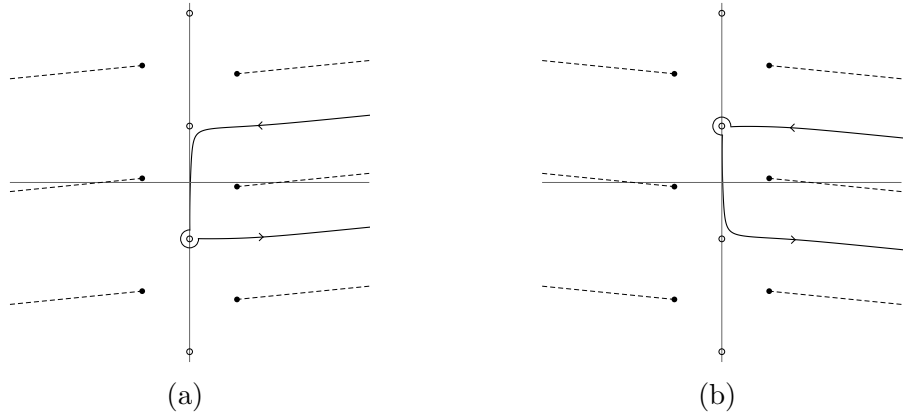

  \centering
  \begin{subfigure}[t]{.3\textwidth}
    \centering
    \ig{0cm}{width=\linewidth}{singular_complex_positive}
    \caption{\footnotesize }
    \label{fig:reflected-under-positive}
  \end{subfigure}
\hspace{2cm}
\begin{subfigure}[t]{.3\textwidth}
  \centering
  \ig{0cm}{width=\linewidth}{singular_complex}
  \caption{\footnotesize }
  \label{fig:reflected-under-negative}
\end{subfigure}
\caption{\footnotesize \bd{Singular Reflected Saddles} ($\mrm{Re}(\eta) <
  w)$. $\eta$ should be shifted off the real axis for the
  saddle-point expansion to be well-defined. Compared to
  Fig. \ref{fig:deformed-contours}, $\eta$ has been given a
  small positive phase on the left and a small negative phase on
  the right.}
\label{fig:reflected-saddles}
\end{figure}

The outcome of this discussion is that, in addition to the
saddles accounted in Eqns. 
\ref{eqn:real-contr}, \ref{eqn:shift-contr}, and \ref{eqn:shift-contr2},
one has singular
saddles of finite action corresponding to contours in the 
$r$-plane that wrap the singularities an integer number of times
and so differ from $\til S_0 $ by residues. They may be thought
of as asymptotic saddles in a fixed topological sector of the
functional integral, where the singular points of the potential
are excised from the plane, and the configuration space of maps
$r \cn [-L,L] \to \{\bbC - \mrm{poles}\}$ is divided into
homotopy classes labeled by their winding numbers around the
punctures. As the deformations of the contour around the poles
shrink away, one asymptotically approaches an exact solution of
the equations of motion, whose action differs from $\til S_0$ in
its imaginary part.

A singular saddle that wraps $N_2$ times
around the pole at $ r  = \frac{\pi i}{2}+ \pi i N_1$ has action 
$\til S_0  - 2\pi i \eta N_1 -2\pi i   w  N_2.$
Figs. \ref{fig:reflected-under-positive} and
\ref{fig:reflected-under-negative} illustrate such saddles with
$N_1 = -1$ and 0, $N_2=-1$ and $1$, and $\eta$ having a small positive
and negative phase, respectively. By summing over saddles
with action of the form $\til S_0 - 2\pi i \eta (N_1+N_2) +2\pi i w N_2 =\til
S_0 - 2\pi i \eta N_1 -2\pi i (\eta-w) N_2  $, one may obtain
both factors of $\csc(\pi k \eta)$ and $\csc(\pi k(w-\eta))$
required in Eqn. \ref{eqn:r-qm-limit}. We give the precise list of
saddles momentarily.

Finally, one must account for the factor of $\sin(\pi k w)$ in
Eqn. \ref{eqn:r-qm-limit}. It is again associated to the singular
saddles, but it is
qualitatively different from the
$\csc$ factors, corresponding to a 2-fold degeneracy of saddles
rather than an infinite geometric series. Written in the form
\begin{align}
  \sin(\pi k w) \propto e^{\pi i k w}(1 - e^{-2\pi i k w}),
\end{align}
we see that we must sum over two sets of singular saddles, identical
except that each contour in one set winds an extra time around
the pole. The relative minus sign is due to the orientation of
the integration contours, which we will not attempt to
determine. 

Having explained the mechanism by which each factor in
Eqn. \ref{eqn:r-qm-limit}
comes about, let us finally give the list
of saddles that reproduces the semi-classical limit for $0 <
\mrm{Re}(\eta) < w$. 

For $\mrm{Im}(\eta) > 0$, the contributing saddles have action
\begin{align}
  \til S_{N_1N_2}
  =\til S_0 - 2\pi i \eta (N_1+N_2) + 2\pi iwN_2,
  \quad &N_1 =  0,1,2,\ldots\\
  &N_2 = 1,2,3,\ldots\nt,
\end{align}
corresponding to a contour that wraps $N_2$ times around the
pole at $r = -\frac{\pi i}{2}+\pi i (N_1+N_2)$. Note that $N_1 +
N_2 \geq 1$, meaning that all the contours are in the upper-half
plane. In addition, one has a second set of saddles with action
\begin{align}
  \til S_{N_1N_2}'
  =\til S_0 - 2\pi i \eta (N_1+N_2) + 2\pi iw(N_2-1),
  \quad &N_1 =  0,1,2,\ldots\\
  &N_2 = 1,2,3,\ldots,\nt
\end{align}
which wrap $N_2 -1$ times instead. The two sets are weighted
with a relative minus sign.

Their contribution to the saddle-point expansion yields at
leading order
\begin{align}
  \sum_{\substack{N_1 \in \bbZ_{\geq 0}\\
  N_2 \in \bbZ_{\geq 1}}}
  &\lp e^{-k \til S_{N_1N_2}} - e^{-k \til S_{N_1N_2}'}\rp\\
  =&\,
     e^{-k \til S_0}
     \sum_{N_1\in \bbZ_{\geq 0}} e^{2\pi i k \eta N_1}
     \sum_{N_2 \in \bbZ_{\geq 1}} e^{2\pi i k (\eta-w)
     N_2}
     \lp 1 - e^{2\pi i k w}   \rp\nt\\
  \propto\,&
           \eta^{-2k \eta}(w+\eta)^{k(w+\eta)}
           (w-\eta)^{k(\eta-w)}
  \sin(\pi k w)
  \csc\lp \pi k \eta\rp
  \csc \lp \pi k (w -\eta) \rp\nt,
\end{align}
reproducing Eqn. \ref{eqn:r-qm-limit}.
Note that the geometric series converge for $\mrm{Im}(\eta) >
0$. For $\mrm{Im}(\eta) < 0$, the required sum is instead
\begin{align}
       e^{-k \til S_0}
     \sum_{N_1\in \bbZ_{\leq 0}} e^{2\pi i k \eta N_1}
     \sum_{N_2 \in \bbZ_{\leq -1}} e^{2\pi i k (\eta-w)
     N_2}
     \lp 1 - e^{-2\pi i k w}   \rp,
\end{align}
with the same result. These are contours that wrap $N_2$ or
$N_2+1$ times around the pole at $r = \frac{\pi i}{2}+\pi i(N_1+N_2)$.
Note that $N_1 + N_2 \leq -1$ implies all of the contours are in
the lower-half plane. 

As forewarned at the beginning of this section, we have not
attempted to explain why these are the 
saddles that contribute to the functional integral, but merely
demonstrated that this is the necessary list to reproduce
the semi-classical limit of the exact answer. In fact, we take
this list as the definition of the contour of the functional
integral that computes the reflection coefficient for the
quantum mechanics, being given by the sum of the corresponding
steepest-descent contours. 

So far we have considered the case $0<\mrm{Re}(\eta) < w$. Next
suppose that $\mrm{Re}(\eta) >w$. In Eqn. \ref{eqn:r-qm-limit},
the bound state factor now disappears, because
$\frac{\eta^2}{2}$ exceeds the height of the potential.
As discussed previously, in light of this abrupt change in the
asymptotic expansion of the reflection coefficient, we expect
that $\mrm{Re}(\eta) = w$ corresponds to a Stokes wall, and that
the set of contributing saddles jumps for $\mrm{Re}(\eta) > w$.

\begin{figure}[h]
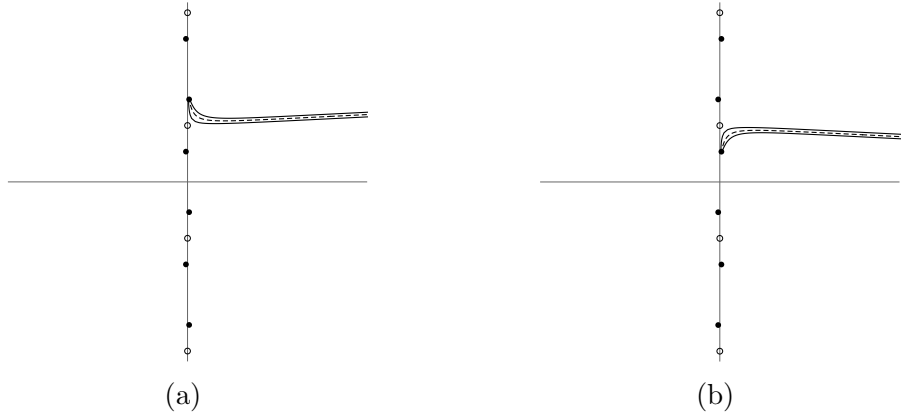

  \centering
  \begin{subfigure}[t]{.3\textwidth}
    \centering
    \ig{0cm}{width=\linewidth}{over-reflected-positive}
    \caption{\footnotesize }
    \label{fig:over-reflected-positive}
  \end{subfigure}
\hspace{2cm}
\begin{subfigure}[t]{.3\textwidth}
  \centering
  \ig{0cm}{width=\linewidth}{over-reflected-negative}
  \caption{\footnotesize }
  \label{fig:over-reflected-negative}
\end{subfigure}
\caption{\footnotesize \bd{Reflected Saddles} ($\mrm{Re}(\eta) >
    w)$. When $\eta$ is real and greater than $w$, the only
    reflected solutions are singular. But one finds non-singular
    solutions for complex $\eta$. On the left $\eta$ has a
    small positive phase and on the right it has a small
    negative phase. The contour is deflected in opposite
    directions by the pole in the two cases.}
\label{fig:over-contours}
\end{figure}

When $\eta$ is real and larger than $w$,
one no longer has a real
solution of the equations of motion. The energy of the
particle is greater than the height of the potential hill and
it rolls over from $r \to \infty$ to $r \to -\infty$,
satisfying the boundary conditions for transmission rather than
reflection. The only reflected trajectories with $\eta$ real and greater
than $w$ are singular. Once again, $\eta$ should be given
a phase, in which case one finds non-singular trajectories. 
Figs. \ref{fig:over-reflected-positive} and
\ref{fig:over-reflected-negative} illustrate reflected solutions
for $\eta$ with a small  positive and a small negative
phase. As is by now familiar, the set of
contributing saddles will depend on whether $\eta$ lies above or
below the real axis.

The action for the contour with $\mrm{Im}(\eta) > 0$ pictured in
Fig. \ref{fig:over-reflected-positive} is
\begin{align}
  \til S_0 = 2\eta \log(\eta)
  -(\eta+w)\log(\eta+w)-(\eta-w)\log(\eta -w)
  -\pi i(\eta + w).
\end{align}
The contributing saddles for $\mrm{Im}(\eta)>0$ are the shifted
saddles of this form in the upper-half plane,
\begin{align}
  \til S_{N_1} = \til S_0 - 2\pi i \eta N_1,\quad N_1 = 0,1,2,\ldots,
\end{align}
as well as a second set that wraps the pole at
$r = \frac{\pi i}{2} + \pi i N_1$ once,
\begin{align}
  \til S'_{N_1} = \til S_{N_1} + 2\pi i w,\quad N_1 = 0,1,2,\ldots.
\end{align}
The two sets are weighted with a relative minus sign, for a
total of
\begin{align}
  \sum_{N_1 \in \bbZ_{\geq 0}}\lp e^{-k \til S_{N_1}}-e^{-k\til S'_{N_1}}\rp
  =&\,
  e^{-k \til S_0} \sum_{N_1\in \bbZ_{\geq 0}} e^{2\pi i k \eta N_1}
  \lp 1- e^{-2\pi i k w} \rp\\
  \propto&\,
  \eta^{-2k\eta}(\eta + w)^{k(\eta+w)}(\eta - w)^{k(\eta -w)}
  \sin(\pi k w) \csc(\pi k \eta).\nt
\end{align}
For $\mrm{Im}(\eta) < 0$, the necessary saddles are of the form
in Fig. \ref{fig:over-reflected-negative}, but in the lower-half
plane. The action for the contour in Fig. \ref{fig:over-reflected-negative}
is
\begin{align}
  \til S_0 = 2\eta \log(\eta)
  -(\eta+w)\log(\eta+w)-(\eta-w)\log(\eta -w)
  -\pi i(\eta - w),
\end{align}
and the saddle-point expansion is
\begin{align}
  e^{-k \til S_0}&\lp \sum_{N_1\in \bbZ_{\leq -1}} e^{2\pi i k \eta N_1}\rp
  \lp 1- e^{2\pi i k w}\rp.
\end{align}
This completes the saddle-point expansion of the reflection
coefficient for the infinite-space quantum mechanics. Next we
turn to the transmission coefficient.

\subsubsection{Transmission Coefficient on the Complex $r$-Plane}
\label{sec:complex-t}

\begin{figure}[h]
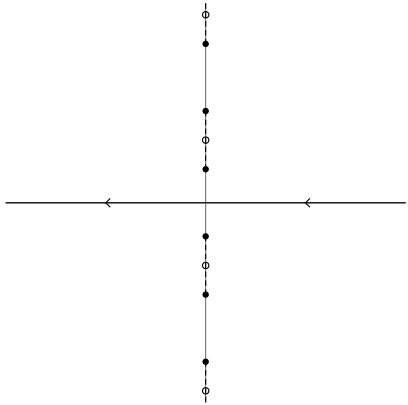

  \centering
  \ig{0cm}{width=\linewidth/3}{over_transmitted}
  \caption{\footnotesize \bd{Transmitted Saddle} ($\eta>w$). For $\eta$ real
    and greater than $w$, there is a real saddle for the
    transmission coefficient, corresponding to a particle that
    rolls over the inverted potential from $r \to \infty$ to $r
    \to -\infty$.}
  \label{fig:over-transmitted}
\end{figure}

In this sub-section we compute
the saddle-point expansion of the transmission
coefficient for the infinite-space quantum mechanics. The action is as
in Eqn. \ref{eqn:r-contour}, but with $r(L) \to -r(L)$ so that
the velocity at late times is fixed to $-\eta$ rather than $\eta$:
\begin{align}
  \label{eqn:t-contour}
  \til{S}[ r ]  = \int_\mathscr{C} \diff  r \,
  \sqrt{\eta^2+2 V ( r )}
  -\eta( -r (L)+ r (-L)). 
\end{align}
The simplest saddle of
Eqn. \ref{eqn:t-contour} is the real trajectory of a particle
with $\eta$ real and greater than $w$ that rolls over the
potential hill, pictured in Fig. \ref{fig:over-transmitted}. 

The action for this saddle is\footnote{Note that the appropriate
branch of the square-root in the integrand of the action should
be negative on the real axis, the velocity of the particle
always being to the left for this trajectory.}
\begin{align}
  \til S_0 = 2\eta \log(\eta)
  -(\eta+w)\log(\eta+w)-(\eta-w)\log(\eta -w),
\end{align}
yielding
\begin{align}
  e^{-k \til S_0}
  = \eta^{-2k\eta}(\eta + w)^{k(\eta+w)}(\eta
  - w)^{k(\eta -w)}. 
\end{align}
This reproduces the transmission coefficient
Eqn. \ref{eqn:t-qm-limit} for $\mrm{Re}(\eta) > w$.
Note that although one again has shifted saddles with constant
imaginary part $\pi N_1$, the action is invariant under the
shift. For the same reason, note that the functional integral
over real $r$ no longer diverges for the transmission
coefficient. Indeed, given that the semi-classical limit of the
transmission coefficient is reproduced by a single real saddle
for real $\eta$, we expect that the contour of integration is
real in this case.

\begin{figure}[h]
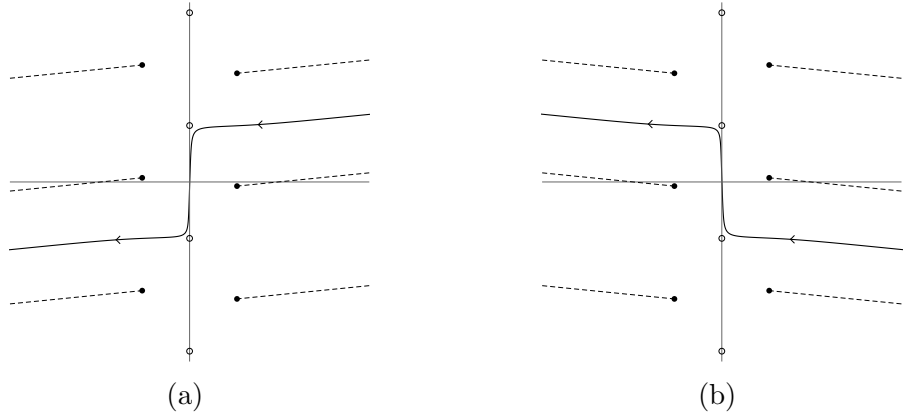

  \centering
  \begin{subfigure}[t]{.3\textwidth}
    \centering
    \ig{0cm}{width=\linewidth}{transmitted-under-positive}
    \caption{\footnotesize }
    \label{fig:transmitted-under-positive}
  \end{subfigure}
\hspace{2cm}
\begin{subfigure}[t]{.3\textwidth}
  \centering
  \ig{0cm}{width=\linewidth}{transmitted-under-negative}
  \caption{\footnotesize }
  \label{fig:transmitted-under-negative}
\end{subfigure}
\caption{\footnotesize \bd{Transmitted Saddles} ($\mrm{Re}(\eta) <
    w)$. When $\eta$ is real and less than $w$, the only
    transmitted saddles are singular. Non-singular solutions are
    found for complex $\eta$. On the left $\eta$ has a small
    positive phase and on the right it has a small negative phase.}
\label{fig:transmitted-under}
\end{figure}

When $\eta$ is real and less than $w$, the
only saddles for the transmission coefficient are singular.
Non-singular saddles are obtained for complex $\eta$, as 
pictured in Fig. \ref{fig:transmitted-under}.

The action for the contour pictured in
Fig. \ref{fig:transmitted-under-positive}, where $\mrm{Im}(\eta)>0$, is
\begin{align}
  \til S_0 = 2\eta \log(\eta)
  -(w+\eta)\log(w+\eta)+(w-\eta)\log(w-\eta)+\pi i(w-\eta ),
\end{align}
which gives
\begin{align}
  e^{-k \til S_0}
  = e^{-\pi i k(w-\eta)}\eta^{-2k\eta}(\eta + w)^{k(\eta+w)}
  (w-\eta)^{k(\eta -w)}. 
\end{align}
The necessary sum is now
\begin{align}
  \label{eqn:transmitted-saddle-expansion}
  e^{-k \til S_0} \sum_{N\in \bbZ_{\geq 0}}
  e^{2\pi ik(\eta - w)N}
  \propto
  \eta^{-2k\eta}(\eta + w)^{k(\eta+w)}
  (w-\eta)^{k(\eta -w)} \csc \lp \pi k (w - \eta) \rp, 
\end{align}
corresponding to saddles with $\til S_N = \til S_0 - 2\pi i \eta
N + 2\pi i w N$. As in the reflected case, the shift by $2\pi i
w N$ is accounted for by singular saddles that wrap the pole $N$
times. The shift by $2\pi i \eta N$ was previously explained by
the change in the boundary action under
$r \to r + \pi i N.$ As noted a moment ago, however, the
boundary action for transmission is invariant under this
shift.

\begin{figure}[h]
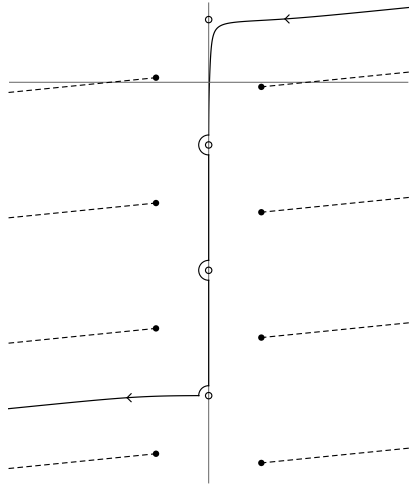

  \centering
  \ig{0cm}{width=\linewidth/3}{transmitted-shifted}
  \caption{\footnotesize \bd{Shifted Transmitted Contour}. When
    $\mrm{Re}(\eta) < w$, one must sum over contours with $r(L)
    \to r(L)-2\pi i N$ in order to reproduce the semi-classical
    limit of the transmission coefficient.}
  \label{fig:transmitted-shifted}
\end{figure}

Instead, one must consider contours with only one end
shifted by $2\pi i N$, as pictured in
Fig. \ref{fig:transmitted-shifted}. The integral along the
shifted contour is identical to that of
Fig. \ref{fig:transmitted-under-positive}; their actions differ
only by the shift of the boundary term. By summing over contours
of this form, where $r(L) \to r(L) - 2 \pi i N$, together with
the $N$-fold wrapping, we obtain the required lattice of actions
for Eqn. \ref{eqn:transmitted-saddle-expansion}.

For $\mrm{Im}(\eta) < 0$, the action of the saddle pictured in
Fig. \ref{fig:transmitted-under-negative} is
\begin{align}
  \til S_0 = 2\eta \log(\eta)
  -(w+\eta)\log(w+\eta)+(w-\eta)\log(w-\eta)-\pi i(w-\eta ),
\end{align}
and the saddle-point expansion is
\begin{align}
  e^{-k \til S_0} \sum_{N\in \bbZ_{\leq 0}}
  e^{2\pi ik(\eta - w)N}.
\end{align}
One has analogous contours as in Fig. \ref{fig:transmitted-shifted},
now with $r(L) \to r(L)-2\pi i N$ shifted upward.

We do not have a clear rationale for why contours of this form
contribute to the saddle-point expansion.
Although the contour pictured in
Fig. \ref{fig:transmitted-shifted}  is constructed out of the
same components as the singular saddles discussed previously,
it is not obtained from a limit of smooth saddles, which are
always confined to an interval of width $\pi i$.
For now we merely observe that this is the set which
reproduces the semi-classical limit of the exact transmission
coefficient, and leave it as an open question
to better understand the rules for determining the set of
contours that contribute to the saddle-point expansion.

\subsection{Bound States and the Cigar Wrapping Saddle}
\label{sec:bound-states}

We conclude this section by returning to the problem of
identifying an asymptotic condition for a bound state
insertion in the functional integral. As explained in
Sec. \ref{sec:cigar-asymptotics}, 
the asymptotic conditions in Eqn. \ref{eqn:ld-circle-green}
describing an operator insertion $\cO_{jnw}$ in the far past on
the cylinder assume a generic value of $j$, with $\mrm{Re}(j)>
\frac{1}{2}$, on which $R(j,n,w)$ is non-singular.
On the bound state spectrum $j_N$, $R$ has simples poles, and
one must be more careful.
In that case, it is the reflected term in
Eqn. \ref{eqn:asymptotic} that dominates in the weak coupling
region, $\frac{1}{R} \cO_{j n w} \to
\cV_{Qj,p_\mrm{L},p_\mrm{R}}$, and the free field asymptotic
condition flips sign,
\begin{align}
  \label{eqn:bound-state-far}
  r \overright{\rho \to-\infty} \frac{2}{k} \lp j -
  \frac{1}{2} \rp \rho.
\end{align}
This asymptotic condition
maps the string out of the free field region and is
inconsistent. This agrees with the fact that the bound states
are normalizable, and do not extend out to infinity in $r$.  

\begin{figure}[t]
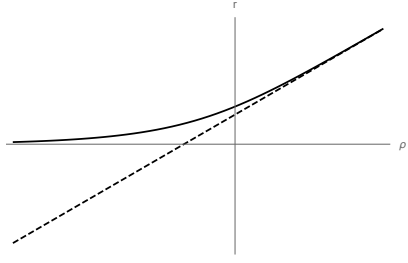

  \centering
  \ig{0cm}{width=\linewidth/3}{cigar-wrapping}
  \caption{\footnotesize \bd{The Cigar Wrapping Saddle.}
  When $j=j_N$ lies on the bound state spectrum
  Eqn. \ref{eqn:real-branch}, the reflection coefficient is
  singular, and it is the otherwise sub-leading term $e^{-2j
    r}$ in Eqn. \ref{eqn:asymptotic} that describes the operator
  $\cO_{j nw}$ in the asymptotic region. The free field
  Green function, shown by the dashed line, maps the string out
  of the free field region 
  and must be modified. The complete solution for $|w| = 1$ and
  $n = 0$ is $\theta = \pm \phi$, $r = \sinh^{-1}(e^\rho)$.
  The neighborhood of the bound state insertion wraps the tip of the cigar, with
  $r\to e^\rho$ asymptotically
  approaching $r = 0$. The leading saddle is independent of $N$,
  which enters in the sub-leading correction to the saddle-point approximation.
}
  \label{fig:cigar-wrapping}
\end{figure}

The linear solution, pictured in Fig. \ref{fig:cigar-wrapping},
is a saddle for the two-point function of the linear-dilaton
primaries $e^{-2(1-j)r}$ and $e^{-2jr},$ sending the string to
$r \to \infty$ and $r \to -\infty$ in the neighborhoods of the
two respective operators. 
In the cigar, of course, the geometry ends at $r = 0$, and
in the vicinity of the tip the free cylinder equations of motion
are modified by the curvature of the cigar.
One would like to understand how
the free trajectory is corrected once the string leaves the free
field region, and
thereby obtain an asymptotic condition for the bound state insertion.
We will argue that the string worldsheet asymptotically wraps
the tip of the cigar.

Because the neighborhood of the bound state insertion is mapped
out of the free field region, the large $r$ expansion $e^{-2jr}$
of the operator is insufficient to determine the requisite
asymptotic condition. The radial dependence of
the vertex operator on the full cigar was
obtained in \cite{Dijkgraaf:1991ba}:\footnote{This expression
  holds for $m \geq \bar m$ (i.e. $n \geq 0$). Note that the
  reflection coefficient is invariant under $w \to -w$. 
  For $m \leq \bar m$ one
  sends $m \to -m$ and $\bar m \to - \bar m$, which is
  equivalent to flipping the signs of $n$ and $w$. Then it is the
  absolute value of $n$ that appears in the reflection
  coefficient, as in Eqn. \ref{eqn:r}.}
\footnote{To obtain the
  wavefunction, one would multiply by $\cosh(r)$ as described in
Sec. \ref{sec:operator-spectrum}.}
\begin{align}
  \label{eqn:dvv}
  &\frac{4^{j-1}}{\Gamma(2j-1)}
  \frac{\Gamma(j+m)\Gamma(j-\bar m)}{\Gamma(1+m-\bar m)}\\
  &\times
    \sinh^{m-\bar m}(r) \cosh^{-(m+\bar m)}(r)
    {}_2F_1 (j-\bar m, 1-j-\bar m; 1+m-\bar m; -\sinh^2(r))\nt\\
  &\quad\overright{r \to \infty}
    e^{-2(1-j)r} + 4^{2j-1}
    \frac{\Gamma(1-2j)}{\Gamma(2j-1)}
    \frac{\Gamma(j+m)\Gamma(j-\bar
    m)}{\Gamma(1-j+m)\Gamma(1-j-\bar m)} e^{-2jr},\nt
\end{align}
where $m = \frac{1}{2}(-kw+n)$ and $\bar m =
\frac{1}{2}(-kw-n)$. Note that the reflection coefficient
reproduces the second line of Eqn. \ref{eqn:r}.\footnote{It is
  unclear to us if the missing factor from the first line of
  Eqn. \ref{eqn:r}, which is of order one in the large $k$
  limit, is due to a non-perturbative correction to 
  the cigar background as has been suggested in the literature
  \cite{Kutasov:2005rr,Giveon:2015cma,Giveon:2013ica}, or if it is a
  perturbative correction.}

In the parent $\mrm{SL}(2,\reals)_k$ WZW
model prior to the coset, $m$ and $\bar m$ correspond to the
eigenvalues\footnote{More precisely, their eigenvalues are $m +
  \frac{kw}{2}$ and $\bar m + \frac{kw}{2}$.} of the
$\what\Lsl_k(2,\reals)_\mrm{L}\oplus 
\what\Lsl_k(2,\reals)_\mrm{R}$ current algebra zero-modes
$J^3_0$ and $\bar J^3_0$. $m - \bar m = n$ is the quantized
angular momentum around the $\mrm{AdS}_3$ cylinder, and $m +
\bar m = -kw$ is the projection condition for the coset, which
gauges $J^3+ \bar J^3$ \cite{Maldacena:2000hw}.
The poles of $\Gamma(j+m)$ in the reflection coefficient correspond
to states in highest-weight discrete series representations of
$\Lsl(2,\reals)$, $m \in -j - \bbN$, while the poles of
$\Gamma(j- \bar m)$ correspond to states in lowest-weight
discrete series representations, $\bar m \in j + \bbN$
\cite{Maldacena:2000hw}. For $w < 0$, the coset bound states
descend from the lowest-weight discrete series states, while for
$w>0$ they descend from the highest-weight states.\footnote{The
  reason being that $m,\bar m \in j + \bbN$ are positive in
  lowest-weight representations since $j$ is positive, while $m,
  \bar m \in -j - \bbN$ are negative in highest-weight
  representations. To satisfy the projection $m + \bar m = -kw$
  therefore requires a highest-weight representation  for $w>0$
  and a lowest-weight representation for $w<0$.}
Likewise, in
Eqn. \ref{eqn:poles} we saw that the coset bound states appear
in the poles of 
$\Gamma \lp j + \frac{|n|\pm k w}{2}\rp$, depending on the
sign of $w$.

Recalling that the hypergeometric series terminates when its
first or second argument is a non-positive integer, we find that
the hypergeometric function is a finite order polynomial in
$\sinh^2(r)$ on the bound state spectrum. On the lowest-weight
states, $j - \bar m =-N$, this is evident from
Eqn. \ref{eqn:dvv}, where the hypergeometric function yields an
order $N$ polynomial in $\sinh^2(r)$. On the highest-weight
states, $j + m = -N,$ it 
becomes evident after applying the hypergeometric fractional
transformation rule
\begin{align}
  &{}_2F_1 (j-\bar m, 1-j-\bar m; 1+m-\bar m; -\sinh^2(r))\\
  &\quad\quad=
  \cosh^{2(m+\bar m)}(r)
  {}_2F_1 (j+ m, 1-j + m; 1+m-\bar m; -\sinh^2(r))\nt.
\end{align}

Let us again restrict our focus to the pure-winding
sector, $n=0$.
On the bound state spectrum $j_N = \frac{k|w|}{2}-N$, one obtains
\begin{align}
  \cO_{j_N,n=0,w} \propto \sech^{k|w|}(r) {}_2F_1\lp
  -N,-k|w|+N+1;1;-\sinh^2(r) \rp. 
\end{align}
For $N = 0$ the hypergeometric function is 1, for $N =
1$ it is $1+ (-k|w|+2) \sinh^2(r)$, and so on.

These operators consist of a heavy
factor $\sech^{k|w|}(r)$, which enters at the same order
as the leading terms in the action, times a light factor,
which is sub-leading.
The heavy factor inserts 
a source in the leading equations of motion 
and therefore affects the form of the saddles. The light factor, by
constrast, is merely evaluated on the leading saddles and
contributes to the sub-leading correction in the saddle-point
expansion. Moreover, since the heavy factor is independent of
$N$, the behavior of the saddle for any bound state insertion is
independent of $N$.

We have assumed here that 
$N$ is of order one in the large $k$ limit, else the order $N$
polynomial in $\sinh^2(r)$ would no longer be a light operator.
In view of the upper-bound $j  < \frac{k-1}{2}$ on
the physical spectrum, we will moreover choose $|w|=1$, such that the
bound may be satisfied for $N$ of order one. Indeed, our principal
interest is in the sine-Liouville
operator Eqn. \ref{eqn:sl-operator}, which is the sum of
operators with $j = \frac{k}{2} - 1$ and $w = \pm 1$. We will
focus below on $w = -1$; the case $w= 1$ is analogous.

Since the asymptotic condition for
$\cO_{j=\frac{k}{2}-N,n=0,w=-1}$ is independent of $N$, we may
set $N = 0$. This state is not part of the physical spectrum, of
course. In fact, it is rather special in the continued
space of states; it is in a sense a reflection 
of the identity operator \cite{Maldacena:2000hw}. Note first of all that its
conformal weight is zero. 
In the coset construction from $\mrm{SL}(2,\reals)_k$, $\cO_{j =
  \frac{k}{2},n=0,w=-1}$ descends from the state 
\begin{align}
  \label{eqn:spectral-flowed}
\ket{j = \frac{k}{2}, m = \frac{k}{2}, \bar m = \frac{k}{2};
  w = -1} \in \what D_{ \frac{k}{2}}^{+,w=-1}\otimes \what D_{
  \frac{k}{2}}^{+,w=-1}.
\end{align}
Here, $D_{\frac{k}{2}}^+$ denotes the spin
$j = \frac{k}{2}$
lowest-weight discrete series representation of the global
$\Lsl(2,\reals)$ sub-algebra, $\what D_{\frac{k}{2}}^+$ denotes
the $\what \Lsl_k(2,\reals)$ current algebra representation
built upon it, and $\what D_{\frac{k}{2}}^{+,w=-1}$ denotes the
spectral-flowed current algebra representation by minus one unit. For
the details of these representations, see
\cite{Maldacena:2000hw}.

$\ket{j=\frac{k}{2},m=\frac{k}{2}, \bar m = \frac{k}{2}}$ is
known as the spectral flow operator
\cite{Maldacena:2000hw,Maldacena:2001km} because its product with
another operator imparts one unit of spectral flow. By flowing
this state backward by one unit as in
Eqn. \ref{eqn:spectral-flowed}, one obtains a trivial operator
of $J^3_0$, $\bar J^3_0$, and conformal weight zero. Under the
isomorphism $\what D_j^{+,w} \simeq \what
D_{\frac{k}{2}-j}^{-,w+1}$ of spectral-flowed discrete series representations,
it maps to the trivial highest-weight state $\ket{j' =0,
  m' =0, \bar m' = 0; w' = 0}.$

To understand the asymptotic condition associated to this
operator, return to the cigar quantum mechanics
Eqn. \ref{eqn:qm-action}, obtained after choosing a pure-winding
configuration $\theta = \phi$ .  The inverted potential $-V(r) =
\frac{1}{2} \sech^2(r)$ is a hill of height $\frac{1}{2}$, as
shown in Fig. \ref{fig:inverted-potential}. For
generic real values of $\eta < 1$, one obtained real
solutions describing a particle that comes in from infinity with
speed $\eta$, rolls partway up the hill to a height of
$\frac{1}{2}\eta^2$, and then rolls back to infinity.
For $\eta =1$, corresponding to $j = \frac{k}{2},$
the particle has just enough energy to asymptotically approach
the top of the potential at $r = 0$. It does not return to
infinity, but rather wraps the tip of the cigar, taking infinite
time to do so.

This cigar wrapping solution is
\begin{align}
  \label{eqn:cigar-wrapping}
  r(\rho) = \sinh^{-1}(e^{\rho}),  
\end{align}
shown in Fig. \ref{fig:cigar-wrapping}, with limiting behavior
\begin{align}
  r(\rho) \to
  \begin{dcases}
    e^\rho & \rho \to -\infty\\
    \rho & \rho \to \infty.
  \end{dcases}
\end{align}
As expected from Eqn. \ref{eqn:bound-state-far} with $j =
\frac{k}{2}$, in the asymptotic region the solution approaches
the free field Green function 
$r\to \rho$. Eqn. \ref{eqn:cigar-wrapping} gives the
completion of the solution beyond the free field region.
The asymptotic condition
\begin{align}
  &r\overright{\rho\to -\infty} e^{\rho}\\
  &\theta \overright{\rho \to -\infty} \phi,
\end{align}
describes a string that wraps the tip of the
cigar. Since the bound state operators for $N\neq 0$ differ at sub-leading
order, we claim that this is the appropriate asymptotic
condition for any $N$ of order one.
Observe that this is simply the holomorphic map that sends the 
worldsheet coordinate $z = e^\rho e^{i\phi}$ 
to the target coordinate $Z = r e^{i\theta}$ in the neighborhood
of the tip of the cigar, where the geometry is $\reals^2$.
As usual, one may shift $\rho$ by a continuous modulus $i
\rho_0$ that changes the angle at which the trajectory
approaches the origin in the complex $r$-plane, and one may
moreover consider solutions shifted by $\pi i \bbZ$.

Note that $\dot{r} \to r$ as $\rho \to -\infty$ and $\dot{r} \to
1$ as $\rho \to \infty$.
This configuration is therefore a saddle of the action
\begin{align}
 S
  = &\frac{k}{4\pi}
      \int\limits_{-L}^L \diff \rho \int\limits_0^{2\pi}\diff\phi\, 
      \bigg( (\partial_\rho r)^2 +
      (\partial_\phi  r)^2
      + \tanh^2(r)
      \lp  (\partial_\rho  \theta)^2 + (\partial_\phi  \theta)^2 \rp
      \bigg)\\
    &+ k\int\limits_0^{2\pi} \frac{\diff
      \phi}{2\pi} \lp  -r|_{\rho =L} +  \frac{1}{2} r^2|_{\rho=-L} \rp     
      \nt\\
    &+k\int\limits_0^{2\pi} \frac{\diff \phi}{2\pi}
      \bigg(
      \sigma_+ \lp \partial_\phi \theta|_{\rho = L} -1 \rp
      +\sigma_- \lp \partial_\phi \theta|_{\rho = -L} -1 \rp
      \bigg)
      + \cO(k^0)\nt,
\end{align}
with radial boundary equations of motion
\begin{subequations}
  \begin{align}
    &\partial_\rho r|_{\rho = L} = 1 \\
    &\partial_\rho r|_{\rho=- L} = r|_{\rho = -L}.
  \end{align}
\end{subequations}
The on-shell action is $S = -k\log(2).$

Thus, one may interpret the cigar wrapping configuration as a saddle for
the two-point function of $\cO_{j=\frac{k}{2}-N,n=0,w=\pm 1}.$
In the special case when $N = 0$, this is a trivial
operator. Then the 
$r^2$ boundary term yields the identity operator in the limit
that it shrinks away, and one may alternatively interpret the
configuration as a saddle for the one-point function. For $N \neq 0$, one
inserts the light factor of the operator at the boundary and
the insertion becomes non-trivial.  

Since the reflection coefficient is singular, the sum over
complex saddles may diverge with the free field boundary
condition specified by the linear boundary term at $\rho =
L$. Our primary interest is not in the cigar wrapping saddle itself,
however, but in the tip wrapping asymptotic condition
$\dot{r} \to r$ for the bound states.

With the asymptotic condition in hand, one may use it to compute
the saddle-point expansion of correlation functions with bound
state insertions. We will not pursue any such calculations
here.\footnote{We conjecture, however, that the relevant
  saddles for the two-point
  function of $\cO_{j=\frac{k}{2}-N,n=0,w=\pm 1}$ computed with
  the tip wrapping asymptotic conditions are given by
  trajectories that asymptote between neighboring maxima of the
  inverted potential on the complex $r$ plane, such as $r = 0$
  and $r = \pi i$.
  On the imaginary axis, pictured in Fig.
  \ref{fig:y-potential}, the potential is singular at $r = \pi
  i/2$. However, in the $k$-corrected potential \cite{Dijkgraaf:1991ba},
  \begin{align}
    V(r) = \frac{1}{2}\lp  \frac{1}{\coth^2(r)-\frac{2}{k}} -
    \frac{k}{k-2} \rp,
  \end{align}
  the double pole at $r = \pi i/2$ splits into a pair of simple
  poles at $\frac{\pi i}{2} \pm \sqrt{\frac{2}{k}} + \cO
  (k^{-3/2}).$ Then the potential on the imaginary axis is
  regular, and one may consider, for example, a trajectory that asymptotes
  between $r = 0$ in the far past and $r = \pi i$ in the far future:
  \begin{figure}[H]
    \centering
  \ig{0cm}{width=\linewidth/3}{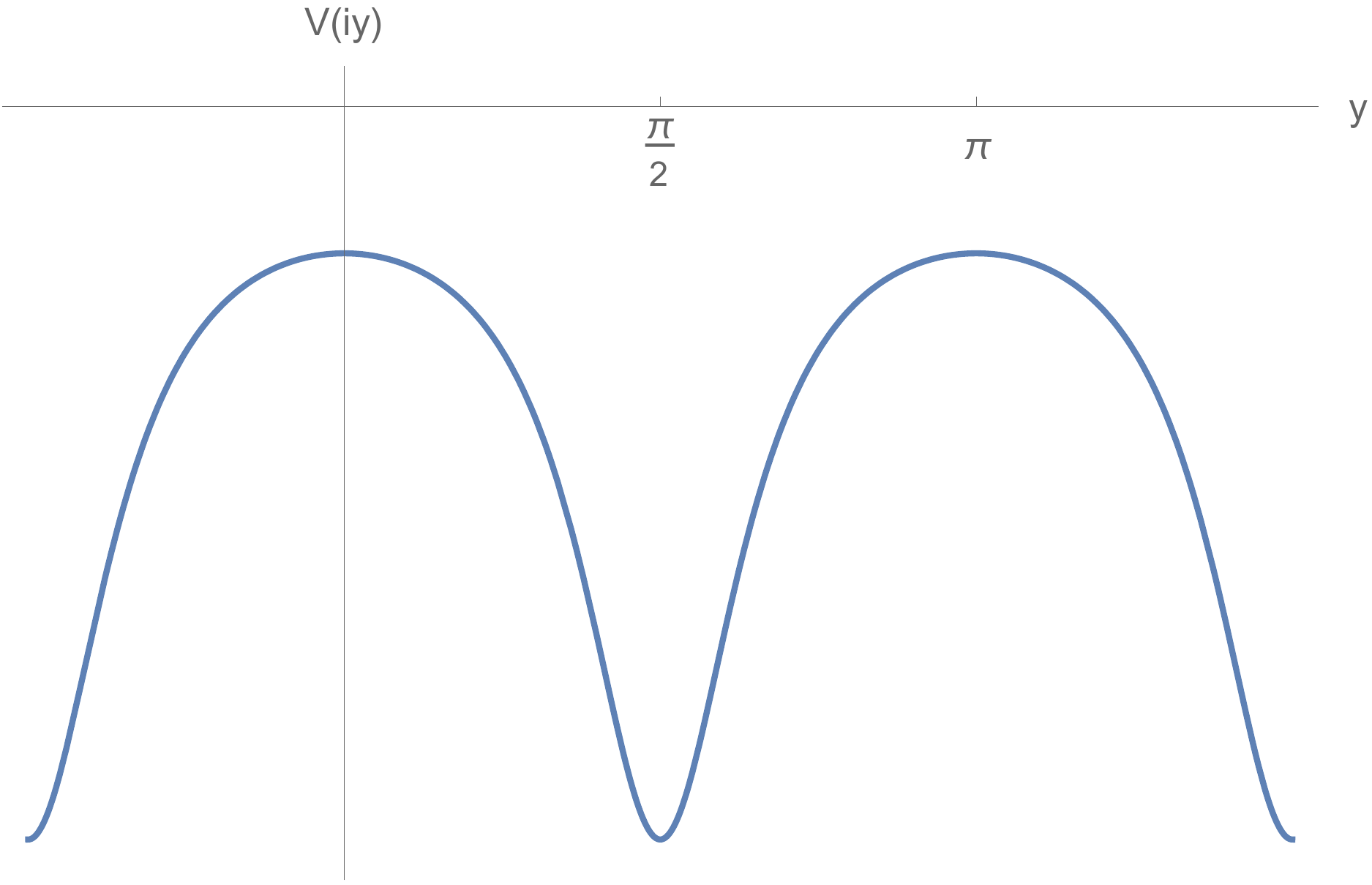}    
  \end{figure}
}
The two-point function of such operators
amounts to a choice of normalization, and the saddle-point
expansion of a three-point function is considerably more
challenging.
Some exact three-point functions with bound state insertions
have been computed exactly, however,
such as the correlator of
$\cO_{j_1=\frac{k}{2}-N,n_1=0,w_1=-1}$,
$\cO_{j_2=\frac{kw_2}{2},n_2=0,w_2}$, and
$\cO_{j_3,n_3=0,w_3=1-w_2}$ in \cite{Giribet:2016eqd}.
The result is independent of $N$ at leading order in the large $k$
limit, in support of the semi-classical picture we have described above. 

It would be very interesting to explore the implications of this
semi-classical definition of the bound state operators for the
infinitesimal version of the FZZ duality, which relates the
sine-Liouville operator to a deformation of the cigar that
shifts the value of the dilaton at the tip \cite{Giveon:2016dxe,epr}.
In particular, in
the Lorentzian continuation of the duality \cite{Giveon:2019gfk,Ben-Israel:2017zyi}
that will be further explored in \cite{epr}, the
above prescription describes a string that crosses the black hole horizon,
as $r=0$ is the bifurcation point. It is interesting that in the
presence of horizons such additional operators are required
beyond the ordinary scattering states to close the OPE, and
we hope to use these semi-classical methods to more
precisely understand string theory in Rindler space \cite{Witten:2018xfj}
and the stringy description of horizon entropy \cite{Susskind:1994sm}.

\section{sine-Liouville Limit}
\label{sec:sL}

The $\mrm{SL}(2,\reals)_k/\mrm{U}(1)$ CFT is defined for $k > 2$.
So far in this note we have focused on the $k \to \infty$
limit, where the cigar sigma-model provides a weakly-coupled
Lagrangian description of the CFT. In the opposite limit, namely $k - 2
\to 0$, the scalar curvature of the cigar diverges, and that
description becomes strongly-coupled.
However, there exists a dual description of the CFT that is 
better suited at small $k$
\cite{FZZ,Kazakov:2000pm}. In this section we consider the
saddle-point expansion in the $k-2\to 0$ limit using the dual
description. 

\begin{figure}[t]
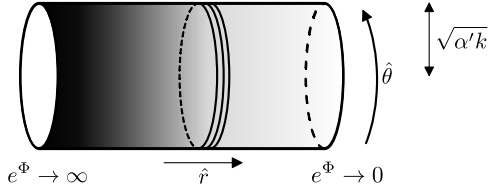

  \centering
    \centering
  \ig{0cm}{scale=.15}{sine-liouville}
  \caption{\footnotesize \bd{The sine-Liouville Background.} According to the
    FZZ duality, the sine-Liouville sigma-model is a dual
     description of the
    $\mrm{SL}(2,\reals)_k/\mrm{U}(1)$ CFT, better suited when $k-2$
    is small. The geometry is an infinite cylinder of radius
    $\sqrt{\alpha' k}$. The dilaton is $\Phi = - Q\hat r$,
    so that the string coupling $e^{\Phi}$ diverges as
    $\hat r \to -\infty$ and vanishes as $\hat r \to \infty$. The
    sine-Liouville potential $e^{- \sqrt{(k-2)/\alpha'}
      \hat r} \mrm{Re}\, e^{i 
    \sqrt{k/\alpha'}(\hat \theta_\mrm{L} - \hat
    \theta_\mrm{R})}$ includes a pure-winding
  mode of $\hat \theta$ (represented by the circles wrapping
  the middle of the cylinder), times a linear-dilaton primary 
  (represented by the color gradient). Alternatively, one may
  T-dualize the geometry to obtain a cylinder of radius
  $\sqrt{\alpha'/k}$. We denote the angular coordinates before
  and after the T-duality 
  by $\hat \theta$ and $\hat\vtheta$.} 
  \label{fig:sine-liouville}
\end{figure}

Recall from Eqn. \ref{eqn:linear-dilaton-circle} that the cigar
sigma-model approaches a free $\text{linear-dilaton}\times S^1$
background in the weak-coupling region, with canonically
normalized coordinates $\hat r$ and $\hat \theta \sim \hat
\theta + 2\pi \sqrt{\alpha' k}$. The asymptotically
linear-dilaton is $\Phi(\hat r) = -Q \hat r$,
where\footnote{Since we are no longer working in the large $k$ limit,
  here we use the exact value of $Q$, compared to
  $Q \overright{k\to\infty} \frac{1}{\sqrt{\alpha' k}}$ in
  Eqn. \ref{eqn:large-k-Q}. See also footnote \ref{foot:Q}.}
\begin{align}
  Q = \frac{1}{\sqrt{\alpha'(k-2)}}.
\end{align}
The $\text{linear-dilaton}\times S^1$ itself, with $\hat r \in
(-\infty,\infty)$ permitted to range over the entire line, is
ill-defined because the string coupling diverges as $\hat r \to
-\infty$. This strong coupling region is eliminated in the cigar
background by ending the geometry.
In the dual description,
the cigar is replaced by a fully infinite
$\text{linear-dilaton}\times S^1$ background, deformed by the
``sine-Liouville'' potential $V_\mrm{sL} \propto e^{-2b_\mrm{sL} \hat
r} \mrm{Re}\, e^{i \sqrt{\frac{k}{\alpha'}} (\hat \theta_\mrm{L} -
\hat \theta_\mrm{R})}$. The potential consists of a
Liouville-like radial factor  $e^{-2b_\mrm{sL}\hat r}$,
together with the unit-winding operator around the $S^1$
direction.
The linear-dilaton momentum,
\begin{align}
  b_\mrm{sL} = \frac{1}{2}\sqrt{\frac{k-2}{\alpha'}},
\end{align}
is chosen such that the potential is of weight $(1,1)$:
\begin{align}
  \alpha' b_\mrm{sL} (Q - b_\mrm{sL}) + \frac{k}{4} = 1.
\end{align}
At large $\hat r$, the potential decays and one recovers the
same asymptotic $\text{linear-dilaton}\times S^1$ theory as for
the cigar.
Note that the presence of the winding operator in the
sine-Liouville potential explicitly breaks the 
winding number symmetry around the cylinder of the free
theory. Likewise, the apparent winding conservation law in the
asymptotic region of the cigar is violated in the interior,
where the string can unwind at the tip. 
One thinks of the sine-Liouville background as being built up of a condensate
of winding strings on top of the cylinder,
as pictured in Fig. \ref{fig:sine-liouville}. The equivalence of
the sine-Liouville and cigar descriptions of the
$\mrm{SL}(2,\reals)_k/\mrm{U}(1)$ CFT is known as the FZZ duality
\cite{FZZ,Kazakov:2000pm}.

In light of the winding operator in the sine-Liouville
potential, the background is better written in terms of the
T-dual coordinate $\hat \vtheta \sim \hat\vtheta + 2\pi
\sqrt{\frac{\alpha'}{k}}.$ The action on a closed worldsheet
$\Sigma$ is 
\begin{align}
  S
  =& \frac{1}{4\pi \alpha'} \int_\Sigma \diff^2\sigma \sqrt{h}
     \lp
     (\del \hat r)^2 + (\del \hat \vtheta)^2
     + 4\pi \lambda e^{-2b_\mrm{sL}\hat r}
     \cos \lp \sqrt{\frac{k}{\alpha'}} \hat\vtheta \rp
     \rp\\
  &-\frac{Q}{4\pi} \int_\Sigma \diff^2 \sigma \sqrt{h}\, \cR[h] \hat r. \nt
\end{align}
The coefficient $\lambda$ is a positive number, analogous to the
coefficient $\mu$ of the Liouville potential
(c.f. Eqn. \ref{eqn:liouville-action}). One is again free to
add a constant mode to the dilaton, but it may be eliminated by
shifting $\hat r$ and rescaling $\lambda$.

As in Liouville, the linear-dilaton factor of the potential
$e^{-2b_\mrm{sL} \hat r}$ is weakly-coupled when $b_\mrm{sL}$ is
small, i.e. when $k$ is near 2. However, neither the original
cylinder radius $\sqrt{\alpha' k}$ nor its T-dual $\sqrt{\alpha'/k}$ is
large in that limit, and so the sine-Liouville background is not
strictly speaking weakly-coupled there. It is a far better
description of the coset for $k$ near 2 than the cigar, however,
which becomes infinitely strongly-coupled in the limit.

Because the asymptotic conditions for the coset operators discussed
in Sec. \ref{sec:cigar-asymptotics} in the cigar description
mapped the neighborhood of the insertion to the free-field
region where the cigar and sine-Liouville backgrounds coincide,
the same apply in sine-Liouville.
T-dualizing Eqn. \ref{eqn:ld-circle-green},
the asymptotic conditions for an insertion of $\cO_{jnw}$ in the far
past on the cylinder are
\begin{subequations}
  \begin{align}
    \label{eqn:r-asymp}
  &\hat r(\rho,\phi) \overright{\rho\to-\infty}
    2\alpha' Q\lp \frac{1}{2} - j \rp \rho + \cO(1)\\
  &\hat \vtheta(\rho,\phi)\overright{\rho\to-\infty}
    i w \sqrt{\alpha' k} \rho + n \sqrt{\frac{\alpha'}{k}} \phi
    + \cO(1).
\end{align}
\end{subequations}
In the previous discussion on the cigar, 
the (asymptotically) linear-dilaton played little role
as $k\to \infty$ because $Q$ vanished in the limit.
By contrast, $Q$ diverges as $k -2 \to 0$.
The background-charge operators, responsible for the shift by
$\frac{1}{2}$ in Eqn. \ref{eqn:r-asymp},  now behave as heavy
operators, scaling with the leading-order terms in the
action. Similarly, the $\cO_{jnw}$ insertion is itself a heavy
operator for $j$ of order one in the $k-2\to 0$ limit.

The action with insertions of $\cO_{jnw}$ in the far past and
$\cO_{j,-n,-w}$ in the far future is 
\begin{align}
  S_{jnw}
  =& \frac{1}{4\pi\alpha'} \int\limits_{-L}^L \diff \rho\,
     \int\limits_0^{2\pi}\diff \phi\,
     \lp (\partial_\rho \hat r)^2 + (\partial_\phi \hat r)^2
     + (\partial_\rho \hat \vtheta)^2 + (\partial_\phi \hat \vtheta)^2
     + 4\pi \lambda e^{-2b_\mrm{sL}\hat r}  \cos \lp
     \sqrt{\frac{k}{\alpha'}} \hat\vtheta \rp 
     \rp\nt\\
   &-\frac{2}{\sqrt{\alpha'(k-2)}}\lp j-\frac{1}{2}\rp
     \int\limits_0^{2\pi}\frac{\diff 
     \phi}{2\pi} \lp \hat r|_{\rho=L} +   \hat r|_{\rho = -L} \rp
     -iw\sqrt{\frac{k}{\alpha'}} \int\limits_0^{2\pi}\frac{\diff \phi}{2\pi}
     \lp \hat\vtheta|_{\rho=L} - \hat\vtheta|_{\rho = -L} \rp      
     \nt\\
   &+\int\limits_0^{2\pi} \frac{\diff \phi}{2\pi}
     \lp
      \sigma_+ \lp \partial_\phi \hat\vtheta|_{\rho = L} 
     -n \sqrt{\frac{\alpha'}{k}} \rp
      +\sigma_- \lp \partial_\phi \hat\theta|_{\rho = -L}  
     -n\sqrt{\frac{\alpha'}{k}} \rp
      \rp\nt\\
   & +\frac{4}{k-2}\lp j-\frac{1}{2} \rp^2 L
     - k w^2 L  - \frac{L}{k} n^2.
\end{align}
Note that the boundary action for $\hat\vartheta$ is
well-defined because $w \in \bbZ$. 

Let us again restrict our attention to the $n = 0$ sector,
where the $k-2 \to 0$ limit of the exact reflection coefficient
Eqn. \ref{eqn:winding-r} yields 
\begin{align}
  \label{eqn:sl-r}
  R(j,w)\overright{k-2 \to 0}
  &2^{4\lp j-\frac{1}{2} \rp}\lp j-\frac{1}{2}\rp
  \frac{\gamma(j+w)\gamma(j-w)}{\gamma(2j)}\\
  &\times\lp \frac{e}{2} \frac
    {k-2}{j-\frac{1}{2}}\rp^{\frac{4}{k-2}\lp j-\frac{1}{2}\rp}
  \csc\lp \frac{2\pi}{k-2} \lp j-\frac{1}{2}  \rp \rp\nt,
\end{align}
where $j = \cO(k^0)$ and $\mrm{Re}(j) > \frac{1}{2}.$
Note that the second line, which is the dominant contribution,
is independent of $w$. 

In this limit, the most interesting factor in
Eqn. \ref{eqn:winding-r} is $\gamma \lp \frac{2j-1}{k-2} 
\rp$, which leads to the csc factor of Eqn. \ref{eqn:sl-r}.
The latter arises in the saddle-point expansion
from the following shift symmetry of the sine-Liouville potential:
\begin{subequations}
\begin{align}
  &\hat r \to \hat r + \frac{\pi i}{2 b_\mrm{sL}}\\
  &\hat\vtheta \to \hat\vtheta + \pi \sqrt{\frac{\alpha'}{k}},
\end{align}
\end{subequations}
under which the linear-dilaton and compact-boson factors of the
potential each transform by a sign. By the same argument as in
Sec. \ref{sec:saddle-point}, the 
functional integral over real $\hat r$ diverges and should 
instead be defined over an appropriate complex cycle. We expect
that the cycle will consist of a sum of steepest-descent
contours associated to saddles related by the shift symmetry.
Under the shift, the action changes by
\begin{align}
  S\to S - \frac{4\pi i}{k-2} \lp j -
  \frac{1}{2} \rp,
\end{align}
due to the boundary terms. Summing over this discrete moduli space will
contribute
\begin{align}
  \sum_{N\in \bbZ_{\geq 0}} e^{ \frac{4\pi i}{k-2} \lp j -
  \frac{1}{2} \rp N}
  =\frac{i}{2}e^{-\frac{2\pi i}{k-2} \lp j - \frac{1}{2}\rp}
  \csc \lp \frac{2\pi}{k-2} \lp j - \frac{1}{2}\rp \rp
\end{align}
for $\mrm{Im}(j) > 0$, reproducing the $\csc$ in
Eqn. \ref{eqn:sl-r}. For $\mrm{Im}(j) < 0$, one sums over $N \in \bbZ_{\leq 0}$.

Because the sine-Liouville Lagrangian is not actually
weakly-coupled, it is more challenging to reproduce the rest of
Eqn. \ref{eqn:sl-r} by the saddle-point 
expansion. To attempt to extract the $\frac{1}{k-2}$ scaling from the action, 
one would define
\begin{align}
  &\til r = \sqrt{\frac{k-2}{\alpha'}} \hat r,\quad\quad
  \til \vtheta = \sqrt{\frac{k-2}{\alpha'}} \hat \vtheta,\quad\quad
  \til \lambda = \frac{k-2}{\alpha'}\lambda,
\end{align}
in terms of which
\begin{align}
  S_{jw}
  = \frac{1}{k-2}
     \Bigg\{&
     \frac{1}{4\pi} \int\limits_{-L}^L \diff \rho\,
     \int\limits_0^{2\pi}\diff \phi\,
     \lp (\partial_\rho \til r)^2 + (\partial_\phi \til r)^2
     + (\partial_\rho \til \vtheta)^2 + (\partial_\phi \til \vtheta)^2
     + 4\pi \til\lambda e^{-\til r}  \cos \lp
     \sqrt{\frac{k}{k-2}} \til\vtheta \rp 
     \rp\nt\\
   &-2\lp j-\frac{1}{2}\rp
     \int\limits_0^{2\pi}\frac{\diff 
     \phi}{2\pi} \lp \til r|_{\rho=L} +   \til r|_{\rho = -L} \rp
     -iw\sqrt{k(k-2)} \int\limits_0^{2\pi}\frac{\diff \phi}{2\pi}
     \lp \til\vtheta|_{\rho=L} - \til\vtheta|_{\rho = -L} \rp      
     \nt\\
   & +4\lp j-\frac{1}{2} \rp^2 L
     - k (k-2)w^2 L \Bigg\} .
\end{align}
Were in the functional in braces $\cO((k-2)^0)$, one could
proceed with the saddle-point expansion as in the preceding
sections. However, the sine-Liouville potential oscillates 
rapidly in this limit, reflecting the fact that the description
is not weakly-coupled.

We will not attempt to reproduce the rest of the semi-classical
limit using the sine-Liouville description. We point out,
however, that the second line of Eqn. \ref{eqn:sl-r} coincides
with the leading terms in the semi-classical limit of the
Liouville reflection coefficient. It was shown in
\cite{Ribault:2005wp} that winding-preserving $n$-point
functions in the $\mrm{SL}(2,\reals)_k/\mrm{U}(1)$ CFT are
reproduced by a sum of $2n-2$ point correlation functions in
Liouville. In particular, the two-point function of the coset is
simply related to the two-point function of Liouville, with a
certain dictionary described in \cite{Ribault:2005wp}, and one
correspondingly finds that their
semi-classical limits are closely related.

\section*{Acknowledgements} We would like to thank Juan Maldacena,  Edward Witten, and Xi Yin for stimulating and helpful discussions. This work was supported in part by NSFCAREER grant PHY-1352084.

\appendix
  \section{Exact Solution of the Cigar Quantum Mechanics}
\label{sec:cigar-qm}

In Sec. \ref{sec:saddle-point} we saw that the large $k$ limit
of the cigar action evaluated on a pure-winding solution $\theta
= -w\phi$ reduced to a quantum mechanics for $r$ with a
potential proportional to $\sech^2(r)$. This quantum mechanics
is exactly solvable,\footnote{See, for
  example, Landau and Lifshitz's Quantum Mechanics (Second
  Edition), Sections 23   and 25. We set $\hbar$ and the mass to
  one. The potential is often referred to as the modified
  P\"oschl-Teller potential.} as we review in this appendix.

\begin{figure}[H]
  \centering
  \ig{0cm}{width=\linewidth/3}{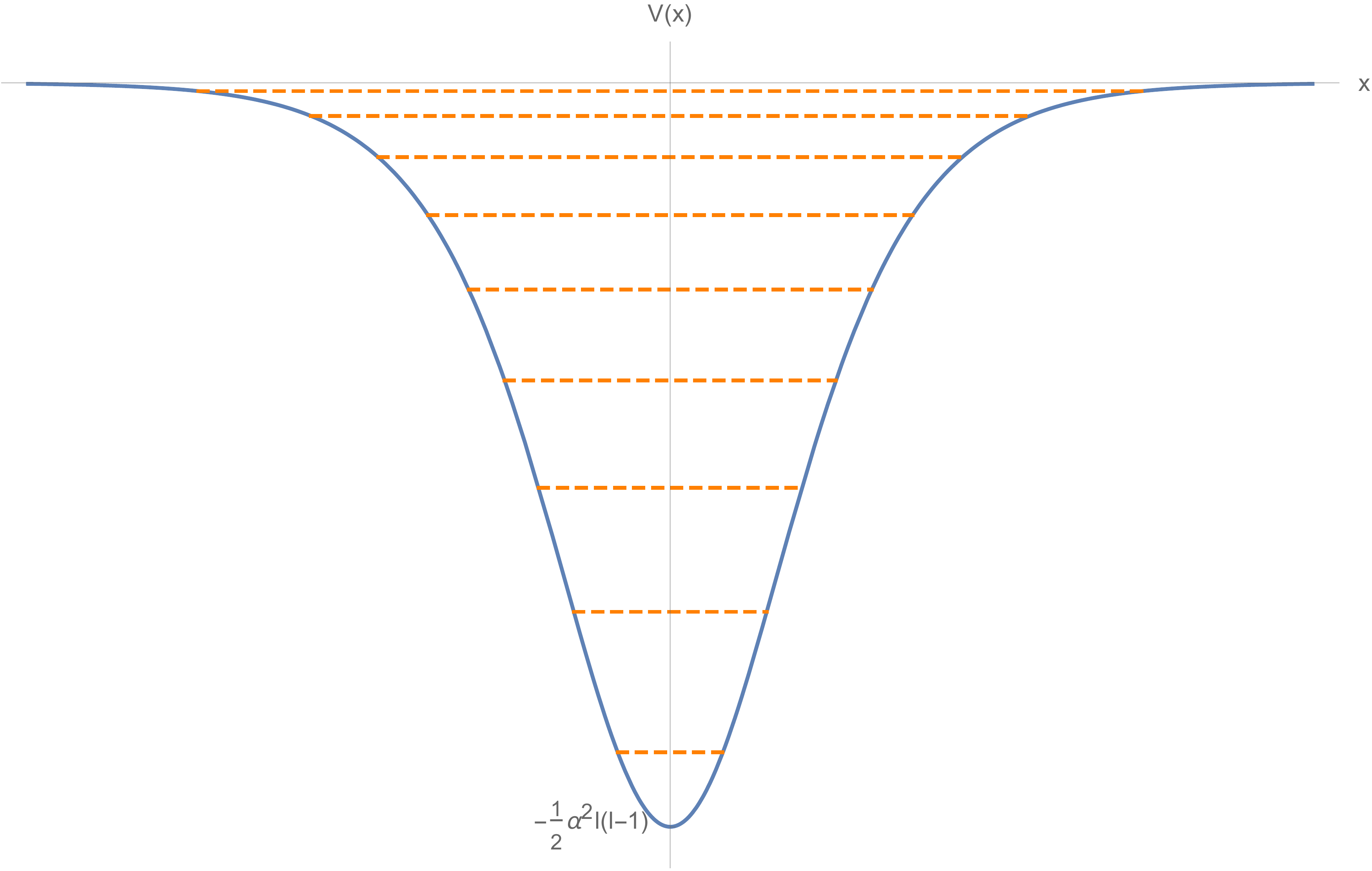}
  \caption{\footnotesize \bd{The Cigar Quantum Mechanics Potential}. }
    \label{fig:p-t-potential}
\end{figure}

The potential is usually written in the form
\begin{align}
  V(x) = - \frac{1}{2} \alpha^2 l(l-1) \mrm{sech}^2(\alpha x).
\end{align}
with $\alpha>0$ and $l>1$. It is a symmetric well of
depth $\frac{1}{2} \alpha^2 l(l-1)$, and it vanishes as $x\to
\pm \infty$, as pictured in Fig. \ref{fig:p-t-potential}. It 
therefore admits both bound states and scattering states. To
begin we consider the quantum mechanics on an infinite line, $x
\in \reals$. The cigar is related to its $\mrm{Z}_2$ quotient $x
\sim -x$. 

Consider first the scattering states. We look for
solutions of
\begin{align}
  \label{eqn:scattering}
  -\frac{1}{2} \psi''(x) + V(x) \psi(x) = \frac{p^2}{2} \psi(x),
\end{align}
behaving asymptotically as
\begin{align}
  \psi(x) \to
  \begin{dcases}
    e^{i p x} + R(p) e^{-i p x}& x \to - \infty\\
    T(p) e^{i p x} & x \to \infty.
  \end{dcases}
\end{align}
The
two linearly independent solutions of this equation are the
associated Legendre polynomials
$P_{l-1}^{ip/\alpha}(\tanh(\alpha x))$ and
$Q_{l-1}^{ip/\alpha}(\tanh(\alpha x))$. The asymptotics of the
$P$ function are
\begin{align}
  P_{l-1}^{ip/\alpha}&(\tanh(\alpha x))
  \\
  &\to\begin{dcases}
    \frac{i \pi \csch\lp \frac{\pi p}{\alpha}\rp}
    {
      \Gamma \lp 1 + \frac{i p}{\alpha} \rp
      \Gamma \lp l - \frac{i p}{\alpha} \rp
      \Gamma \lp 1- l - \frac{i p}{\alpha} \rp
    }
    e^{i p x}
    -
    \frac{i \sin(\pi l) \csch\lp \frac{\pi p}{\alpha} \rp}
    {\Gamma \lp 1 - \frac{i p}{\alpha}\rp}
    e^{-i p x}
    & x \to -\infty\\
    \frac{1}{\Gamma \lp 1 - \frac{ip}{\alpha} \rp}
    e^{i p x}
    &x \to \infty.
  \end{dcases}\nt
\end{align}
The asymptotics of the $Q$ function, on the other hand, contain
$e^{i px}$ and $e^{-i px}$ at both limits, and must
be discarded. The scattering wavefunction
is then
\begin{align}
  \psi(x;p)
  =-\frac{i}{\pi}
  \sinh\lp \frac{\pi p}{\alpha}\rp
  \Gamma \lp 1 + \frac{i p}{\alpha} \rp
  \Gamma \lp l - \frac{i p}{\alpha} \rp
  \Gamma \lp 1- l - \frac{i p}{\alpha} \rp
  P_{l-1}^{ip/\alpha}&(\tanh(\alpha x)),
\end{align}
yielding the reflection and transmission coefficients
\begin{align}
  \label{eqn:qm-r}
  R(p) = -\frac{1}{\pi} \sin(\pi l)
  \frac{\Gamma \lp 1 + \frac{i p}{\alpha} \rp}
  {\Gamma \lp 1 - \frac{i p}{\alpha} \rp}
  \Gamma \lp l - \frac{i p}{\alpha} \rp
  \Gamma \lp 1- l - \frac{i p}{\alpha} \rp
\end{align}
and
\begin{align}
  \label{eqn:qm-t}
  T(p)= 
  -\frac{i}{\pi} \sinh\lp \frac{\pi p}{\alpha}\rp
  \frac{\Gamma \lp 1 + \frac{i p}{\alpha} \rp}
  {\Gamma \lp 1 - \frac{i p}{\alpha} \rp}
  \Gamma \lp l - \frac{i p}{\alpha} \rp
  \Gamma \lp 1- l - \frac{i p}{\alpha} \rp.
\end{align}
Note this potential has the remarkable property that it is
reflectionless when $l$ is an integer;
$R(p)$ vanishes 
due to the factor of $\sin(\pi l)$. In that case, the
transmission coefficient may be written
\begin{align}
  T(p)\bigg|_{l \in \bbZ}
  =
  \prod_{n=1}^{l-1} \frac{l-n-\frac{ip}{\alpha}}{n-l-\frac{ip}{\alpha}}
\end{align}
by repeatedly applying the factorial property of the Gamma function, $\Gamma(z+1)
= z \Gamma(z)$. In particular, $T(p)$ is a pure phase,
\begin{align}
  |T(p)|^2\bigg|_{l\in \bbZ} = \prod_{n=1}^{l-1}
  \frac{(l-n)^2+\frac{p^2}{\alpha^2}}{(n-l)^2+\frac{p^2}{\alpha^2}}=1,
\end{align}
as required by probability conservation, $|R|^2+|T|^2 = 1$.

Meanwhile, the bound states are solutions of 
\begin{align}
  -\frac{1}{2} \psi''(x) + V(x) \psi(x) = E\psi(x),
\end{align}
with $-\frac{1}{2} \alpha^2 l(l-1)<E < 0$. They may be obtained
from the scattering solutions by continuing $p = i
\sqrt{2|E|}\in i \reals_+,$
so that  $\psi(x;i\sqrt{2|E|}) \overright{x \to \infty}  T
e^{-\sqrt{2|E|}x}$ decays. As $x \to -\infty$, 
\begin{align}
  \psi(x; i \sqrt{2|E|}) \overright{x \to -\infty}
  e^{-\sqrt{2|E|} x}+ Re^{\sqrt{2|E|}x},
\end{align}
which generically diverges, unless $p = i \sqrt{2|E|}$ is such
that $R$ has a pole. At those discrete points, one may hope to
find a normalizable bound state proportional to $\frac{1}{R} \psi(x;
i \sqrt{2|E|}).$

$R(p)$ has three sets of simple poles due to the three Gamma
functions in its numerator. The first, $\Gamma \lp 1 + \frac{ip}{\alpha}
\rp$, has poles for $p = i \alpha (n + 1)$, with $n$ a natural
number. These do not correspond to bound states, however,
because in the $x \to -\infty$ limit of the $P$ function it is
the ratio
$\csch \lp \frac{\pi p}{\alpha}\rp/\Gamma \lp 1 +
  \frac{ip}{\alpha} \rp$ that appears,
which is regular. The second Gamma function in $R(p)$, $\Gamma \lp l -
\frac{ip}{\alpha} \rp$, has poles for $p = -i\alpha(l+n)$, but
these do not belong to the domain $i \reals_+$, which was
necessary for convergence at large $x$.

It is instead the last Gamma function which is responsible for
the bound states, $\Gamma\lp 1 - l - \frac{ip}{\alpha} \rp$. The
poles are found at
\begin{align}
  p_n = i \alpha (l-1-n),
\end{align}
which belong to $i \reals_+$ provided $n < l-1$. Thus we find
the spectrum of bound state energies
\begin{align}
  E_n = \frac{p_n^2}{2} = -\frac{\alpha^2}{2}(l-1-n),\quad 0\leq
  n < l-1,
\end{align}
with wavefunctions
\begin{align}
  \label{eqn:bound-state-wavefunctions}
  \psi_n(x) = P_{l-1}^{-(l-1-n)}(\tanh(\alpha x)).
\end{align}
Next consider the semi-classical limit. Define
\begin{align}
  \til{x} \equiv \alpha x,\quad
  \til{l} \equiv \alpha^2 \sqrt{l(l-1)},
\end{align}
in terms of which the Hamiltonian may be written
\begin{align}
  H = \frac{1}{\alpha^2}
  \lp \frac{1}{2} \lp \td{\til{x}}{t}\rp^2
  +\til{V}(\til{x})  \rp,
\end{align}
where
\begin{align}
  \til{V}(\til{x}) \equiv -\frac{1}{2} \til{l}^2
  \mrm{sech}^2(\til{x}).
\end{align}
Comparing to Eqn. \ref{eqn:qm-potential}, we find the same
quantum mechanics as the pure-winding sector of the cigar CFT,
with the dictionary
$w=\til l $ and $k = \frac{1}{\alpha^2}$.

The semi-classical limit is $\alpha \to 0$ with $\til{l}$
fixed. In this limit the bound state spectrum is
\begin{align}
  E_n \to - \frac{\alpha^2}{2} \lp n- \frac{\til{l}}{\alpha^2} \rp^2.
\end{align}
The same semi-classical spectrum may be obtained from the WKB
approximation, which says that
\begin{align}
  \int_{-x_*}^{x_*}\diff x\,  \sqrt{2(E_n - V(x))} = \pi n,
\end{align}
where
\begin{align}
  x_* = \frac{1}{\alpha} \cosh^{-1} \lp \sqrt{ \frac{\frac{1}{2}
  \alpha^2 l(l-1)}{-E_n} } \rp
\end{align}
is the classical turning point, $V(\pm x_*) = E_n$. The integral is 
\begin{align}
  \int_{-x_*}^{x_*}\diff x\,  \sqrt{2(E_n - V(x))}
  = \pi \lp \sqrt{l(l-1)} - \frac{1}{\alpha}\sqrt{-2E_n} \rp,
\end{align}
from which we obtain
\begin{align}
  E_n \approx - \frac{\alpha^2}{2} \lp n - \sqrt{l(l-1)} \rp^2,
\end{align}
reproducing the semi-classical limit of the exact spectrum. 

As for the scattering states, define
\begin{align}
  p \equiv i\frac{\eta}{\alpha},
\end{align}
in terms of which the exact reflection and transmission
coefficients may be written
\begin{align}
  R(\eta) =
  \frac{1}{\pi} \frac{\til l - \eta}{\eta }
  \frac{\sin(\pi k \til l)}{\gamma(k \eta)}
  \Gamma(k (\til l + \eta)) \Gamma(-k(\til l - \eta))
\end{align}
and
\begin{align}
  T(\eta) =
  -\frac{1}{\pi} \frac{\til l - \eta}{\eta }
  \frac{\sin(\pi k \eta)}{\gamma(k \eta)}
  \Gamma(k (\til l + \eta)) \Gamma(-k(\til l - \eta)),
\end{align}
where again  $k \equiv \frac{1}{\alpha^2}$.
Applying Eqns. \ref{eqn:Gamma-asymp}-\ref{eqn:gamma-asymp} we
find in the semi-classical limit
\begin{align}
  \label{eqn:qm-r-semi}
  R(\eta) \overright{k\to\infty}
  & \eta^{-2k \eta}(\til l+\eta)^{k(\til l+\eta)}
    \sin(\pi k \til l)
    \csc\lp \pi k \eta\rp\\
  &\times
    \begin{dcases}
      -\frac{1}{2}(\til l-\eta)^{k(\eta-\til l)}
    \csc \lp \pi k (\til l -\eta) \rp
    &0< \mrm{Re}(\eta) < \til l\\
    -i (\eta-\til l)^{k(\eta-\til l)}
   &\mrm{Re}(\eta) > \til l
 \end{dcases}\nt\\
  &\times  \sqrt{\frac{\til l - \eta}{\til l + \eta}}\nt
\end{align}
and
\begin{align}
  \label{eqn:qm-t-semi}
  T(\eta) \overright{k\to\infty}
  & \eta^{-2k \eta}(\til l+\eta)^{k(\til l+\eta)} \\
  &\times
    \begin{dcases}
      \frac{1}{2}(\til l-\eta)^{k(\eta-\til l)}
    \csc \lp \pi k (\til l -\eta) \rp
    &0< \mrm{Re}(\eta) < \til l\\
    i (\eta-\til l)^{k(\eta-\til l)}
   &\mrm{Re}(\eta) > \til l
 \end{dcases}\nt\\
  &\times  \sqrt{\frac{\til l - \eta}{\til l + \eta}}.\nt
\end{align}
The bound states now correspond to the poles of the $\csc(\pi k
(\til{l}-\eta))$ factors.

So far we have considered the quantum mechanics on a
fully-infinite line. However, the cigar quantum mechanics
obtained in Sec. \ref{sec:saddle-point} by setting $\theta = -w
\phi$ was defined on a half-line. Returning to
Eqn. \ref{eqn:scattering}, the scattering solutions on a half-line are now the
linear combinations of Legendre polynomials that vanish at the
origin:
\begin{align}
  \psi_{1/2}(x;p)
  =
  &2
  \frac{
  \Gamma \lp l - \frac{i p}{\alpha} \rp
  \Gamma \lp 1- l - \frac{i p}{\alpha} \rp}
  {\Gamma \lp -\frac{i p}{\alpha} \rp}
  \cos^2 \lp \frac{\pi}{2} \lp l + \frac{i p}{\alpha} \rp \rp\\
  &\times\lp
  P_{l-1}^{ip/\alpha}(\tanh(\alpha x))
  - \frac{2}{\pi}
  \tan \lp \frac{\pi}{2} \lp l + \frac{i p}{\alpha} \rp \rp
  Q_{l-1}^{ip/\alpha}(\tanh(\alpha x))
  \rp\nt.
\end{align}
It behaves as
\begin{align}
  \psi_{1/2}(x;p) \overright{x \to - \infty} e^{i p x} +
  R_{1/2}(p) e^{-i p x},
\end{align}
where the reflection coefficient for the half-line problem is
\begin{align}
  \label{eqn:half-r}
  R_{1/2}(p) =
  &2\frac{
  \Gamma\lp l - \frac{ip}{\alpha} \rp
  \Gamma \lp 1 - l - \frac{ip}{\alpha} \rp}
  {\Gamma \lp 1 -
  \frac{ip}{\alpha} \rp
  \Gamma \lp -\frac{ip}{\alpha} \rp}\\
  &\times
  \sin\lp \frac{\pi}{2} \lp l - \frac{ip}{\alpha} \rp \rp
  \cos\lp \frac{\pi}{2} \lp l + \frac{ip}{\alpha} \rp \rp
  \csc\lp \frac{\pi i p}{\alpha} \rp.\nt
\end{align}
Alternatively, having already solved the theory on a line, the
solution on the half-line is given by its quotient with respect
to the reflection symmetry $x \sim - x$. The reflection
coefficient $R_{1/2}(p)$ is then the difference of the reflection and
transmission coefficients $R(p)$ and $T(p)$,
\begin{align}
  R_{1/2}(p) = R(p) - T(p),
\end{align}
as can be checked for Eqns. \ref{eqn:qm-r}, \ref{eqn:qm-t}, and
\ref{eqn:half-r}, and the bound state spectrum is given by
the odd solutions
\begin{align}
  E_n = -\frac{\alpha^2}{2}(l-1-n)^2,\quad
  n=1,3,5,\ldots < l-1. 
\end{align}
Previously, we identified the bound states with the poles
$p_n = i \alpha(l-1-n)$ of
$\Gamma\lp 1 - l -\frac{ip}{\alpha} \rp$. Now we find
\begin{align}
  R_{1/2}(p_n) =
  \frac{
  \Gamma\lp 2l-1-n \rp
  }
  {\Gamma(l-n)\Gamma \lp l-1-n \rp}
  \Gamma \lp -n \rp\lp (-)^n -1 \rp. 
\end{align}
Whereas $R(p_n)\supset\Gamma(-n)$ was singular for all $n = 0,1,2,\ldots$, 
the additional factor of $(-)^n-1$ in $R_{1/2}(p_n)$ eliminates
the poles with even $n$.

Finally, let us compute the semi-classical limit of
$R_{1/2}$. With the same notation as before we may write
\begin{align}
  R_{1/2}(\eta) =
  &2 \frac{\til l - \eta}{\eta}
  \frac{\Gamma(k(\til l + \eta)) \Gamma(-k(\til l -
  \eta))}{\Gamma(k \eta)^2}\\
   &\times
     \sin\lp \frac{\pi}{2} k (\til l + \eta) \rp
     \cos\lp \frac{\pi}{2} k (\til l - \eta) \rp
     \csc\lp \pi k \eta \rp.\nt
\end{align}
At large $k$ we obtain
\begin{align}
  \label{eqn:half-semi}
  R_{1/2}(\eta) \overright{k\to\infty}
  & \eta^{-2k \eta}(\til l+\eta)^{k(\til l+\eta)}
    \csc\lp \pi k \eta\rp
    \sin\lp \frac{\pi}{2} k (\til l + \eta) \rp\\
  &\times
    \begin{dcases}
      -\frac{1}{2}(\til l-\eta)^{k(\eta-\til l)}
    \csc \lp \frac{\pi}{2} k (\til l -\eta) \rp
    &0< \mrm{Re}(\eta) < \til l\\
    -2i (\eta-\til l)^{k(\eta-\til l)}
    \cos\lp \frac{\pi}{2} k (\til l - \eta) \rp
   &\mrm{Re}(\eta) > \til l
 \end{dcases}\nt\\
  &\times  \sqrt{\frac{\til l - \eta}{\til l + \eta}}\nt.
\end{align}
Compared to Eqn. \ref{eqn:qm-r-semi}, the factor of $\csc (\pi k(\til
l - \eta))$ has been replaced by $\csc\lp \frac{\pi}{2} k (\til
l - \eta) \rp$, reflecting the smaller set of bound states. 

Looking back at the large $k$ limit of the exact coset
reflection coefficient Eqn. \ref{eqn:cigar-semi-r}, we find 
agreement with Eqn. \ref{eqn:half-semi} to order $e^{k}$.\footnote{Note
that the discrepancy between $\cos\lp \frac{\pi}{2} k (\til l -
\eta) \rp$ and $\sin\lp \frac{\pi}{2} k (\til l - \eta) \rp$ is
order $e^{k^0}$.}
We conclude
that the restriction to the cigar quantum mechanics in
Sec. \ref{sec:saddle-point} is sufficient to extract the
saddle-point expansion of the reflection coefficient,
i.e. $\theta = - w \phi$ is the only saddle of the $\theta$
equations of motion that contributes to the expansion.


\bibliographystyle{utphys}

\bibliography{bib_cigar}

\end{document}